\documentclass[pdflatex,sn-mathphys-num]{sn-jnl}

\usepackage{graphicx}%
\usepackage{multirow}%
\usepackage{amsmath,amssymb,amsfonts}%
\usepackage{amsthm}%
\usepackage{mathrsfs}%
\usepackage[title]{appendix}%
\usepackage{xcolor}%
\usepackage{textcomp}%
\usepackage{manyfoot}%
\usepackage{booktabs}%
\usepackage{algorithm}%
\usepackage{algorithmicx}%
\usepackage{algpseudocode}%
\usepackage{listings}%
\usepackage{amsmath}
\usepackage{array}
\usepackage{ amssymb }
\usepackage{braket}
\usepackage{qcircuit}
\usepackage{soul}
\usepackage{braket}

\theoremstyle{thmstyleone}%
\theoremstyle{thmstyletwo}%

\theoremstyle{thmstylethree}%

\begin{document}

\title{The emptiness formation probability, and representations for nonlocal correlation functions, of the 20-vertex model}
\author{Pete Rigas \footnote{Cornell AEP and Mathematics Departments, pbr43@cornell.edu, ORCID: 0009-0003-1053-9720}}
 
\date{\today}

\abstract{
             We study the emptiness formation probability, along with various representations for nonlocal correlation functions, of the 20-vertex model. In doing so, we leverage previous arguments for representations of nonlocal correlation functions for the 6-vertex model, under domain-wall boundary conditions, due to Colomo, Di Giulio, and Pronko, in addition to the inhomogeneous, and homogeneous, determinantal representations for the 20-vertex partition function due to Di Francesco, also under domain-wall boundary conditions. By taking a product of row configuration probabilities, we obtain a desired contour integral representation for nonlocal correlations from a determinantal representation. Finally, a counterpart of the emptiness formation probability is introduced for the 20-vertex model. }

\maketitle

\noindent \textbf{{Keywords}}: Statistical physics, square ice, ice rule, triangular ice, twenty-vertex model, quantum inverse scattering, Poisson structure, integrability, action-angle variables

\bigskip

\noindent \textbf{MSC Class}: 34L25; 60K35

\section{Introduction}

\subsection{Overview}

Integrability has long remained a topic of investigation for vertex {\color{blue}[34]}, and closely related, classes of models, with efforts being devoted towards combinatorics {\color{blue}[15]}, limit shapes {\color{blue}[35]}, exact solvability {\color{blue}[2]}, the Bethe ansatz {\color{blue}[12},{\color{blue}20]}, interactions with statistical and mathematical physics {\color{blue}[9]}, the free energy landscape {\color{blue}[11]}, encoding of boundary conditions {\color{blue}[13},{\color{blue}14]}, Hamiltonian methods {\color{blue}[16]}, conformal invariance {\color{blue}[17]}, inhomogeneities {\color{blue}[19]}, Riemann-Hilbert approaches {\color{blue}[28]}, Heisenberg, and $D^{(2)}_3$, spin chains {\color{blue}[25},{\color{blue}27},{\color{blue}42]}, the existence of phase transitions {\color{blue}[39]}, discrete holomorphicity {\color{blue}[40]}, symmetry {\color{blue}[29},{\color{blue}44]}, amongst a myriad of other topics {\color{blue}[1},{\color{blue}3},{\color{blue}4},{\color{blue}8},{\color{blue}9},{\color{blue}18},{\color{blue}21},{\color{blue}23]}. To quantify the impact of staggering on vertex models within the quantum inverse scattering formalism, as investigated in {\color{blue}[23]} for results on the continuum limit, besides two and three dimensional approaches provided in {\color{blue}[41]}, and in {\color{blue}[45]}, from integrability of inhomogeneous limit shapes {\color{blue}[23]}, one can also further examine determinantal structures arising within Integrable probability. In comparison to notions within Discrete Probability that can be used to study vertex models, from Russo-Seymour-Welsh, and crossing probabilities {\color{blue}[37},{\color{blue}38]}, under sufficiently flat, or other encodings, of boundary conditions {\color{blue}[10]} that can also be studied in terms of sloped boundary conditions {\color{blue}[36]}, in addition to the Ashkin-Teller, generalized random-cluster, and $\big( q_{\sigma} , q_{\tau} \big)$-spin models, other intriguing aspects of vertex models can be examined from the quantum inverse scattering method, and Integrable Probability, which are realized through algebraic, geometric, and combinatorial qualities of three-dimensional L-operators, up to Dynkin automorphism {\color{blue}[5},{\color{blue}6]}. Besides well characterized aspects of Schur, and Markov, processes which can be expressed in terms of determinants, such observations play an extremely important role in seminal work of Di Francesco - not only for obtaining a homogenized determinantal representation for the 20-vertex partition function under domain wall boundary conditions over the pentagonal lattice, but also for relating this partition function to "U-turn" boundary conditions for the 6-vertex model in which neighboring horizontal lines of $\textbf{Z}^2$ are connected together {\color{blue}[15]}. In comparison to domain wall boundary conditions for the 6-vertex model, such boundary conditions for the 20-vertex model, over the triangular lattice, admit determinantal representations, with the determinant being taken of a matrix that is a function of the 20-vertex weights {\color{blue}[15]}, which could potentially shed light on possible behaviors for correlations {\color{blue}[26]}, entropy, besides a seminal computations for $a \equiv b \equiv c \equiv 1$, {\color{blue}[30]}, amongst 'quantizations' of integrability, through quantum integrability {\color{blue}[31]}.

For the 6-vertex model under domain wall boundary conditions, over the finite lattice integral representations have been previously obtained in {\color{blue}[7]}, which predominantly rely upon the computation of top, and bottom, partition functions. Equipped with a two-dimensional L-operator expressed in terms of Pauli basis elements, the authors of the aforementioned work make use of two partition functions to recursively compute, as a summation of products, finite representations of the correlation functions. As is the case in previous works on vertex models that were cited in the previous paragraph, knowing the composition of finite representation for such correlation functions has implications for exact solvability, amongst several other characteristics. In terms of the quantum inverse scattering method, seminal work for Hamiltonian systems, {\color{blue}[16]}, has remained of great interest for potential extensions that were suggested in the last sentence of {\color{blue}[24]}, not only in terms of inhomogeneities for the 6-vertex model, {\color{blue}[41]}, but also for the 20-vertex model {\color{blue}[45]}. In the setting of the 20-vertex model, it is attractive to not only determine how the composition of such correlations, as companions to their 6-vertex counterparts obtained in {\color{blue}[7]}, but also to make use of Di Francesco's determinantal representation for the partition function.

For such an effort, one not only needs to make use of inhomogeneities for the determinantal representation of the partition function, as the 6-vertex partition function described in {\color{blue}[7]} is inhomogeneous, but also to compute additional partition functions rather than the top and bottom ones that are used to recursively define correlations in finite volume. As such, beginning in the next section, as an overview of the objects for the 6-vertex, and 20-vertex, models, we define objects associated with each vertex model and aspects of the corresponding determinantal representations which differ. Despite the fact that such objects that will be introduced in the next section have been examined within the quantum inverse scattering method, they have not yet been examined within the context of obtaining integral representations for correlation functions.

In general, determining closed-form representations, through contour integrals, for correlations of the 20-vertex model is of significance for many reasons, including the fact that: (1) three-dimensional representations for correlations of the 20-vertex model, in comparison to two-dimensional representations for correlations of the 6-vertex model, rely upon determinantal representations of the partition function that are of combinatorial relevance for the Azetc diamond; (2) determinantal representations for partition functions of the 20-vertex model, in comparison to those of the 6-vertex model, have more intricate commutation relations, inherently related to the underlying Yang-Baxter algebra that has previously been characterized within the QISM framework by the author, {\color{blue}[45]}; (3) being able to leverage contour integral formulas for correlations, to a three-dimensional analog of the emptiness formation probability; (4) broadening the scope of vertex models, and admissible boundary conditions, in which the transfer and quantum monodromy matrices play a fundamental role. With all of these considerations in mind, the forthcoming contour integral representation that is obtained for correlation functions of the 20-vertex model is valuable for: (1) determining how the structure of said correlations can differ for other classes of boundary conditions, rather than for domain-wall boundary conditions; (2) elucidating how higher-dimensional analogues of 6-vertex correlation functions are expected to generalize; (3) contributing to the possible range of behaviors of the 6-vertex, and te20-vertex, models, in respective phase diagrams for each model. As such, determining whether additional modifications to the contour integral representations for correlations obtained in this work for the 20-vertex model can be transformed into correlation functions of other models remains of great interest to explore. For such classes of correlation functions, one can expect that the collection of singularities which are introduced for evaluating contour integral representations in correlation functions for the 20-vertex model can be related to those for other models, including the Solid-on-Solid, models, amongst other possibilities.

\begin{figure}
\begin{align*}
\includegraphics[width=0.75\columnwidth]{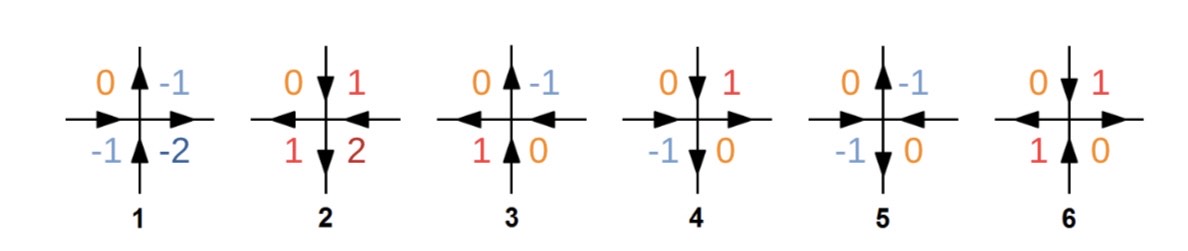}
\end{align*}
\caption{One depiction of each possible vertex for the six-vertex model, adapted from {\color{blue}[11]}.}
\end{figure}

\begin{figure}
\begin{align*}
\includegraphics[width=0.75\columnwidth]{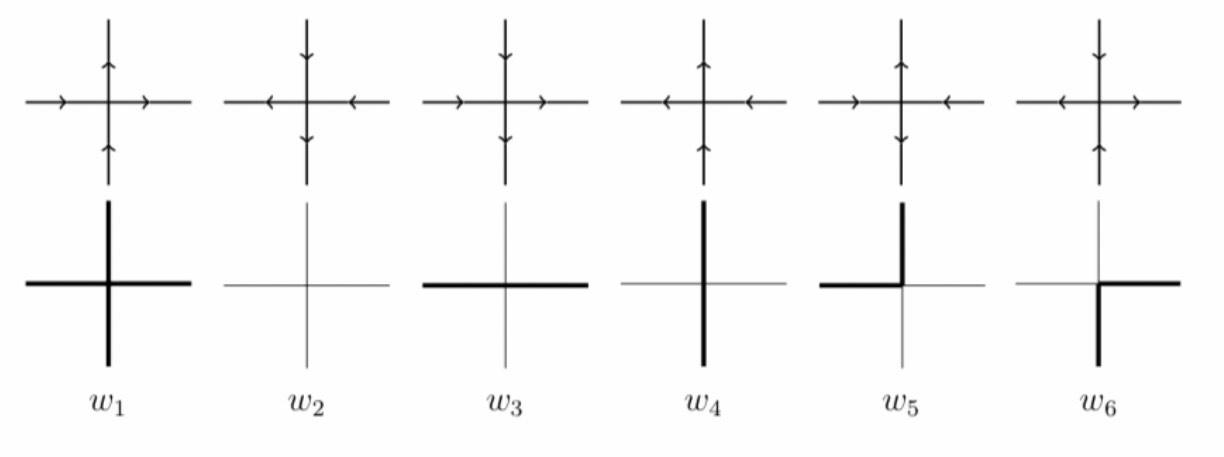}
\end{align*}
\caption{Another depiction of each possible vertex for the six-vertex model, adapted from {\color{blue}[24]}.}
\end{figure}

\subsection{This paper's contributions}

\noindent This paper provides expressions, from contour integral representations, of correlation functions for the 20-vertex model. Given the fact that the 20-vertex model is a generalization of the 6-vertex model, it is of great interest, along the lines of many arguments in Statistical Mechanics, to determine whether correlation functions, and different phase of behavior, of vertex models can be related to each other. For the 6-vertex and 20-vertex models, domain-wall boundary conditions are classes of boundary conditions that have been examined thoroughly in past work. Such boundary conditions are particularly helpful, for expressing correlation functions supported over the triangular lattice for the 20-vertex model. Furthermore, the existence of a suitable encoding of domain-wall boundary conditions for the 20-vertex model, as is the case for the 6-vertex model, makes it possible to relate correlation functions of each model to each other. 

Besides being able to establish several associations between correlation functions of vertex models, the construction of such functions reflects upon the respective phase diagrams of each model. While ferrorelectric, and antiferroelectric, regions of the conjectured phase diagram for the 6-vertex model relate to classical properties of the Ising model, determining how residues of correlation functions, which are manipulated in contour integral representations, depend upon the underlying state space of the model. For the 20-vertex model, while significant developments are still needed to characterize the phase diagram, and possible behaviors for correlation functions, amongst other observables, the contour integral representations in this work provide a glimpse into connections between Mathematical Physics, and Statistical Physics. Between such fields is a novel adaptation of the QISM framework, described extensively in another work oof the author, {\color{blue}[45]}, which underlies the computation of the correlation function. While previous adaptations have made use of the transfer matrix, and quantum monodromy, matrix, for obtaining representations for correlation functions oft he 6-vertex model under domain-wall boundary conditions, {\color{blue}[7]}, the fact that such computations remain applicable for other vertex models leaves open the possibility of being able to explore more connections between Integrable, and Discrete, Probability.

\subsection{Objects from Integrable Probability}

\noindent With the following objects from Integrable Probability, we discuss how the probability measure for the 6-vertex model informs structures of its correlation functions. From the 6-vertex probability measure, correlation functions were obtained in the previous work, {\color{blue}[7]}, by restricting the support of $\textbf{Z}^2$ to either top, or bottom, partition functions which respectively correspond to partitions of $\textbf{Z}^2$ into a top, and bottom, half. After introducing the 6-vertex probability measure, the higher-dimensional counterpart over the triangular lattice, with the 20-vertex model, the support of the correlation functions depends upon three partitions of the triangular lattice. The appropriate partition of the lattice is reflected through a product of three partition functions, $Z^{\mathrm{Bottom}} Z^{\mathrm{Top}} Z^{\mathrm{Side}}$, which is introduced later for a function that is integrated against with contour integrals for obtaining expressions for correlation functions, and the emptiness formation probability.

To this end, fix an instance of domain wall boundary conditions $\xi$, along with parameters $a_1, a_2, b_1, b_2, c_1, c_2 >0$, and the number of vertices of six different types, $n_1$, $n_2$, $n_3$, $n_4$, $n_5$, and $n_6$, respectively. Over the torus $\textbf{T}_{N} \equiv \big( V \big( \textbf{T}_N \big) , E \big( \textbf{T}_N \big) \big)$, the six-vertex model can be defined through the probability measure,

\begin{align*}
 \textbf{P}^{\xi}_{\textbf{T}_N} \big[ \omega \big] \equiv \textbf{P}^{\xi} \big[ \omega \big] \equiv     \frac{w \big( \omega \big)}{Z^{\xi}_{\textbf{T}_N}}           \text{ } \text{ , } 
\end{align*}

\noindent where $\omega$ is a \textit{six-vertex configuration} determined by the six possible configurations (see Figure 1 and Figure 2 for the set of possible configurations of the 6-vertex model, and Figure 3 and Figure 4 for the set of possible configurations of the 20-vertex model), with the weight in the numerator of the probability measure taking the form,

\begin{align*}
  w_{\mathrm{6V}} \big( \omega \big) \equiv w \big( \omega \big) \equiv a_1^{n_1} a_2^{n_2} b_1^{n_3} b_2^{n_4} c_1^{n_5} c_2^{n_6}   \text{ } \text{ , }
\end{align*}

\noindent for $a_1 , a_2 , b_1 , b_2 , c_1 , c_2 \geq 0$, with the partition function,

\begin{align*}
 Z^{\xi}_{\textbf{T}_N} \big( \omega , \Omega \big)  \equiv Z^{\xi}_{\textbf{T}_N}  \equiv Z^{\xi}_N = \underset{\omega \in \Omega ( \textbf{T}_N ) }{\sum} w \big( \omega \big)    \text{ } \text{ . } 
\end{align*}

\noindent Under the isotropic parameter assumption, the weight function that is normalized by the partition expression in the 6-vertex probability measure above can be expressed as,

\begin{align*}
  w_{6V}(\omega) \equiv w(\omega) =        a_1^{n_1} a_2^{n_2} b_1^{n_3} b_2^{n_4} c_1^{n_5} c_2^{n_6}  \underset{c_1 \equiv c_2 \equiv c}{\underset{b_1 \equiv b_2 \equiv b}{\underset{a_1 \equiv a_2 \equiv a}{\overset{\mathrm{Isotropic}}{\Longleftrightarrow}} }}          a^{n_1+n_2}  b^{n_3+n_4}  c^{n_5+n_6}    \text{. }   
\end{align*}

\noindent Although isotropic parameters remain useful for analyzing several aspects of the 6-vertex model, from crossing probability estimates to Russo-Seymour-Welsh that were initially developed in seminal work for sufficiently flat boundary conditions {\color{blue}[10]}, in addition to connections with the Ashkin-Teller, generalized random-cluster, and $\big( q_{\sigma}, q_{\tau} \big)$-spin models, {\color{blue}[36]}, other possible parametrizations for the weights include, {\color{blue}[24]},

 \begin{align*}
   a_1 \equiv      a \text{ }  \mathrm{exp} \big(  H + V \big)   \text{, } \\ a_2 \equiv   a  \text{ }  \mathrm{exp} \big( - H - V \big)   \text{, } \\  b_1 \equiv  \text{ }  \mathrm{exp} \big( H - V \big)  \text{, } \\ b_2 \equiv \text{ }  \mathrm{exp} \big( - H + V \big)    \text{, } \\ c_1 \equiv  c \lambda \text{, } \\ c_2 \equiv c \lambda^{-1}  \text{, } 
\end{align*}

\noindent in the presence of two external fields, $H$ and $V$, and $\lambda \geq 1$, corresponding to the non-symmetric parametrization. Equipped with this parametrization, from a suggestion of the authors of {\color{blue}[24]} whose work demonstrated that inhomogeneous limit shapes are integrable from a variational principle and solutions to the Euler-Lagrange equation, in a previous work of the author integrability of a Hamiltonian flow in the presence of inhomogeneities was also shown to be integrable, {\color{blue}[41]}, with the Hamiltonian flow taking the following form that is related to the semigrand canonical free energy, {\color{blue}[24]},

\begin{align*}
  \mathcal{H}_u \big( q , H \big)  \equiv  \mathcal{H}_u = \mathrm{log} \big[   Z_{\textbf{T}_{MN}}^n \big( u , H \big)       \big]                  \text{, } 
\end{align*}

\noindent for the partition function,

\begin{align*}
 Z^n_{\textbf{T}_{MN}} \equiv \text{6-vertex Partition function over the } M \times N \text{ torus under periodic boundary conditions}     \text{. }
\end{align*}

\noindent Alternatively, the semigrand canonical free energy can be expressed as the maximum over $\pm$, with,

\begin{align*}
  \mathcal{H}_u \big( q , H \big) \equiv \underset{\pm}{\mathrm{max}}    \text{ } \mathcal{H}^{\pm}_u \big( q , H \big) \equiv \underset{\pm}{\mathrm{max}}  \big\{     \pm H + l_{\pm} + \int_C  \psi^{\pm}_u \big( \alpha \big) \rho \big( \alpha \big)   \text{ } \mathrm{d} \alpha   \big\}  \text{, } 
\end{align*}

\noindent with $l_{-} \equiv \mathrm{log} \text{ } \mathrm{sinh} \big( \eta - u \big)$, $l_{+} \equiv \mathrm{log} \text{ } \mathrm{sinh} \text{ }  u$, contour $C$,

\begin{align*}
  C \equiv    \underset{\text{paths}}{\bigcup} \big\{ \text{orientable paths in the upper half plane} \big\}   \text{, }       
\end{align*}

\noindent and density,

\begin{align*}
  \rho \big( \alpha \big) \equiv       \# \big\{    \text{Bethe roots along contours } C        \big\}   \equiv \underset{\alpha > 0 }{\bigcup}  \big\{\text{roots } \alpha : \alpha \cap C \neq \emptyset \big\} \equiv  \bigg|  \big\{ \alpha : \alpha \cap C \neq \emptyset \big\}      \bigg| \text{. } 
\end{align*}

\noindent Fix the test functionals $F$, $G$, $A$, and $B$, which are manipulated in the following defintions for the Poisson bracket, and tensor products of the Poisson bracket. The mathematical problem of determining whether a probabilistic model is integrable can be formulated in terms of computations with the Poisson bracket, which takes the form, {\color{blue}[16]},

\begin{align*}
 \big\{ F ,  G \big\}   \equiv    i \int_{[-L,L]} \bigg[          \frac{\delta F}{\delta \psi} \frac{\delta G}{\delta \bar{\psi}} - \frac{\delta F}{\delta \bar{\psi}} \frac{\delta G}{\delta \psi }             \bigg]  \text{ } \mathrm{d} x      \text{, } 
\end{align*}

\noindent and also,

\begin{align*}
 \big\{  A \overset{\bigotimes}{,} B \big\} \equiv     i \int_{[-L,L]} \bigg[  \frac{\delta A}{\delta \psi} \bigotimes \frac{\delta B}{\delta \bar{\psi}} - \frac{\delta A}{\delta \bar{\psi}} \bigotimes  \frac{\delta B}{\delta \psi }                     \bigg]    \text{ } \mathrm{d} x     \text{, } 
\end{align*}

\noindent corresponding to the tensor product of the Poisson bracket, over $\big[ - L , L \big] \subsetneq \textbf{R}$, for some $L > 0$. The function $\psi$, with respect to which the derivative is taken in both the definition of the Poisson bracket and tensor products of the Poisson bracket, is given by solutions to the nonlinear Schrodinger's equation,

\begin{align*}
   i \frac{\partial \psi}{\partial t} = - \frac{\partial^2 \psi}{\partial x^2} + 2 \chi \big| \psi \big|^2 \psi      \text{, } 
\end{align*}

\noindent under well-posed initial data, and some $\chi > 0$. Combinatorially, the objects above form completely different relations in being related to the partition function of the number of domino tilings {\color{blue}[15]}. In this work, the partition function was shown to be in correspondence with the partition function of the 20-vertex model with domain wall boundary conditions, which as an adaptation of two-dimensional domain wall boundary conditions, equals,

\begin{align*}
 Z^{\mathrm{20V}}_n  = \underset{0 \leq u, v \leq n-1}{\mathrm{det}}   \bigg[ \frac{\big( 1 + u^2 \big) \big( 1 + 2u - u^2 \big)}{\big(  1 - u^2 v \big) \big[ \big( 1 -u \big)^2 - v \big( 1 + u \big)^2 \big]} \bigg] \text{, }
\end{align*}

\noindent for some $n>0$, and spatial parameters $u,v$. Besides the connection, and perspective, offered within this work that is related to the number of ways in which regions of $\textbf{T}$ can be tiled, for the 6-vertex model, over $\textbf{Z}^2$, a different determinantal representation was used, which takes the form, {\color{blue}[7]},

\begin{align*}
  Z^{\mathrm{6V}}_N = \frac{\underset{1 \leq \alpha \leq k}{\prod} \text{ } \underset{1 \leq k \leq N}{\prod} a \big( \lambda_{\alpha} , \nu_k \big) b \big( \lambda_{\alpha} , \nu_k \big) }{\underset{1 \leq a < \beta \leq N}{\prod} d \big( \lambda_{\beta} , \lambda_{\alpha} \big) \text{ } \underset{1 \leq i < k \leq N}{\prod} d \big( \nu_j , \nu_k \big)}   \mathrm{det} \mathcal{M} \text{, }
\end{align*}

\noindent for some $N>0$, corresponding to a determinant of Izergin-Korepin type, for the matrix,

\begin{align*}
  \mathcal{M}_{\alpha k} \equiv  \frac{c}{a \big( \lambda_{\alpha} , \nu_k \big) b \big( \lambda_{\alpha} , \nu_k \big) } \text{, }
\end{align*}

\noindent and weights $a \equiv a \big( \lambda_{\alpha} , \nu_k \big)$, $b \equiv b \big( \lambda_{\alpha} , \nu_k \big)$, with corresponding spectral parameters, $ \lambda_{\alpha} , \lambda_{\beta} ,  \nu_j 
 , \nu_k$. In the expression that follows below, fix some strictly positive $\epsilon$ that is taken to be sufficiently large. To transform the determinantal representation to a contour integral representation, we adapt the identity, {\color{blue}[7]},

\begin{align*}
                 f \big( \omega \big( \epsilon \big) \big) \bigg|_{\epsilon =0 } \propto   \frac{1}{2 \pi i} \underset{\mathscr{S}^{\prime}_0}{\oint} \frac{\big( z-1 \big)^{N-1}}{z^N} h_N \big( z \big) f \big( z \big) \mathrm{d} z        \text{, }
\end{align*}

\noindent for positively oriented surfaces containing $\big(0,0\big)$, and a smooth enough test function $f$, and suitably defined functions $\omega$ that are dependent upon weights of the 6-vertex model. The two determinant representations for the 20-vertex, and 6-vertex, partition functions above is very intriguing, and draws our attention to several characteristics of vertex models, from integrability, algebraic properties, whether through Grothendieck polynomials and related objects, amongst many other possible connections within the quantum inverse scattering framework {\color{blue}[41},{\color{blue}45]}. Over the lattice in {\color{blue}[7]}, the authors performed computations for top and bottom partition functions of finite volumes over the square lattice, which enabled them to obtain integral representations for nonlocal correlations. In comparison to the more combinatorial perspective provided in the enumeration of the number of tilings through the determinant for the partition function of the 20-vertex model, {\color{blue}[15]}, the main focus of the determinantal representation,

\begin{align*}
   Z^{20V}_N = \frac{\big[ \mathrm{sin} \big( \lambda - \nu \big)  \mathrm{sin} \big( \lambda + \nu \big) \big]^{N^2}}{\underset{1 \leq n \leq N-1}{\prod} \big( n! \big)^2} \mathrm{det} \mathcal{N}  \text{, }
\end{align*}

\noindent where,

\begin{align*}
  \mathcal{N}_{\alpha k} =  \partial^{a + k -2}_{\lambda} \bigg[   \frac{\mathrm{sin} \big( 2 \eta \big) }{\mathrm{sin} \big( \lambda - \eta \big) \mathrm{sin } \big( \lambda + \nu \big) }      \bigg] \text{, }
\end{align*}

\noindent which is introduced as a special case of the Izergin-Korepin type determinant in {\color{blue}[7]}, is within the quantum inverse scattering framework. This framework continues to remain of great interest for not only determining which models are integrable, {\color{blue}[41},{\color{blue}45]}, but also for the 20-vertex model, in possibly being able to investigate the conditions under which nonlocal integral correlations exist. Previous work of the author significantly pursued this theme, in which information from integrability of inhomogeneous limit shapes, and a Hamiltonian flow, for two-dimensional vertex models in the presence of inhomogeneities can be explored for three-dimensional vertex models. Over three-dimensions over the triangular lattice, in comparison to two-dimensions over the square lattice, the 20-vertex model does not possess such integrability properties, stemming from the lack of existence of action angle coordinates which have vanishing Poisson bracket, which could also be seen as being related to Poisson anticommutativity of Hamiltonians.

From a representation of the two-dimensional transfer matrix, {\color{blue}[41]},

\[
\begin{bmatrix}
       A \big( \lambda_{\alpha} \big)   & B \big( \lambda_{\alpha} \big)   \\
    C \big( \lambda_{\alpha} \big)  & D \big( \lambda_{\alpha} \big)  \text{ }  
  \end{bmatrix} \text{, }
\]

\noindent that can be expressed in terms of a product of L-operators,

\begin{align*}
     \overset{N-1}{\underset{i=0}{\prod}}   L_{\alpha , N - i  } \big( \lambda_{\alpha} , v_{N-i} \big)       \text{, } 
\end{align*}

\noindent for some $N>0$, where each L-operator is defined in terms of two spectral parameters, {\color{blue}[7]},

\[
       L_{\alpha , k   } \big( \lambda_{\alpha} , v_{k} \big)    \equiv 
  \begin{bmatrix}
     \mathrm{sin} \big( \lambda_{\alpha} - v_k + \eta \sigma^z_k \big)       &    \mathrm{sin} \big( 2 \eta \big) \sigma^{-}_k    \\
      \mathrm{sin} \big( 2 \eta \big) \sigma^{+}_k     &   \mathrm{sin}  \big( \lambda_{\alpha} - v_k - \eta \sigma^z_k \big)     
  \end{bmatrix} \text{, } 
\]

\noindent corresponding to the to the $k$ th horizontal line, and $\alpha$ th vertical line, with $L_{\alpha,k} \curvearrowright \bigg(  \text{vertical space } \bigotimes$ $\text{horizontal space } \bigg)$, and Pauli basis elements $\sigma^z_k$, $\sigma^{+}_k$, and $\sigma^{-}_k$ (see the definitions provided in {\color{blue}[7]}). To extend the arguments provided in {\color{blue}[7]} for obtaining nonlocal correlation functions for the 6-vertex model with domain wall boundary conditions, we pursue the following program:

\begin{itemize}
    \item[$\bullet$] \underline{(1)}, \textit{Computation of top, bottom, and side, partition functions for the 20-vertex model}. In the same manner that the authors of {\color{blue}[7]} perform computations of top and bottom partition functions over the square lattice, over the triangular lattice, we straightforwardly extend their methods to obtain three partition functions, instead of two.

 \item[$\bullet$] \underline{(2)}, \textit{Manipulation of the top and bottom partition functions}. Over the triangular lattice, in comparison to over the square lattice, computations using the top, bottom, and side, partition functions for the 20-vertex model are related to the evaluation of the three-dimensional product representation of L-operators. 

  \item[$\bullet$] \underline{(3)}, \textit{Transforming the determinantal representation to a contour integral representation}. From the inhomogeneous representation of the Izergin-Korepin type determinant, the desired integral representation for nonlocal correlations is obtained by expressing the original determinant into a smaller one, with suitable orthogonal polynomials. Suitable polynomials satisfying orthogonality relations pertaining to the 20-vertex model, and its own intrinsic correlation structure, can be obtained as have been for the 6-vertex model by evaluating the larger determinant to obtain a smaller one.
    
\end{itemize}

\begin{figure}
\begin{align*}
\includegraphics[width=0.78\columnwidth]{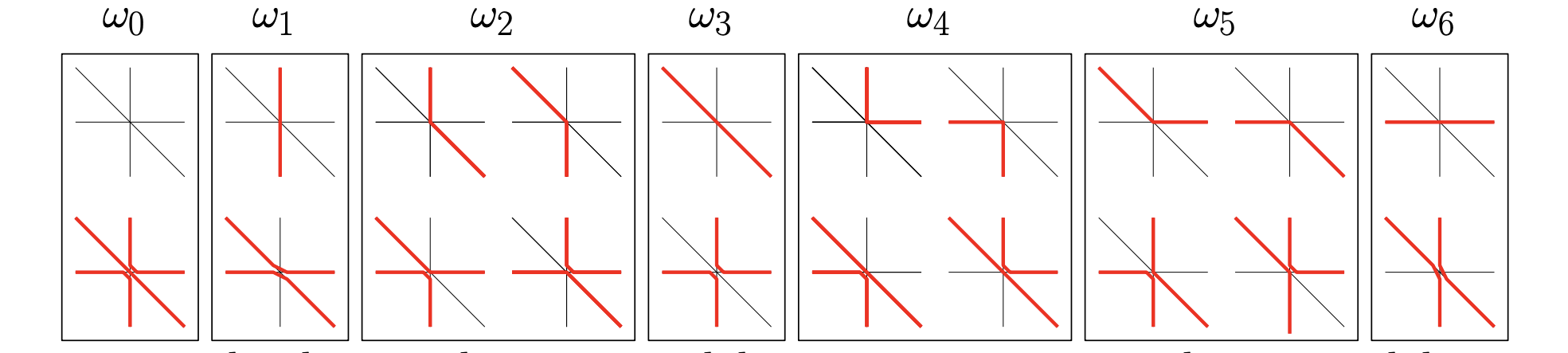}
\end{align*}
\caption{A depiction of each possible vertex for the triangular, or three dimensional, six-vertex model, adapted from {\color{blue}[15]}.}
\end{figure}

\begin{figure}
\begin{align*}
\includegraphics[width=0.75\columnwidth]{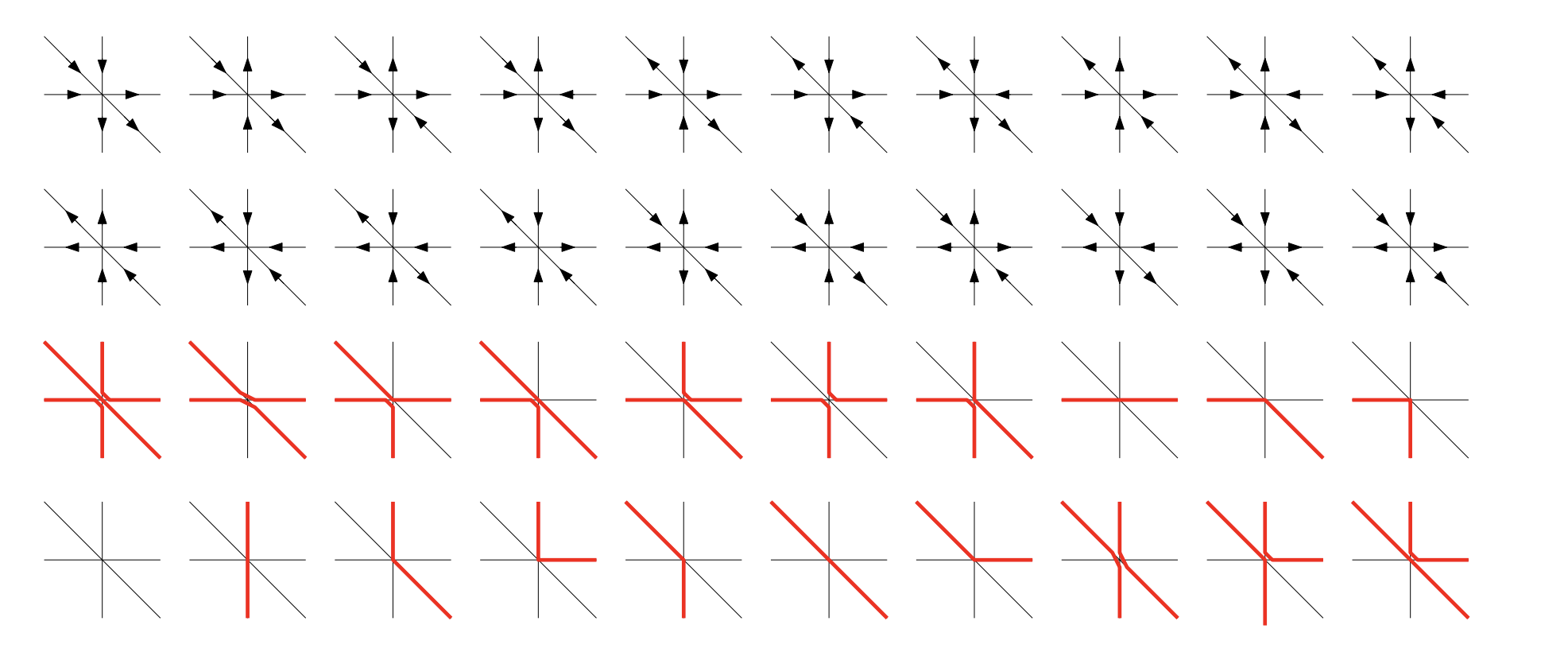}
\end{align*}
\caption{A depiction of each Boltzman weight for the triangular, or three dimensional, six-vertex model, also adapted from {\color{blue}[15]}.}
\end{figure}

\subsection{Statement of Main Results}

\noindent We provide statements of the results which are provided in the next section, which together are used to obtain contour integral representations for correlation functions, and also for the Emptiness formation probability. As it will be further described before \textit{3.1}, we introduce a suitable partition of $\textbf{T}$, along each degree of freedom, for obtaining contour integral representations for three partition functions, each if which is taken under domain-wall boundary conditions over $\textbf{Z}^2$. With the three contour integral representations for the partition functions, we readily obtain the contour integral representation for correlation functions of the 20-vertex model, supported over all of $\textbf{T}$. We take a homogenization limit of spectral parameters of the inhomogeneous contour integral representation so that the prefactor to the determinantal representation of the partition function is identically $1$.

\bigskip

\noindent \textbf{Lemma} \textit{1} (\textit{contour integral representation for the restriction of the 20-vertex partition function to the Side sublattice of the triangular lattice}). The Side partition function of the 20-vertex model,

\begin{align*}
 Z^{\mathrm{20V}}_n \bigg|_{\mathrm{Side}} \equiv Z^{\mathrm{Side},\mathrm{20V}} \equiv Z^{\mathrm{Side}} \big( r^{\prime}_1 , \cdots , r^{\prime}_{s^{\prime}} , r^{\prime\prime}_1 , \cdots , r^{\prime\prime}_{s^{\prime\prime}} \big)  \equiv   Z^{\mathrm{Side}}_{r^{\prime}_1 , \cdots , r^{\prime}_{s^{\prime}} , r^{\prime\prime}_1 , \cdots , r^{\prime\prime}_{s^{\prime\prime}}}        \text{,}
\end{align*}

\noindent has the following contour integral representation,

\begin{align*}
    \underline{Z^{\mathrm{Side}}_{r^{\prime}_1 , \cdots , r^{\prime}_{s^{\prime}} , r^{\prime\prime}_1 , \cdots , r^{\prime\prime}_{s^{\prime\prime}}}}  =  \bigg[         c^{s^{\prime}+s^{\prime\prime}} a^{( s^{\prime} + s^{\prime\prime} )  (2N-2)}          \bigg[ \underset{1 \leq j \leq s^{\prime}}{\prod}   t^{r^{\prime}_j}              \bigg] \bigg[ \underset{1 \leq k \leq s^{\prime\prime}}{\prod}      \big( t^{\prime}\big)^{r^{\prime\prime}_k}      \bigg]   \bigg] \bigg\{ \underset{\mathscr{S}^{\prime}_1}{\oint}  \times   \cdots \times     \underset{\mathscr{S}^{\prime}_{s^{\prime}+s^{\prime\prime}}}{\oint}       \\ \times      \bigg\{   \bigg[ \underset{1\leq j \leq s^{\prime}}{\prod}     \frac{w^{r^{\prime}_j-1}}{\big( w_j - 1 \big)^s}                \bigg] 
   \bigg[ \underset{1\leq k \leq s^{\prime\prime}}{\prod}               \frac{\big( w^{\prime}\big)^{r^{\prime\prime}_k-1}}{\big( w^{\prime\prime}_k - 1 \big)^{s^{\prime\prime}}}     \bigg]             \bigg[ \underset{1 \leq j < k \leq s^{\prime}}{\prod}     \big[ \big( w_j - w_k \big) \big( t^2 w_j w_k \\ - 2 \Delta t w_j +1 \big)  \big]               \bigg]   \bigg[ \underset{1 \leq j^{\prime} < k^{\prime} \leq s^{\prime\prime}}{\prod}         \big[ \big( w^{\prime}_j - w^{\prime}_k \big)  \big( t^2 w^{\prime}_j w^{\prime}_k - 2 \Delta t w^{\prime}_j   +1 \big)  \big]              \bigg] \bigg\}  \frac{d^{s^{\prime}} w}{\big( 2 \pi i \big)^s} \\ \times   \frac{d^{s^{\prime\prime}} w^{\prime} }{\big( 2 \pi i \big)^{s^{\prime}}}  \bigg\}  \text{, }
\end{align*}

\noindent for the collection of contours,

\begin{align*}
        \mathscr{S}_{s^{\prime}+s^{\prime\prime}} \equiv  \big\{ \text{surface } \mathcal{S} \text{ containing the residue at } \big(s^{\prime}+s^{\prime\prime} , s^{\prime}+s^{\prime\prime} \big)  \big\}   \text{. } \end{align*}

\noindent \textbf{Lemma} \textit{2} (\textit{contour integral representation for the restriction of the 20-vertex partition function to the Top sublattice of the triangular lattice}). The Top partition function of the 20-vertex model,

\begin{align*}
   Z^{\mathrm{20V}}_n \bigg|_{\mathrm{Top}} \equiv   Z^{\mathrm{Top},\mathrm{20V}}  \equiv  Z^{\mathrm{Top}} \big( r_1 , \cdots , r_s , r^{\prime}_1 , \cdots , r^{\prime}_{s^{\prime}} \big)   \equiv Z^{\mathrm{Top}}_{r_1 , \cdots , r_s , r^{\prime}_1 , \cdots , r^{\prime}_{s^{\prime}}}   \text{,}
\end{align*}

\noindent has the following contour integral representation,

\begin{align*}
    \underline{Z^{\mathrm{Top}}_{r_1 , \cdots , r_s , r^{\prime}_1 , \cdots , r^{\prime}_{s^{\prime}}}}  =  \bigg[         c^{s+s^{\prime}} a^{( s + s^{\prime} )  (2N-2)}          \bigg[ \underset{1 \leq j \leq s}{\prod}   t^{r_j}              \bigg] \bigg[ \underset{1 \leq k \leq s^{\prime}}{\prod}      \big( t^{\prime}\big)^{r^{\prime}_k}      \bigg]   \text{ } \bigg] \bigg\{  \underset{\mathscr{S}_1}{\oint}  \times \cdots \times     \underset{\mathscr{S}_{s+s^{\prime}}}{\oint}   \\ \times        \bigg\{   \bigg[ \underset{1\leq j \leq s}{\prod}     \frac{w^{r_j-1}}{\big( w_j - 1 \big)^s}                \bigg]      \bigg[ \underset{1\leq k \leq s^{\prime}}{\prod}               \frac{\big( w^{\prime}\big)^{r^{\prime}_k-1}}{\big( w^{\prime}_k - 1 \big)^{s^{\prime}}}     \bigg]               \bigg[ \underset{1 \leq j < k \leq s}{\prod}     \big[ \big( w_j - w_k \big) \big( t^2 w_j w_k \\ - 2 \Delta t w_j +1 \big)  \big]               \bigg]   \bigg[ \underset{1 \leq j^{\prime} < k^{\prime} \leq s^{\prime}}{\prod}         \big[ \big( w^{\prime}_j - w^{\prime}_k \big)  \big( t^2 w^{\prime}_j w^{\prime}_k - 2 \Delta t w^{\prime}_j   +1 \big)  \big]              \bigg] \bigg\}  \frac{d^s w}{\big( 2 \pi i \big)^s} \\ \times \frac{d^{s^{\prime}} w^{\prime} }{\big( 2 \pi i \big)^{s^{\prime}}}  \bigg\}   \text{, }
\end{align*}

\noindent for the collection of contours,

\begin{align*}
        \mathscr{S}_{s+s^{\prime}} \equiv  \big\{ \text{surface } \mathcal{S} \text{ containing the residue at } \big( s+s^{\prime} , s+s^{\prime} \big)  \big\}   \text{. } \end{align*}

\noindent \textbf{Lemma} \textit{3} (\textit{contour integral representation for the restriction of the 20-vertex partition function to the Bottom sublattice of the triangular lattice}). The Bottom partition function of the 20-vertex model,

\begin{align*}
   Z^{\mathrm{20V}}_n \bigg|_{\mathrm{Bottom}} \equiv    Z^{\mathrm{Bottom}, \mathrm{20V}} \equiv 
   Z^{\mathrm{Bottom}} \big( r_1 , \cdots , r_s , r^{\prime}_1 , \cdots , r^{\prime}_{s^{\prime}} \big) \equiv Z^{\mathrm{Bottom}}_{r_1 , \cdots , r_s , r^{\prime}_1 , \cdots , r^{\prime}_{s^{\prime}}}    \text{,}
\end{align*}

\noindent has the following contour integral representation,

\begin{align*} \underline{Z^{(\mathrm{Side})_2, 20V}_{r_1, \cdots , r_s, r^{\prime}_1, \cdots, r^{\prime}_{s^{\prime}}}} =        \mathcal{D}^{\prime}_{\text{Inhomogeneous}}  \bigg[    \underset{1 \leq j^{\prime}\leq s^{\prime}}{\underset{1 \leq j \leq s}{\prod}}      v_{\lambda,r^{\prime}} \big( i,j \big)  \bigg] \bigg[         \underset{1 \leq j^{\prime} < k^{\prime} \leq s^{\prime}}{\underset{1 \leq j < k \leq s}{\prod}}   \bigg[              \mathrm{sin}^{-1} \big( \epsilon_j - \epsilon_k + 2 \eta \big)                   \bigg] \\ \times  \bigg[     \mathrm{sin}^{-1}  \big( \epsilon_{j^{\prime}} - \epsilon_{k^{\prime}} + 2 \eta \big)       \bigg]      \bigg]  
\text{,} \end{align*}

\noindent for the inhomgeneous determinant,

\begin{align*}
  \frac{\mathcal{D}_{\text{Inhomogeneous}}}{       \bigg[        \underset{1 \leq i \leq N}{\prod}  z^{n-1}_i \bigg] \bigg[ \underset{1 \leq i < j \leq N}{\prod} ( z_i - z_j ) ( w_j - w_i )  \bigg] \bigg[   \underset{1 \leq i < j \leq N}{\prod} ( 1 - z_i z_j ) ( 1 - w_i w_j ) \bigg] }    \\ \\  \equiv       \bigg| \begin{smallmatrix} 
\frac{1}{a ( z_1 , w_2 ) b ( z_1 , w_2 )} - \frac{1}{a ( 1 , z_1 w_2 ) b ( 1 , z_1 w_2 )} & \cdots & \frac{1}{a ( z_N , w_{N+1} ) b ( z_N , w_{N+1} )} - \frac{1}{a ( 1 , z_N w_{N+1} ) b ( 1 , z_N w_{N+1} )}  \\ \frac{1}{a ( z_1 , w_3 ) b ( z_1 , w_3 )} - \frac{1}{a ( 1 , z_1 w_3 ) b ( 1 , z_1 w_3 )}  & \cdots & \frac{1}{a ( z_N , w_{N+2} ) b ( z_N , w_{N+2} )} - \frac{1}{a ( 1 , z_N w_{N+2} ) b ( 1 , z_N w_{N+1} )}  \\ \vdots & \cdots & \vdots \\ \frac{1}{a ( z_1 , w_{N+1} ) b ( z_1 , w_{N+1} )} - \frac{1}{a ( 1 , z_1 w_{N+1} ) b ( 1 , z_1 w_{N+1}  )}  & \cdots & \frac{1}{a ( z_N , w_{2N} ) b ( z_N , w_{2N} )} - \frac{1}{a ( 1 , z_N w_{2N} ) b ( 1 , z_N w_{2N} )} 
\end{smallmatrix} \bigg|  \\ \times   \bigg[ \bigg[        \underset{1 \leq i \leq N}{\prod}  z^{n-1}_i \bigg] \bigg[ \underset{1 \leq i < j \leq N}{\prod} ( z_i - z_j ) ( w_j - w_i )  \bigg] \bigg[   \underset{1 \leq i < j \leq N}{\prod} ( 1 - z_i z_j ) ( 1 - w_i w_j ) \bigg]  \bigg]^{-1}     \text{. }
\end{align*}

\noindent \textbf{Corollary} (\textit{constants of proportionality for each of the contour integral representations for the partition functions}). Given the contour integral representations for partition functions provided in the previous three results above, one has,

\begin{align*}
     \underline{Z^{\mathrm{Bottom}}_{r_1 , \cdots , r_s , r^{\prime}_1 , \cdots , r^{\prime}_{s^{\prime}}}}  \propto    \text{ } \underset{\mathscr{S}_1}{\oint}  \times \cdots \times     \underset{\mathscr{S}_{s+s^{\prime}}}{\oint}         \bigg\{  \bigg[ \underset{1\leq j \leq s}{\prod}     \frac{w^{r_j-1}}{\big( w_j - 1 \big)^s}                \bigg]     \bigg[ \underset{1\leq k \leq s^{\prime}}{\prod}               \frac{\big( w^{\prime}\big)^{r^{\prime}_k-1}}{\big( w^{\prime}_k - 1 \big)^{s^{\prime}}}     \bigg]            \bigg[ \underset{1 \leq j < k \leq s}{\prod}     \big[ \big( w_j - w_k \big) \\ \times  \big( t^2 w_j w_k - 2 \Delta t w_j +1 \big)  \big]               \bigg]     \bigg[ \underset{1 \leq j^{\prime} < k^{\prime} \leq s^{\prime}}{\prod}         \big[ \big( w^{\prime}_j - w^{\prime}_k \big) \big( t^2 w^{\prime}_j w^{\prime}_k - 2 \Delta t w^{\prime}_j   +1 \big)  \big]              \bigg] \bigg\}  \\ \times  \frac{d^s w}{\big( 2 \pi i \big)^s} \frac{d^{s^{\prime}} w^{\prime} }{\big( 2 \pi i \big)^{s^{\prime}}}            \text{, }
\end{align*}

\noindent where the constant of proportionality for which there is an exact equality takes the form,

\begin{align*}
                c^{s+s^{\prime}} a^{( s + s^{\prime} )  (2N-2)}          \bigg[ \underset{1 \leq j \leq s}{\prod}   t^{r_j}              \bigg] \bigg[ \underset{1 \leq k \leq s^{\prime}}{\prod}      \big( t^{\prime}\big)^{r^{\prime}_k}      \bigg]           \text{, }
\end{align*}

\noindent corresponding to the contour integral representation provided in \textbf{Lemma} \textit{3}. For the restriction of the 20-vertex partition function to the Side sublattice,

\begin{align*}
     \underline{Z^{\mathrm{Side}}_{r^{\prime}_1 , \cdots , r^{\prime}_{s^{\prime}} , r^{\prime\prime}_1 , \cdots , r^{\prime\prime}_{s^{\prime\prime}}} } \propto    \text{ } \underset{\mathscr{S}_1}{\oint}  \times \cdots \times     \underset{\mathscr{S}_{s+s^{\prime}}}{\oint}         \bigg\{  \bigg[ \underset{1\leq j \leq s^{\prime}}{\prod}     \frac{w^{r_j-1}}{\big( w_j - 1 \big)^{s^{\prime}}}                 \bigg]     \bigg[ \underset{1\leq k \leq s^{\prime\prime}}{\prod}               \frac{\big( w^{\prime}\big)^{r^{\prime}_k-1}}{\big( w^{\prime}_k - 1 \big)^{s^{\prime\prime}}}     \bigg]             \bigg[ \underset{1 \leq j < k \leq s}{\prod}     \big[ \big( w_j - w_k \big) \\ \times  \big( t^2 w_j w_k  - 2 \Delta t w_j +1 \big)  \big]               \bigg] \bigg[ \underset{1 \leq j^{\prime} < k^{\prime} \leq s^{\prime\prime}}{\prod}         \big[ \big( w_{j^{\prime}} - w_{k^{\prime}} \big) \big( t^2 w_{j^{\prime}} w_{k^{\prime}} - 2 \Delta t w_{j^{\prime}}   +1 \big)  \big]              \bigg] \bigg\}  \frac{d^{s^{\prime}} w}{\big( 2 \pi i \big)^s} \\ \times  \frac{d^{s^{\prime\prime}} w^{\prime} }{\big( 2 \pi i \big)^{s^{\prime\prime}}}            \text{, }
\end{align*}

\noindent where the constant for which the proportionality above is an equality takes the form,

\begin{align*}
            c^{s^{\prime}+s^{\prime\prime}} a^{( s^{\prime} + s^{\prime\prime} )  (2N-2)}          \bigg[ \underset{1 \leq j^{\prime} \leq s^{\prime}}{\prod}   t^{r_{j^{\prime}}}              \bigg] \bigg[ \underset{1 \leq k^{\prime} \leq s^{\prime\prime}}{\prod}      \big( t^{\prime}\big)^{r^{\prime}_{k^{\prime}}}       \bigg]         \text{, }
\end{align*}

\noindent corresponding to the contour integral representation provided in \textbf{Lemma} \textit{1}. For the restriction of the partition to the remaining degree of freedom, one has the countour integral representation,

\begin{align*}
    \underline{Z^{\mathrm{Side}}_{r^{\prime}_1 , \cdots , r^{\prime}_{s^{\prime}} , r^{\prime\prime}_1 , \cdots , r^{\prime\prime}_{s^{\prime\prime}}}}  \propto \underset{\mathscr{S}^{\prime}_1}{\oint}  \times   \cdots \times     \underset{\mathscr{S}^{\prime}_{s^{\prime}+s^{\prime\prime}}}{\oint}        \bigg\{   \bigg[ \underset{1\leq j \leq s^{\prime}}{\prod}     \frac{w^{r^{\prime}_j-1}}{\big( w_j - 1 \big)^s}                \bigg]    
   \bigg[ \underset{1\leq k \leq s^{\prime\prime}}{\prod}               \frac{\big( w^{\prime}\big)^{r^{\prime\prime}_k-1}}{\big( w^{\prime\prime}_k - 1 \big)^{s^{\prime\prime}}}     \bigg]         \\ \times        \bigg[ \underset{1 \leq j < k \leq s^{\prime}}{\prod}     \big[ \big( w_j - w_k \big) \big( t^2 w_j w_k  - 2 \Delta t w_j +1 \big)  \big]               \bigg]   \bigg[ \underset{1 \leq j^{\prime} < k^{\prime} \leq s^{\prime\prime}}{\prod}         \big[ \big( w^{\prime}_j - w^{\prime}_k \big)  \big( t^2 w^{\prime}_j w^{\prime}_k \\ - 2 \Delta t w^{\prime}_j   +1 \big)  \big]              \bigg] \bigg\}  \frac{d^{s^{\prime}} w}{\big( 2 \pi i \big)^s}   \frac{d^{s^{\prime\prime}} w^{\prime} }{\big( 2 \pi i \big)^{s^{\prime}}}   \text{, }
\end{align*}

\noindent where the constant for which the proportionality above is an equality takes the form,

\begin{align*}
c^{s^{\prime}+s^{\prime\prime}} a^{( s^{\prime} + s^{\prime\prime} )  (2N-2)}          \bigg[ \underset{1 \leq j \leq s^{\prime}}{\prod}   t^{r^{\prime}_j}              \bigg] \bigg[ \underset{1 \leq k \leq s^{\prime\prime}}{\prod}      \big( t^{\prime}\big)^{r^{\prime\prime}_k}      \bigg]  \text{, }
\end{align*}

\noindent corresponding to the contour integral representation provided in \textbf{Lemma} \textit{2}.

\bigskip

\noindent \textbf{Lemma} \textit{4} (\textit{symmetrization of  contour integral representations for correlation functions, for obtaining the representation for the emptiness formation probability}). The final contour integral representation, corresponding to the Emptiness Formation Probability, takes the form,

\begin{align*}
 \underset{\mathscr{S}^{\prime}_1}{\oint}  \times \cdots \times   \underset{\mathscr{S}^{\prime}_1}{\oint}        \bigg[   \text{ }  \underset{\mathscr{S}^{\prime}_0}{\oint}  \times \cdots \times   \underset{\mathscr{S}^{\prime}_0}{\oint}        \bigg[  \underset{1 \leq j^{\prime} \leq s^{\prime}}{\underset{1 \leq j \leq s}{\prod}}                \frac{w^r_{j}}{\big( w_j - 1 \big)^s z^{r-s+j}_j \big(  \underset{1 \leq l \leq j}{\prod}  \big[ w_l - z_l \big]     \big) }   \\ \times                          \frac{w^r_{j^{\prime}}}{\big( w_{j^{\prime}} - 1 \big)^{s^{\prime}} z^{r^{\prime}-s^{\prime}+j^{\prime}}_{j^{\prime}} \big(  \underset{1 \leq l \leq j^{\prime}}{\prod}  \big[ w_l - z_l \big]     \big) }                             \bigg]     \\ \\ \times \bigg[    \underset{1 \leq j^{\prime} < k^{\prime} \leq s^{\prime}}{\underset{1 \leq j < k \leq s}{\prod}}                 \frac{\big( w_j - w_k \big) \big( t^2 w_j w_k - 2 \Delta t w_j + 1 \big) \big( z_k - z_j \big) }{t^2 z_j z_k - 2 \Delta t z_j + 1}   \\ \times   \frac{\big( w_{j^{\prime}} - w_{k^{\prime}} \big) \big( t^2 w_{j^{\prime}} w_{k^{\prime}} - 2 \Delta t w_{j^{\prime}} + 1 \big) \big( z_{k^{\prime}} - z_{j^{\prime}} \big) }{t^2 z_{j^{\prime}} z_{k^{\prime}} - 2 \Delta t z_{j^{\prime}} + 1}                 \bigg]   \bigg]  \\ \times H_{N,s} \big( z_1, \cdots, z_s \big) H_{N,s^{\prime}} \big( z^{\prime}_1, \cdots, z^{\prime}_{s^{\prime}} \big)        \frac{\mathrm{d}^s w + \mathrm{d}^s z}{\big( 2 \pi i \big)^s}   \frac{\mathrm{d}^{s^{\prime}} z^{\prime} + \mathrm{d}^{s^{\prime}} w }{\big( 2 \pi i \big)^{s^{\prime}} }    \text{. }
\end{align*}

\noindent The main result, provided below, provides a contour integral representation for the correlation function.

\bigskip

\noindent \textbf{Theorem} (\textit{nonlocal correlations}). The representation for nonlocal correlation functions of the 20-vertex model under domain-wall boundary conditions takes the form,

\begin{align*}
    Z^{\mathrm{Top}}_{r_1 , \cdots , r_s , r^{\prime}_1 , \cdots , r^{\prime}_s } =  \mathscr{P}    \bigg\{ \underset{\mathscr{S}_1  }{\oint}  \times \cdots \times     \underset{\mathscr{S}_1 }{\oint}   \bigg[ \mathscr{P}_1 \mathscr{P}_2   h^{\prime}_{N,s,s^{\prime}}         \frac{\mathrm{d}^{s^{\prime}} z^{\prime}}{ \big( 2 \pi i \big)^{s^{\prime}}}   \mu_0                  \bigg]      \bigg\}         \text{, } \tag{\textit{*}}
\end{align*}

\noindent where,

\begin{align*}
        \mathscr{S}_1 \equiv  \big\{ \text{surface } \mathcal{S} \text{ containing the residue at } \big( 1 , 1 \big)  \big\}   \text{, } \\  \\     \mathscr{P}       \equiv       \mathscr{P} \big( a,  b , c , N \big) =   \bigg[   \frac{Z^{20V}_N}{a^{\frac{s ( 2N -s +1) }{2}+ \frac{s^{\prime} ( 2N -s^{\prime} +1) }{2}} b^{\frac{s(s-3)}{2}+ \frac{s^{\prime}(s^{\prime}-3)}{2}}      c^{s+s^{\prime}} }                      \bigg]         \bigg[ \frac{a}{b} \bigg]^{r_1 + \cdots + r_s + r^{\prime}_1 + \cdots + r^{\prime}_{s^{\prime}}}    \text{, } \\  \\ \mathscr{P}_1     \equiv   \underset{1 \leq j \leq s^{\prime}}{\underset{1 \leq i \leq s}{\prod} }  \frac{1}{ z_{r_i} z^{\prime}_{r_j} } \text{, } \\ \\   \mathscr{P}_2   \equiv    \bigg[ \underset{1 \leq j < k \leq s}{\prod} \frac{z_j - z_k}{ t^2 z_j z_k - 2 \Delta t z_j + 1 }  \bigg] \bigg[ \underset{1 \leq j^{\prime} < k^{\prime} \leq s^{\prime}}{\prod}  \frac{z^{\prime}_j - z^{\prime}_k}{ t^2 z^{\prime}_j z^{\prime}_k - 2 \Delta t z^{\prime}_j + 1 } \bigg]                                   \text{, }   \\ \\ h^{\prime}_{N,s,s^{\prime}}  \equiv     h^{\prime}_{N,s,s^{\prime}} \big( z_1 , \cdots , z_s , z^{\prime}_1 , \cdots , z^{\prime}_{s^{\prime}} \big) =  Z^{\mathrm{20V}}_n  = \underset{0 \leq u, v \leq n-1}{\mathrm{det}}   \bigg[ \frac{\big( 1 + u^2 \big) \big( 1 + 2u - u^2 \big)}{\big(  1 - u^2 v \big) \big[ \big( 1 -u \big)^2 - v \big( 1 + u \big)^2 \big]} \bigg] \text{, }  \\ \\  \mu_0 \equiv      \frac{\mathrm{d}^s z  }{ \big( 2 \pi i \big)^s}   \text{, }
\end{align*}

\noindent A closely related representation for the side partition function can be obtained with the same method.

\bigskip

\noindent The main results stated, \textbf{Lemma} \textit{1}-\textit{3}, \textbf{Theorem}, and \textbf{Corollary}, are of interest to formulate, and characterize, within Statistical Physics for many reasons. First, correlation functions of vertex models are related to different ranges of possible behaviors for a model of interest, which for the 6-vertex and 20-vertex models alike can exhibit ferroelectric, or antiferroelectric, behaviors. Given possible behaviors of other models in Statistsical Physics, for example the Ising model which can, or cannot, be spontaneously magnetized, determining similar connections between regions of a phase diagram for a model of interest, is extremely valuable. For the phase diagram of the 6-vertex model, in comparison to that of the 20-vertex model, possible behaviors of the model can be encoded with domain-wall boundary conditions. Domain-walls stipulate that arrows of configurations incident to the boundary of a finite volume point inwards, while configurations lying along the top and bottom configurations that are incident to the boundary point outwards. This encoding of boundary conditions, despite the fact that the 20-vertex model is a three-dimensional vertex model, significantly determines the composition of representations for correlation functions.

\subsection{Paper organization}

With the objects defined in the two-dimensional quantum inverse scattering method for the inhomogeneous 6-vertex model, along with corresponding determinantal representations of the partition functions for the 6-vertex, and 20-vertex, models, in the next section we describe how the product representation for the three-dimensional transfer matrix, studied in {\color{blue}[45]}, can be leveraged for computing the desired nonlocal correlations. Given the fact that previous computations of correlation functions for the 6-vertex model were dependent upon a partition of domains over $\textbf{Z}^2$ into top, and bottom, components, over $\textbf{T}$, we introduce, and extensively manipulate, partition functions for the 20-vertex model over the top, bottom, and side of domains. As such, we introduce an additional partition of the triangular lattice, rather than the top and bottom portions as introduced for the partition function of the 6-vertex model over $\textbf{Z}^2$, for the additional degree associated with each vertex of $\textbf{T}$, in comparison to the two degrees of freedom associated with each vertex of $\textbf{Z}^2$.

\section{Obtaining nonlocal correlation representations from the quantum inverse scattering method}

As mentioned in the introduction, in order to extend notions obtained for integral representations of nonlocal correlations for the 6-vertex model to the 20-vertex model, one must not only manipulate higher dimensional objects, but also relate such objects within the quantum inverse scattering approach. In the following subsection, objects from a previous work of the author, {\color{blue}[45]}, on the 20-vertex model are introduced.

\subsection{Quantum inverse scattering for the 20-vertex model under domain-wall boundary conditions}

We provide a discussion similar to that of \textit{1.5} in {\color{blue}[45]}. First, for boundary conditions $\xi$, such as those introduced over the pentagonal lattice, {\color{blue}[15]}, namely a sublattice $\big( \mathscr{ST} \big)$ over the triangular lattice, such that,

\begin{align*}
\big(  \mathscr{ST}  \big) \cap \textbf{T} \neq \emptyset    \text{, }
\end{align*}

\noindent which is spanned by the vertices of a polygon with side lengths $n-1$, $k$, and $n-k-1$, for $0 \leq k \leq n-1$, the probability measure takes the form,

\begin{align*}
   \textbf{P}^{20V , \xi}_{\textbf{T}} \big[ \cdot \big] \equiv  \textbf{P}^{20V}_{\textbf{T}} \big[ \cdot \big]   \text{, }
\end{align*}

\noindent which is explicitly given by the ratio of the vertex weight function and the partition function,

\begin{align*}
\textbf{P}^{20V}_{\textbf{T}}[      \omega         ]   \equiv \textbf{P}^{20V}[   \omega     ]     =  \frac{w_{20V}(\omega)}{Z^{20V}_{\textbf{T}}} \equiv \frac{w(\omega)}{Z_{\textbf{T}}} \text{, }
\end{align*}

\noindent for some vertex configuration $\omega \in \Omega^{20V}$ - the 20-vertex sample space, and weights similar to those introduced in the previous section for the 6-vertex model, namely, {\color{blue}[15]},

\begin{align*}
     w_0 \equiv   a_1 a_2 a_3   \text{, } 
\\
    w_1 \equiv   b_1 a_2 b_3   \text{, } \\    w_2    \equiv   b_1 a_2 c_3    \text{, } \\
   w_3 \equiv       a_1 b_2 b_3 + c_1 c_2 c_3  \text{, } \\     w_4 \equiv    c_1 a_2 a_3        \text{, } \\               w_5 \equiv   b_1 c_2 a_3   \text{, } \\  w_6 \equiv  b_1 b_2 a_3    \text{, } 
\end{align*}

\noindent and the partition function,

\begin{align*}
 Z_{\textbf{T}} \equiv \underset{\omega \in \Omega^{20V}}{\sum}  w \big( \omega \big) \text{. }
\end{align*}

\noindent Second,  observe that in a three dimensional, versus two dimensional, state space for the six-vertex model, L-operators have been shown to take a myriad of other forms from, {\color{blue}[6]},

\[
\hat{L} \big( \xi \big) \equiv L^{3D}_1 =  \mathrm{exp} \big( \lambda_3 ( q^{-2 } \xi^s ) \big)    \begin{bmatrix}
        q^{D_1}       &    q^{-2} a_1 q^{-D_1-D_2} \xi^{s-s_1}        &   a_1 a_2 q^{-D_1 - 3D_2} \xi^{s - s_1 - s_2}  \\ a^{\dagger}_1 q^{D_1} \xi^{s_1} 
             &      q^{-D_1 + D_2} - q^{-2} q^{D_1 -D_2} \xi^{s}     &     - a_2 q^{D_1 - 3D_2} \xi^{s-s_2}  \\ 0  &    a^{\dagger}_2 q^{D_2} \xi^{s_2} &  q^{-D_2} \\   
  \end{bmatrix}  \text{, }         
\]

\noindent in addition to, {\color{blue}[6]},

\[
L^{3D}_2 =     \frac{\mathrm{exp} \big( - \lambda_3 ( q^{-2} \xi^{-s} ) \big)    }{1 - \xi^s}   \bigg[ \begin{smallmatrix}
     q^2 q^{D_1} - q^{-D_1} \xi^s   &  a_1 q^{D_1} \xi^{s_1}  &  q^{-1} a_1 a_2 \xi^{s_1 + s_2} \\                 
         a^{\dagger}_1 q^{-D_1 - D_2} \xi^{s-s_1}    &    - q^{D_1- D_2} \xi^s &    - a_2 q^{-D_2} \xi^{s_2} \\ - a_1^{\dagger} a^{\dagger}_2 q^{-D_1 - D_2} \xi^{s-s_1 - s_2} &    a^{\dagger}_2 q^{D_1 - D_2} \xi^{s-s_2} &  q^{-D_2} - q^{D_2} \xi^s \\   
  \end{smallmatrix} \bigg] \text{. }         
\]

\noindent In {\color{blue}[45]}, to perform computations with the L-operator above for studying asymptotic properties of the transfer and quantum monodromy matrices, one must approximate $81$ Poisson brackets appearing in the three-dimensional Poisson structure, from the tensor product of transfer matrices, {\color{blue}[45]},

\[
\big\{ \textbf{T} \big( \underline{u}  , \big\{ u_i \big\} ,  \big\{ v_j \big\} , \big\{ w_k \big\}  \big) \overset{\bigotimes}{,}  \textbf{T} \big( \underline{u^{\prime}} , \big\{ u^{\prime}_i \big\} ,  \big\{ v^{\prime}_j \big\} , \big\{ w^{\prime}_k \big\}   \big)   \big\}\]  \[ = \bigg\{    \begin{bmatrix}
 A \big( \underline{u} 
 \big) & D \big( \underline{u} \big)  & G \big( \underline{u}  \big) \\ B \big( \underline{u}  \big) & E \big( \underline{u}  \big) & H \big( \underline{u}  \big)  \\ C \big( \underline{u}  \big)  &  F \big( \underline{u}  \big) & I \big( \underline{u}  \big) 
\end{bmatrix}\overset{\bigotimes}{,} \begin{bmatrix}
 A \big( \underline{u^{\prime}} \big) & D \big( \underline{u^{\prime}} \big)  & G \big( \underline{u^{\prime}} \big) \\ B \big( \underline{u^{\prime}}\big) & E \big( \underline{u^{\prime}} \big) & H \big( \underline{u^{\prime}} \big)  \\ C \big( \underline{u^{\prime}} \big)  &  F \big( \underline{u^{\prime}} \big) & I \big( \underline{u^{\prime}} \big) 
\end{bmatrix} \bigg\} \text{, } 
\]

\noindent for collections of spectral parameters $\big\{ u_i \big\}$, $\big\{ v_j \big\}$, $\big\{ w_k \big\}$, $\big\{ u^{\prime}_i \big\}$, $\big\{ v^{\prime}_j \big\}$, and $\big\{ w^{\prime}_k \big\}$, where the matrices in each entry of the tensor product of the Poisson bracket above corresponds the three-dimensional product representation of L-operators for the 20-vertex model, {\color{blue}[45]}, given spectral parameters, $\underline{u}$ and $\underline{u^{\prime}}$,

\begin{align*}
\underline{u} \equiv  \begin{bmatrix} \underline{u}_1 \\ \underline{u}_2 \\ \underline{u}_3  \end{bmatrix}
 \text{, } \\ \underline{u^{\prime}} \equiv  \begin{bmatrix}  \underline{u^{\prime}_1} \\ \underline{u^{\prime}_2} \\ \underline{u^{\prime}_3}  \end{bmatrix}  \text{, }
\end{align*}

\noindent with,

\[
      \begin{bmatrix}
 A \big( \underline{u} \big) & D \big( \underline{u} \big)  & G \big( \underline{u} \big) \\ B \big(\underline{u} \big) & E \big( \underline{u} \big) & H \big( \underline{u} \big)  \\ C \big( \underline{u} \big)  &  F \big( \underline{u} \big) & I \big( \underline{u} \big) 
\end{bmatrix}   \text{,  } 
\]

\noindent and with,

\[\begin{bmatrix}
 A \big( \underline{u^{\prime}} \big) & D \big( \underline{u^{\prime}} \big)  & G \big( \underline{u^{\prime}} \big) \\ B \big( \underline{u^{\prime}}\big) & E \big( \underline{u^{\prime}} \big) & H \big( \underline{u^{\prime}} \big)  \\ C \big( \underline{u^{\prime}} \big)  &  F \big( \underline{u^{\prime}} \big) & I \big( \underline{u^{\prime}} \big) 
\end{bmatrix} \text{. } \]

\noindent Under a choice of suitable parameters of the L-operator in order for the equality, given $M,N$ to be integers taken sufficiently large,

\[
T  \big(   \underline{M} , N ,  \underline{\lambda_{\alpha}} , u , v , w \big) \equiv    \overset{\underline{M}}{\underset{\underline{j}=0}{\prod}}  \text{ }  \overset{0}{\underset{i=-N}{\prod}}  \bigg\{ \mathrm{exp} \big( \lambda_3 ( q^{-2} \xi^{s_i} ) \big)  \bigg[   \begin{smallmatrix}     q^{D_i}       &    q^{-2} a_i q^{-D_i-D_j} \xi^{s-s_i}        &   a_i a_{j} q^{-D_i - 3D_j} \xi^{s - s_i - s_j}  \\ a^{\dagger}_i q^{D_i} \xi^{s_i} 
             &      q^{-D_i + D_j} - q^{-2} q^{D_i -D_j} \xi^{s}     &     - a_j q^{D_i - 3D_j} \xi^{s-s_j}  \\ 0  &    a^{\dagger}_j q^{D_j} \xi^{s_j} &  q^{-D_j} \\    \end{smallmatrix} \bigg]    \bigg\} \text{, } 
\]

\noindent In the same manner in which taking products of two-dimensional L-operators, three-dimensional L-operators of the type above, which hold up to Dynkin automorphism, can be used to approximate the quantum monodromy matrix from the mapping,

\begin{align*}
 T^{3D}_{a,b} \big(    \big\{ u_i \big\} , \big\{ v^{\prime}_j  \big\} , \big\{ w^{\prime\prime}_k \big\}     \big) :   \textbf{C}^3 \otimes \big( \textbf{C}^3 \big)^{\otimes ( |N| + ||M||_1 )}  \longrightarrow   \textbf{C}^3 \otimes \big( \textbf{C}^3 \big)^{\otimes ( |N| + ||M||_1 )}  \\   \mapsto  \overset{0}{\underset{j=-N}{\prod}} \text{ }  \overset{\underline{M}}{\underset{k=0}{\prod}} \bigg[  \mathrm{diag} \big( \mathrm{exp} \big(  \alpha \big(i, j,k \big)  \big)   , \mathrm{exp} \big(  \alpha \big( i,j,k\big)  \big)   , \mathrm{exp} \big(  \alpha \big( i, j,k \big)  \big) \big) \\ \times  R_{ia,jb,kc} \big( u - u_i ,  u^{\prime} - v^{\prime}_j , w-w^{\prime\prime}_k \big) \bigg]  \text{, } 
\end{align*}

\noindent for $N<0$, a suitable function $\alpha \big( i , j , k \big)$, and $R \equiv \mathcal{R}$ denotes the universal R-matrix (a discussion of $\mathcal{R}$, in terms of four factors, from {\color{blue}[5]}, is reproduced in {\color{blue}[45]}). A factorization of $\mathcal{R}$ is extremely significant for determining the quantities which satisfy the Yang-Baxter relation, which together with action-angle variables, provide characterizations of integrability. In the same way that a two-dimensional product representation of L-operators was analyzed in {\color{blue}[41]},

\[
\begin{bmatrix}
       A \big( \lambda_{\alpha} \big)   & B \big( \lambda_{\alpha} \big)   \\
    C \big( \lambda_{\alpha} \big)  & D \big( \lambda_{\alpha} \big)  \text{ }  
  \end{bmatrix}
\]

\noindent where each block of the representation is parametrized in spectral parameters described in {\color{blue}[7]}, for the 20-vertex model the three-dimensional product takes the form, {\color{blue}[45]},

\[
      \begin{bmatrix}
 A \big( \underline{u} \big) & D \big( \underline{u} \big)  & G \big( \underline{u} \big) \\ B \big(\underline{u} \big) & E \big( \underline{u} \big) & H \big( \underline{u} \big)  \\ C \big( \underline{u} \big)  &  F \big( \underline{u} \big) & I \big( \underline{u} \big) 
\end{bmatrix}   \text{, } 
\]

\noindent Albeit some differences between the notation for the two, and three, dimensional blocks of the two product representations above, the three-dimensional product representation exhibits many dependencies not in the two-dimensional product representation, from mappings into a unital associative algebra, amongst other algebraic quantities, which do not appear in the spectral parameters, Pauli basis elements, and other components, of two-dimensional L-operators.

For some vector $\underline{M} \equiv \big( M_1 , M_2 \big)$ over $\textbf{R}^2$, a negative real $N$, and natural $j,k$, the three-dimensional transfer matrix takes the form,

\begin{align*}
 \textbf{T}^{3D} \big( \underline{\lambda} \big)   \equiv    \underset{ N \longrightarrow -  \infty}{\underset{\underline{M} \longrightarrow + \infty}{\mathrm{lim}}} \mathrm{tr} \bigg\{     \overset{\underline{M}}{\underset{j=0}{\prod}}  \text{ }  \overset{0}{\underset{k=-N}{\prod}} \mathrm{exp} \big( \lambda_3 ( q^{-2} \xi^{s^j_k} ) \big)     \bigg[ \begin{smallmatrix}     q^{D^j_k}       &    q^{-2} a^j_k q^{-D^j_k -D^j_{k+1}} \xi^{s-s^k_j}        &  *_2  \\ \big( a^j_k \big)^{\dagger} q^{D^j_k} \xi^{s^j_k} 
             &    *_1     &     - a^j_k q^{D^j_k - 3D^j_{k+1}} \xi^{s-s^j_k}  \\ 0  &    a^{\dagger}_j q^{D^j_k} \xi^{s^j_k} &  q^{-D^j_k} \\    \end{smallmatrix} \bigg]      \bigg\} 
\text{, }
\end{align*}

\noindent where,

\begin{align*}
 *_1 \equiv   q^{-D^j_k + D^j_{k+1}} - q^{-2} q^{D^j_k -D^j_{k+1}} \xi^{s}    \text{, }\\  *_2 \equiv a^j_k a^j_{k+1} q^{-D^j_k - 3D^j_{k+1}} \xi^{s - s^j_k - s^j_{k+1}}  \text{, }
\end{align*}

\noindent which in previous work of the author in {\color{blue}[45]}, is relabled as,

\begin{align*}
L^{3D}_2 \big( \xi_k \big)  \equiv  \begin{bmatrix}     q^{D^j_k}       &    q^{-2} a^j_k q^{-D^j_k -D^j_{k+1}} \xi^{s-s^k_j}        &  a^j_k a^j_{k+1} q^{-D^j_k - 3D^j_{k+1}} \xi^{s - s^j_k - s^j_{k+1}}  \\ \big( a^j_k \big)^{\dagger} q^{D^j_k} \xi^{s^j_k} 
             &      q^{-D^j_k + D^j_{k+1}} - q^{-2} q^{D^j_k -D^j_{k+1}} \xi^{s}     &     - a^j_k q^{D^j_k - 3D^j_{k+1}} \xi^{s-s^j_k}  \\ 0  &    a^{\dagger}_j q^{D^j_k} \xi^{s^j_k} &  q^{-D^j_k} \\    \end{bmatrix}   \text{. }
\end{align*}

\noindent The representation of the three-dimensional transfer matrix above was studied extensively in {\color{blue}[45]}, in which several recursive relations from L-operators were obtained.

\subsection{Preliminaries for representations of nonlocal 20-vertex domain-wall correlations}

All of the objects introduced in this subsection, within the quantum inverse scattering framework, can be leveraged for computing nonlocal correlations for the 20-vertex model by following the three items, \underline{(1)}, \underline{(2)}, and \underline{(3)}, from computations of the authors in {\color{blue}[7]} provided at the end of \textit{1.2} in the previous section. However, despite the fact that there are some similarities between two, and three, dimensional quantum inverse scattering methods, such as those previously examined by the author in {\color{blue}[41]} for the inhomogeneous 6-vertex model, and in {\color{blue}[45]} for the 20-vertex model, performing comptuations to obtain closed formed representations of nonlocal correlations for the 20-vertex model presents differences to correlations that have been obtained over the finite lattice for the 6-vertex model, including: lacking notions of exact solvability, integrability, Hamiltonian flows, and Poisson commutativity, from solutions to the Euler Lagrange equations {\color{blue}[24},{\color{blue}45]}; a broader set of relations, 81, corresponding to the three-dimensional Poisson structure, in comparison to 16 relations for the two-dimensional Poisson structure; different asymptotic properties of the transfer and quantum monodromy matrices. 

To make use of the determinantal representation for the 20-vertex partition function derived in {\color{blue}[15]}, recall that in the case of the 6-vertex model, taking the homogeneous limit of all spectral parameters, leads to the contour integral representation for the bottom partition function, $Z^{\mathrm{Bottom}}$, {\color{blue}[7]},

\begin{align*}
  \underline{Z^{\mathrm{Bottom}}_{r_1 , \cdots , r_s}} \equiv Z_N \frac{\underset{1 \leq j \leq s}{\prod} t^{j-r_j}}{a^{s(N-1)} c^s} \bigg\{  \underset{C_0}{\oint}  \times \cdots \times     \underset{C_0}{\oint}  \bigg[   \underset{1 \leq j \leq s}{\prod}    \frac{1}{z^{r_j}_j}  \underset{1 \leq j < k \leq s}{\prod}  \frac{z_k - z_j}{t^2 z_j z_k - 2 \Delta t z_j + 1}  \\ \times   h_{N,s} \big( z_1 , \cdots , z_s \big) \bigg]   \frac{\mathrm{d}^s s}{\big( 2 \pi i \big)^s}     \bigg\}      \text{, } \tag{1}
\end{align*}

\noindent where, besides the "full" partition function $Z_N$ over a finite volume in the square lattice of some positive length $N$, $s$, $C_0$, and $t$, $\Delta t$, respectively denote a strictly positive parameter, a contour surrounding the origin of the finite lattice, and another two strictly positive parameters; explicitly, the function $h$ preceding the measure $\mathrm{d}^s s $ for the above countour integral representation equals, {\color{blue}[7]}.

\begin{align*}
 h_{N,s} \big( z_1 , \cdots , z_s \big) \equiv  \frac{\underset{1 \leq j , k \leq s}{\mathrm{det}}  \big\{    z^{s-j}_k \big( z_k - 1 \big)^{j-1} h_{N-s+j } \big( z_k \big)  \big\} }{ \underset{1 \leq j < k \leq s}{\prod} \big( z_k - z_j \big) }  \text{, }
\end{align*}

\noindent As objects over $\textbf{T}$ for the 20-vertex model one must instead define, and manipulate, the  partition functions, $Z^{\mathrm{Bottom}}_{\textbf{T}} \equiv Z^{\mathrm{Bottom}}$, $Z^{\mathrm{Top}}_{\textbf{T}} \equiv Z^{\mathrm{Top}}$, and $Z^{\mathrm{Side}}_{\textbf{T}} \equiv Z^{\mathrm{Side}}$, corresponding to the tensor product decomposition of the spaces, for some finite volume in $\textbf{T}$ that is of strictly positive length $N>0$,

\begin{align*}
    \text{First degree of freedom} \equiv \ket{ \Uparrow_{1 , \cdots , N} } = \underset{1 \leq k \leq N}{\bigotimes}  \ket{\uparrow_k } \text{, } \\ 
    \text{Second degree of freedom}  \equiv  \ket{\Downarrow_{1 , \cdots , N}} = \underset{1 \leq k \leq N}{\bigotimes} \ket{\downarrow_k} \text{, }  \\  \text{Third degree of freedom} \equiv \ket{\Rightarrow_{1 , \cdots , N} } = \underset{1 \leq k \leq N}{\bigotimes}  \ket{\rightarrow_k} \text{. }
\end{align*}

\noindent In arguments that follow, one can also introduce decompositions of projections of each degree of freedom to horizontal, or vertical, components only, in which,

\begin{align*}
    \text{Horizontal restriction of first degree of freedom} \equiv \ket{ \Uparrow^H_{1 , \cdots , N} } = \underset{1 \leq k \leq N}{\bigotimes}  \ket{\uparrow^H_k } \text{, } \\     \text{Horizontal restriction of 
 second degree of freedom}  \equiv  \ket{\Downarrow^H_{1 , \cdots , N}} = \underset{1 \leq k \leq N}{\bigotimes} \ket{\downarrow^H_k} \text{, } \\   \text{Horizontal restriction of  third degree of freedom} \equiv \ket{\Rightarrow^H_{1 , \cdots , N} } = \underset{1 \leq k \leq N}{\bigotimes}  \ket{\rightarrow^H_k} \text{, }
\end{align*}

\noindent and similarly for the vertical component. Each one of the states above coincides with the basis of the triangular lattice, $\mathrm{span} \big\{ \ket{\Uparrow_{1 , \cdots , N}}, \ket{\Downarrow_{1 , \cdots , N}}, \ket{\Rightarrow_{1 , \cdots , N}} \}$, where,

\begin{align*}
\ket{\Uparrow_{1 , \cdots , N}} = \begin{bmatrix}
   \frac{\sqrt{3}}{2} \\ \frac{1}{2} \\ 0  
 \end{bmatrix}   \text{, } \\  \ket{\Rightarrow_{1 , \cdots , N}}  =    \begin{bmatrix}
   \frac{\sqrt{3}}{2} \\ - \frac{1}{2} \\ 0  
 \end{bmatrix}   \text{, } \\ \ket{\Downarrow_{1 , \cdots , N}}  =    \begin{bmatrix}
 0  \\ 0 \\ 1  
 \end{bmatrix}   \text{. } 
\end{align*}

\noindent From the tensor product decomposition into $N$ components above, the three-dimensional transfer matrix, denoted in the previous section with $T^{3D}$, admits each corresponding factorization,

\begin{align*}
    \big( T_1 \big)_{1, \cdots , N} \equiv T_1 = 
    \begin{bmatrix}
 \big( A_1 \big)_{1 , \cdots , N} \big( \underline{u} \big) & \big( D_1 \big)_{1 , \cdots , N} \big( \underline{u} \big)  & \big( G_1 \big)_{1 , \cdots , N} \big( \underline{u} \big) \\\big( B_1 \big)_{1 , \cdots , N} \big(\underline{u} \big) & \big( E_1 \big)_{1 , \cdots , N } \big( \underline{u} \big) &  \big( H_1 \big)_{1 , \cdots , N} \big( \underline{u} \big)  \\ \big( C_1 \big)_{1 , \cdots , N} \big( \underline{u} \big)  & \big(  F_1 \big)_{1 , \cdots , N } \big( \underline{u} \big) &  \big( I_1 \big)_{1 , \cdots , N} \big( \underline{u} \big) 
\end{bmatrix}   \\  = \underset{1 \leq k \leq N}{\prod}  
    \begin{bmatrix}
\big(   A_1 \big)_k   \big( \underline{u} \big) & \big(   D_1   \big)_k \big( \underline{u} \big)  & \big( G_1 \big)_k  \big( \underline{u} \big) \\ \big( B_1 \big)_k \big(\underline{u} \big) & \big( E_1 \big)_k \big( \underline{u} \big) & \big(  H_1 \big)_k \big( \underline{u} \big)  \\ \big(  C_1 \big)_k\big( \underline{u} \big)  & \big(  F_1 \big)_k \big( \underline{u} \big) & \big( I_1 \big)_k \big( \underline{u} \big) 
\end{bmatrix}   \text{, } \\   \big( T_2 \big)_{1 , \cdots , N} \equiv T_2 =    \begin{bmatrix}
\big(  A_2 \big)_{1 , \cdots , N} \big( \underline{u} \big) &  \big( D_2 \big)_{1 , \cdots , N} \big( \underline{u} \big)  &  \big( G_2 \big)_{1 , \cdots , N} \big( \underline{u} \big) \\ \big( B_2 \big)_{1 , \cdots , N} \big(\underline{u} \big) &  \big( E_2 \big)_{1 , \cdots , N} \big( \underline{u} \big) &  \big( H_2 \big)_{1 , \cdots , N} \big( \underline{u} \big)  \\ \big( C_2 \big)_{1 , \cdots , N} \big( \underline{u} \big)  & \big(  F_2 \big)_{1 , \cdots , N} \big( \underline{u} \big) &  \big( I_2 \big)_{1 , \cdots , N} \big( \underline{u} \big) 
\end{bmatrix}  \\ =   \underset{1 \leq k \leq N}{\prod}   \begin{bmatrix}
\big(  A_2  \big)_k \big( \underline{u} \big) &  \big( D_2 \big)_k \big( \underline{u} \big)  & \big( G_2 \big)_k \big( \underline{u} \big) \\ \big( B_2  \big)_k \big(\underline{u} \big) & \big( E_2 \big)_k \big( \underline{u} \big) & \big( H_2  \big)_k \big( \underline{u} \big)  \\ \big(  C_2 \big)_k\big( \underline{u} \big)  & \big( F_2 \big)_k \big( \underline{u} \big) & \big(  I_2 \big)_k  \big( \underline{u} \big) 
\end{bmatrix}     \text{, } \\ 
    \big( T_3 \big)_{1,\cdots , N} \equiv  T_3 =    \begin{bmatrix}
 \big( A_3 \big)_{1 , \cdots , N} \big( \underline{u} \big) & \big( D_3  \big)_{1 , \cdots , N} \big( \underline{u} \big)  & \big( G_3  \big)_{1 , \cdots , N} \big( \underline{u} \big) \\ \big( B_3 \big)_{1 , \cdots , N} \big(\underline{u} \big) & \big( E_3  \big)_{1 , \cdots , N} \big( \underline{u} \big) & \big( H_3  \big)_{1 , \cdots , N}  \big( \underline{u} \big)  \\ \big( C_3  \big)_{1 , \cdots , N} \big( \underline{u} \big)  &  \big( F_3  \big)_{1 , \cdots , N} \big( \underline{u} \big) & \big( I_3  \big)_{1 , \cdots , N} \big( \underline{u} \big) 
\end{bmatrix}  \\ = \underset{1 \leq k \leq N}{\prod}    \begin{bmatrix}
\big(  A_3 \big)_k \big( \underline{u} \big) & \big( D_3 \big)_k  \big( \underline{u} \big)  &  \big( G_3 \big)_k \big( \underline{u} \big) \\ \big(  B_3 \big)_k \big(\underline{u} \big) & \big( E_3 \big)_k \big( \underline{u} \big) & \big( H_3 \big)_k \big( \underline{u} \big)  \\ \big(  C_3 \big)_k \big( \underline{u} \big)  &  \big( F_3 \big)_k \big( \underline{u} \big) & \big( I_3 \big)_k  \big( \underline{u} \big) 
\end{bmatrix}   \text{, }
 \end{align*}

\noindent into $T_1$, $T_2$, and $T_3$, respectively, for the block representations $\big( A_1 \big)_k, \cdots , \big( A_3 \big)_k, \cdots , \big(I_3\big)_k$ of the three-dimensional transfer matrix. Introducing such a decomposition for the product of rows of the transfer matrix representation, is useful for computations of three-dimensional correlation functions for the 20-vertex model, as: (1) local representations for correlations in the 6-vertex, and 20-vertex, models alike can be factorized into two, or one, dimensional interactions; (2) contour integral representations for correlation functions, and the emptiness formation probability, are dependent upon the collection of singularities of a collection of suitable surfaces over one, and two, dimensions; (3) an asymmetrization procedure, which involves a computation of a contour integral representation with a prefactor dependent upon a product of partition functions.

Besides decompositions for the product of the transfer matrix along each degree of freedom, from the determinantal representation for the nonlocal correlation in the 6-vertex model reproduced at the beginning of the section from {\color{blue}[7]}, it is expected that the representation for the nonlocal correlations of the 20-vertex is proportional to the determinant,

\begin{align*}
 \mathrm{det} \bigg[ \frac{c}{a \big( \lambda_{\alpha} , \nu_k \big) b \big( \lambda_{\alpha} , \nu_k \big) }  \bigg]_{\alpha , k} \text{, }
\end{align*}

\noindent from the homogenized representation {\color{blue}[15]}. Relatedly, another object that one examines,

\begin{align*}
  H^{(r_1 , \cdots , r_s , r^{\prime}_1 , \cdots , r^{\prime}_s)} \equiv  \frac{Z^{\mathrm{Bottom}} Z^{\mathrm{Top}} Z^{\mathrm{Side}}}{Z_N} \text{, }
\end{align*}

\noindent by taking the ratio of the three partiton functions, with the partition function over the entire volume of some length $N>0$, which is a function of spectral parameters $r_1 , \cdots , r_s , r^{\prime}_1 , \cdots , r^{\prime}_s$, corresponding to incoming arrows of domain wall boundary conditions for the 20-vertex model; as reflected in determinantal formulae for nonlocal, and local, correlations of the 6-vertex in past work, the determinantal representation for the partition function over a smaller finite volume of the lattice is related to the partition function over a larger finite volume of the lattice, in addition to being dependent upon more spectral parameters than $r_1 , \cdots , r_s$ for the 6-vertex model. Irrespective of the product decomposition for each degree of freedom over $k$, to study behaviors of the emptiness formation probability for the 20-vertex model, one examines products of the form,

\begin{align*}
{\underset{1 \leq \vec{r} \leq \bar{s} }{\prod}} L^{3D}_2 \big( \xi_i \big)   =   \underset{1 \leq j \leq s_2}{\underset{1 \leq i \leq s_1}{\prod}} L^{3D}_2 \big( \xi_i \big)             \text{, }
\end{align*}

\noindent corresponding to the weak infinite volume approximation of the three-dimensional transfer matrix up to lines $s_1$ and $s_2$, with spectral parameters $\xi_i$ for every $1 \leq i \leq s$. The remaining terms in the approximation of the product representation over finite volume,

\begin{align*}
  \underset{s_2 + 1 \leq j \leq N}{\underset{s_1 + 1 \leq i \leq N}{\prod}} L^{3D}_2 \big( \xi_i \big)             \text{, }
\end{align*}

\noindent are incorporated into the bottom partition function, ie in the complementary space over the same finite volume. In computations for obtaining the representation of nonlocal correlations for the 20-vertex model in the next section, one must consider products of operators of the following form,

\begin{align*}
  \bigg[  \underset{1 \leq k \leq s}{\prod}  A_k \big( \underline{u} \big) \bigg] \ \ket{\Uparrow_{1 , \cdots , N} }  \propto   \bigg[  \underset{1 \leq k \leq s}{\prod}  \big( A_{1,2,3} \big)_k \big( \underline{u} \big) \bigg] \ket{\Uparrow_{1 , \cdots , N} }          \text{, } \\            \bigg[  \underset{1 \leq k \leq s}{\prod}  B_k \big( \underline{u} \big) \bigg] \ \ket{\Uparrow_{1 , \cdots , N} }  \propto   \bigg[  \underset{1 \leq k \leq s}{\prod}  \big( B_{1,2,3} \big)_k \big( \underline{u} \big) \bigg] \ket{\Uparrow_{1 , \cdots , N} }                        \text{, } \\     \vdots    \\       \bigg[  \underset{1 \leq k \leq s}{\prod}  I_k \big( \underline{u} \big) \bigg] \ \ket{\Uparrow_{1 , \cdots , N} }  \propto   \bigg[  \underset{1 \leq k \leq s}{\prod}  \big( I_{1,2,3} \big)_k \big( \underline{u} \big) \bigg] \ket{\Uparrow_{1 , \cdots , N} }                     \text{, } 
\end{align*}

\noindent in which operators $A_k \big( \underline{u} \big) , \cdots , I_k  \big( \underline{u} \big)$ act on the up state, $\ket{\Uparrow_{1, \cdots , N}}$, for,

\begin{align*}
  \big( A_{1,2,3} \big)_k \big( \underline{u} \big)  \equiv \big\{ \big( A_1 \big)_k \big( \underline{u} \big), \big( A_2 \big)_k \big( \underline{u} \big), \big( A_3 \big)_k \big( \underline{u} \big) \big\}, \cdots ,   \big( I_{1,2,3} \big)_k \big( \underline{u} \big) \equiv \big\{ \big( I_1 \big)_k  \big( \underline{u} \big), \big( I_2 \big)_k \big( \underline{u} \big) \\ , \big( I_3 \big)_k \big( \underline{u} \big) \big\}  \text{. }
\end{align*}

\noindent Besides the observation that similar products can be formed for studying the action of the same collection of operators for $\ket{\Downarrow_{1,\cdots,N}}$, and also for $\ket{\Rightarrow_{1,\cdots,N}}$, the fundamental identity, for domain-wall boundary conditions in the 6-vertex model, which is used to compute the bottom partition function from the top partition function states, {\color{blue}[7]},

\begin{align*}
     A \big( \lambda_r \big) \bigg[ \underset{1 \leq \beta \leq r-1}{\prod}  B \big( \lambda_{\beta} \big) \bigg]  =   \bigg[  \underset{1 \leq \alpha \leq r}{\sum}  \frac{g \big( \lambda_{\alpha} , \lambda_r \big)}{f \big( \lambda_{\alpha} , \lambda_r \big)} \bigg] \bigg[ \underset{\beta \neq \alpha}{\underset{1 \leq \beta \leq r}{\prod}}   f \big( \lambda_{\alpha} , \lambda_{\beta} \big)      \bigg]   \bigg[ \underset{\beta \neq \alpha}{\underset{1 \leq \beta \leq r}{\prod}} B \big( \lambda_{\beta}    \big)          \bigg]  A \big( \lambda_{\alpha} \big)     \text{, }
\end{align*}

\noindent for spectral parameter $\lambda_r$ corresponding to the $r$th horizontal line of the finite lattice, and, {\color{blue}[7]},

\begin{align*}
  f \big( \lambda_{\alpha} , \lambda_r \big) \equiv               \frac{\mathrm{sin} \big( \lambda_r - \lambda_{\alpha}  + \eta \big) }{\mathrm{sin} \big( \lambda_r - \lambda_{\alpha} \big) }     \text{, }  \\   g \big( \lambda_{\alpha} , \lambda_r \big) \equiv            \frac{\mathrm{sin} \big( 2 \eta \big)}{\mathrm{sin} \big( \lambda_r  - \lambda_{\alpha} \big) }            \text{. }
\end{align*}

\noindent The above identity, from operators in the two-dimensional representation, 

\[
\begin{bmatrix}
       A \big( \lambda_{\alpha} \big)   & B \big( \lambda_{\alpha} \big)   \\
    C \big( \lambda_{\alpha} \big)  & D \big( \lambda_{\alpha} \big)  \text{ }  
  \end{bmatrix} \text{, }
\]

\noindent with spectral parameters $\lambda_{\alpha}$ and $\lambda_{\beta}$, will also be used to relate the computation of one partition function for the 20-vertex model to the computation of another partition function. As the weak finite volume exhausts $\textbf{T}$, the three-dimensional product representation of L-operators, as extensively characterized in {\color{blue}[45]}, is dependent upon obtaining a recursive set of relations for each one of the blocks $A \big( \underline{u} \big) , \cdots , I \big( \underline{u} \big)$. As such, the candidate representation for nonlocal correlations is of the form,

\begin{align*}
    Z^{\mathrm{Top}}_{r_1 , \cdots , r_s , r^{\prime}_1 , \cdots , r^{\prime}_s } =  \mathscr{P}    \bigg\{ \underset{\mathscr{S}_1  }{\oint}  \times \cdots \times     \underset{\mathscr{S}_1 }{\oint}   \bigg[ \mathscr{P}_1 \mathscr{P}_2   h^{\prime}_{N,s,s^{\prime}}         \frac{\mathrm{d}^{s^{\prime}} z^{\prime}}{ \big( 2 \pi i \big)^{s^{\prime}}}   \mu_0                  \bigg]       \bigg\}        \text{, } \tag{\textit{*}}
\end{align*}

\noindent given the same objects defined in the statement of the main result, $\textbf{Theorem}$, in \textit{1.4}, where $z$ and $z^{\prime}$ denote coordinates of the triangular lattice. In the next section, we perform computations to approximate each of the partition functions over $\textbf{T}$.

To demonstrate that the desired representation above holds for the partition function, which then implies that the same desired contour integral representation holds for the correlation functions holds, we establish several parallels between the 6-vertex, and 20-vertex, models. The most salient characteristics that these two vertex models share in common include:

\begin{itemize}
\item[$\bullet$] \textit{Encoding of domain-wall boundaries}. Domain-wall boundary conditions exist for both vertex modoels, which significantly allows for comparisons between contour integral representations for the partition function, and the correlation function.
\item[$\bullet$] \textit{Determinantal representations for the partition function}. Both vertex models have determinantal representations for the partition function, which also enters into several computations with contour integral representations.
\item[$\bullet$] \textit{Collection of residues over which the support of contour integrals is defined}. Contour integral representations for the partition, and correlation, functions alike depend upon residues of surfaces. 
\end{itemize}

\section{Computation of nonlocal correlations}

\noindent In the following, we implement several computations for obtaining the desired contour integral representation for 20-vertex correlation functions. In the presence of domain-wall boundary conditions, the contour integral representation is dependent upon two-dimensional representations of the partition function. In comparison to the computation of correlation functions for the 6-vertex model under domain-wall boundary conditions, {\color{blue}[7]}, correlation functions over three-dimensions are obtained from the partition functions $Z_{\mathrm{Top}}$, $Z_{\mathrm{Bottom}}$, and $Z_{\mathrm{Side}}$. As representatives of the summation over the sample spaces, $\Omega_{\mathrm{Top}}$, $\Omega_{\mathrm{Bottom}}$, and $\Omega_{\mathrm{Side}}$, namely the sample spaces of configurations for the 20-vertex model, which satisfy,

\begin{align*}
  \Omega_{\mathrm{20} \text{V}} \equiv \Omega_{\mathrm{Top}} \cup   \Omega_{\mathrm{Bottom}} \cup    \Omega_{\mathrm{Side}}      \text{. }
\end{align*}

\noindent Given the fact that each one of the sample spaces introduced for the decomposition of the \textit{entire} sample space of configurations for the 20-vertex model above can be expressed from the set of linear combinations,

\begin{align*}
  \Omega_{\mathrm{Top}} \equiv   \mathrm{span} \big\{ \text{edges } e: e \text{ is the restriction of edges to the first degree of freedom of } \textbf{T} \big\} \text{,} \\ \\ \Omega_{\mathrm{Bottom}} \equiv  \mathrm{span} \big\{ \text{edges } e: e \text{ is the restriction of edges to the second degree of freedom of } \textbf{T} \big\} \text{,}  \\ \\ \Omega_{\mathrm{Side}}     \equiv \mathrm{span} \big\{ \text{edges } e: e \text{ is the restriction of edges to the third degree of freedom of } \textbf{T} \big\}  \text{.} 
\end{align*}

\noindent Given the three samples spaces above, representations for one-dimensional correlation functions of the 6-vertex model can be generalized to obtain two-dimensional correlation functions which appear in the representation for correlation functions supported over $\textbf{T}$. In comparison to the underlying Yang-Baxter algebra which is used for computing the product of blocks of two-dimensional representations for the 6-vertex transfer matrix, for the three-dimensional 20-vertex transfer matrix a generalization of relations for the underlying Yang-Baxter algebra is provided in the next section. As input, the relations can be obtained from three three-dimensional transfer matrices of the 20-vertex model. More generally, the fact that products of block representations of transfer matrices for the 6-vertex and 20-vertex models can be expressed as relations of Yang-Baxter algebras relates to computations with the Poisson bracket {\color{blue}[41, 45]}.

\subsection{Overview}

\noindent We confirm that the representation for nonlocal correlations in the 20-vertex model agrees with that given at the end of the previous section. To this end, we manipulate the inhomogeneous determinantal representation for the 20-vertex model, which in, \textit{2.5}, {\color{blue}[15]}, is shown to be in correpondence with the determinantal representation for the 20-vertex domain-wall partition function,

\begin{align*}
 Z^{\mathrm{6V, \text{U-turn}}}_{\text{Inhomogeneous}}  \propto              \mathrm{det}     \mathcal{M}^{\prime}                     \text{, }
\end{align*}

\noindent where,

\begin{align*}
  \mathcal{M}^{\prime}_{i,j} \equiv      \mathcal{M}^{\prime} \big( i , j\big)  =       \frac{1}{a \big(  z_i ,  w_j  \big)  b \big( z_i , w_j   \big) } - \frac{1}{a \big( 1 , z_i w_j 
 \big)  b \big( 1 , z_i w_j   \big) }              \text{, }
\end{align*}

\noindent where $\big( a,b, c \big) \big( z , w \big) = \big( a_3 , b_3 , c_3 \big) \big( z,w \big) = \big( A,B,C \big) \big( qz , q^{-1} w\big)$, before taking the homogeneous limit. The prefactor for the U-turn 6-vertex partition function, besides the determinantal factor, includes several complicated product factors dependent upon the same parameters that $\mathcal{M}^{\prime}$ is. Furthermore, we also manipulate the bottom partition function, from the expression obtained for the top partition function provided in {\color{blue}[7]}. However, the relation that we introduce for performing intertwinning like operations on $A$, or $B$, ie, two block representations from the two-dimensional product representation of the transfer matrix, instead takes the form,

\begin{align*}
     A \big( \lambda_r , \lambda_{r^{\prime}} \big)      \bigg[  \underset{1 \leq \beta^{\prime} \leq r^{\prime}-1}{\underset{1 \leq \beta \leq r-1}{\prod}}    B \big(   \lambda_{\beta} , \lambda_{\beta^{\prime}} \big)   \bigg]  =  \bigg[  \underset{1 \leq \alpha^{\prime} \leq r^{\prime}}{\underset{1 \leq \alpha r}{\sum}}   \frac{g \big( \lambda_{\alpha} , \lambda_{r} , \lambda_{r^{\prime}} \big) }{f \big( \lambda_{\alpha} , \lambda_{r} , \lambda_{r^{\prime}} \big)}      \bigg]    \bigg[          \underset{1 \leq \beta^{\prime} \leq r^{\prime}-1}{\underset{1 \leq \beta \leq r}{\underset{\beta^{\prime} \neq \alpha^{\prime}}{\underset{\beta \neq \alpha}{\prod} }}}                 f \big( \lambda_{\beta} , \lambda_r , \lambda_{r^{\prime}} \big)            \bigg]         \\ \times      \bigg[    \underset{1 \leq \beta^{\prime} \leq r^{\prime}-1}{\underset{1 \leq \beta \leq r-1}{\underset{\beta^{\prime} \neq \alpha^{\prime}}{\underset{\beta \neq \alpha}{\prod}}}}     B \big( \lambda_{\beta} , \lambda_{\beta^{\prime}} \big)                 \bigg]  A \big( \lambda_{\alpha} , \lambda_{\alpha^{\prime}} \big)            \text{, }
\end{align*}

\noindent for the functions,

\begin{align*}
f \big( \lambda_{\alpha} , \lambda_{r} , \lambda_{r^{\prime}} \big) \equiv     \frac{\mathrm{sin} \big( \lambda_{r^{\prime}} - \lambda_r - \lambda_{\alpha} + 2 \eta  \big)}{\mathrm{sin} \big( \lambda_{r^{\prime}} - \lambda_r - \lambda_{\alpha } \big) }                    \text{, } \\ 
g \big( \lambda_{\alpha} , \lambda_{r} , \lambda_{r^{\prime}} \big) \equiv   \frac{\mathrm{sin} \big( 2 \eta \big)}{\mathrm{sin} \big( \lambda_{r^{\prime}} - \lambda_r - \lambda_{\alpha} \big)}                  \text{, }
\end{align*}

\noindent and, 

\begin{align*}
 A \big( \underline{u} \big)   \equiv   A \big( u , \underline{\lambda} \big) \equiv B \big( u , \lambda_r , \lambda_{r^{\prime}} \big) \equiv A \big( \lambda_r , \lambda_{r^{\prime}} \big)          \text{, } \\ B \big( \underline{u} \big)  \equiv B \big( u , \underline{\lambda} \big) \equiv B \big( u , \lambda_{\beta} , \lambda_{\beta^{\prime}} \big) \equiv B \big( \lambda_{\beta} , \lambda_{\beta^{\prime}}  \big)  \text{, }
\end{align*}

\noindent which are the blocks of the product representation for the three-dimensional transfer matrix, {\color{blue}[45]},

\[
      \begin{bmatrix}
 A \big( \underline{u} \big) & D \big( \underline{u} \big)  & G \big( \underline{u} \big) \\ B \big(\underline{u} \big) & E \big( \underline{u} \big) & H \big( \underline{u} \big)  \\ C \big( \underline{u} \big)  &  F \big( \underline{u} \big) & I \big( \underline{u} \big) 
\end{bmatrix}   \text{. } 
\]

\subsection{Yang-Baxter algebra}

\noindent Underlying the intertwinning operation for operators $A$ and $B$ above is a higher-dimensional analog of the Yang-Baxter algebra, the lower dimensional version of which is discussed in various applications of the quantum inverse scattering type method, including {\color{blue}[7]} for the two-dimensional Yang-Baxter algebra. Over $\textbf{T}$, in comparison to over $\textbf{Z}^2$, the three-dimensional Yang-Baxter algebra partly consists of relations,

\begin{align*}
         \underline{G \big( \underline{u} \big) E \big( \underline{u^{\prime}} \big) C \big( \underline{u^{\prime\prime}} \big) } =       f \big( \lambda_{\alpha} , \lambda_r , \lambda_{r^{\prime}} \big)   f \big( \lambda , \lambda^{\prime} \big)     C \big( \underline{u^{\prime\prime}} \big) E \big( \underline{u^{\prime}} \big)  G \big( \underline{u} \big)     +    f \big( \lambda_{\alpha} , \lambda_r , \lambda_{r^{\prime}} \big) g \big( \lambda^{\prime} , \lambda \big) \\ \times     C \big( \underline{u^{\prime}}    \big) E \big( \underline{u^{\prime\prime}} \big)    G \big( \underline{u} \big)                
         +        g \big( \lambda_{\alpha} , \lambda_r , \lambda_{r^{\prime}} \big)      f \big(   \lambda , \lambda^{\prime} \big)  C \big( \underline{u^{\prime}} \big) E \big( \underline{u} \big)         G \big( \underline{u^{\prime\prime} } \big)     \\          +  g \big( \lambda_{\alpha} , \lambda_r , \lambda_{r^{\prime}} \big)      g \big( \lambda^{\prime} , \lambda \big)   C \big(  \underline{u}     \big) E \big( \underline{u^{\prime}}       \big)                      G \big( \underline{u^{\prime\prime} } \big)                \text{, }    \\ \\  \underline{ I \big( \underline{u} \big) H \big( \underline{u^{\prime}} \big) G \big( \underline{u^{\prime\prime}} \big)}  =           f \big( \lambda_{\alpha} , \lambda_r , \lambda_{r^{\prime}} \big)       f \big( \lambda , \lambda^{\prime} \big)                G \big( \underline{u^{\prime\prime}} \big) H \big( \underline{u^{\prime}} \big)        I \big( \underline{u} \big)     +   f \big( \lambda_{\alpha} , \lambda_r , \lambda_{r^{\prime}} \big)   g \big( \lambda^{\prime} , \lambda \big)   \\ \times  G \big( \underline{u^{\prime}} \big) H \big( \underline{u^{\prime\prime}} \big)                          I \big( \underline{u} \big)    +  g \big( \lambda_{\alpha} , \lambda_r  , \lambda_{r^{\prime}} \big)    f \big( \lambda , \lambda^{\prime} \big) G \big( \underline{u^{\prime\prime}} \big) H \big(   \underline{u^{\prime}} \big) I \big( \underline{u^{\prime}}  \big) \\   +  g \big( \lambda_{\alpha} , \lambda_r  , \lambda_{r^{\prime}} \big)          g \big( \lambda^{\prime} , \lambda \big)                       G \big( \underline{u^{\prime}} \big)  H \big( \underline{u^{\prime\prime}} \big) I \big( \underline{u^{\prime}}  \big)        \text{, }     \\ \\   \underline{  A \big( \underline{u }  \big) D \big( \underline{u^{\prime}} \big) G \big( \underline{u^{\prime\prime}} \big) }   =          f \big( \lambda_{\alpha}  , \lambda_r , \lambda_{r^{\prime}} \big)               f \big( \lambda , \lambda^{\prime} \big)                       G \big(    \underline{u^{\prime\prime}}   \big)    D \big( \underline{u^{\prime}} \big)       A \big( \underline{u} \big)   +  f \big( \lambda_{\alpha}  , \lambda_r , \lambda_{r^{\prime}} \big)    g \big( \lambda^{\prime} , \lambda \big)  \\ \times   G \big( \underline{u^{\prime}} \big) D \big( \underline{u^{\prime\prime}} \big)                               A \big( \underline{u} \big)       +   g \big( \lambda_{\alpha} , \lambda_r , \lambda_{r^{\prime}} \big)           f \big( \lambda , \lambda^{\prime} \big)   G \big( \underline{u} \big)  D \big( \underline{u^{\prime\prime}} \big)      A \big( \underline{u} \big)      \\ +  g \big( \lambda_{\alpha} , \lambda_r , \lambda_{r^{\prime}} \big)   g \big( \lambda^{\prime} , \lambda \big)  G \big( \underline{u^{\prime\prime}} D \big( \underline{u} \big)          A \big( \underline{u} \big)     \text{, }    \end{align*}

         \begin{align*}  
         \underline{A \big( \underline{u } \big) E \big( \underline{u^{\prime}} \big) I \big( \underline{u^{\prime\prime}} \big)}       =           f \big( \lambda_{\alpha}  , \lambda_r , \lambda_{r^{\prime}} \big)        f \big( \lambda , \lambda^{\prime} \big)     I \big( \underline{u^{\prime\prime}} \big) E \big( \underline{u^{\prime}} \big)  A \big( \underline{u} \big)    +    f \big( \lambda_{\alpha}  , \lambda_r , \lambda_{r^{\prime}} \big)    g \big( \lambda^{\prime} , \lambda \big)       \\ \times          I \big( \underline{u^{\prime}} \big) E \big( \underline{u^{\prime\prime}} \big)          A \big( \underline{u} \big)     +   g \big( \lambda_{\alpha} , \lambda_r , \lambda_{r^{\prime}} \big)        f \big( \lambda , \lambda^{\prime} \big)     I \big( \underline{u}  \big) E \big( \underline{u^{\prime\prime}} \big)               A \big( \underline{u^{\prime}} \big)    \\   + g \big( \lambda_{\alpha} , \lambda_r , \lambda_{r^{\prime}} \big)  g \big( \lambda^{\prime} , \lambda \big)                     I \big( \underline{u^{\prime\prime}} \big)           E \big( \underline{u}  \big)      A \big( \underline{u^{\prime}} \big)             \text{, } 
\end{align*}

\noindent which respectively correspond to the product of block representations for the transfer matrix which are either: (1) along the antidiagonal, (2) third column, (3) first row, and (4) diagonal, of the product representation for the transfer matrix. The first expression from the collection above in the three-dimensional Yang-Baxter algebra is related to the relation in the lower two-dimensional algebra, each of which takes the form, $B \big( \lambda \big) B \big( \lambda^{\prime} \big)$, $C \big( \lambda \big) C \big( \lambda^{\prime} \big)$, $A \big( \lambda \big) B \big( \lambda^{\prime} \big)$ {\color{blue}[7]}. Before stating the final identity that will be extrapolated to obtain the system of relations for the three-dimensional Yang-Baxter algerbra given above, introduce two more transfer matrices, the first of which is,

\[
      \begin{bmatrix}
 A \big( \underline{u^{\prime}} \big) & D \big( \underline{u^{\prime}} \big)  & G \big( \underline{u^{\prime}} \big) \\ B \big(\underline{u^{\prime}} \big) & E \big( \underline{u^{\prime}} \big) & H \big( \underline{u^{\prime}} \big)  \\ C \big( \underline{u^{\prime}} \big)  &  F \big( \underline{u^{\prime}} \big) & I \big( \underline{u^{\prime}} \big) 
\end{bmatrix}   \text{. } 
\]

\noindent and the second of which is,

\[
      \begin{bmatrix}
 A \big( \underline{u^{\prime\prime}} \big) & D \big(\underline{u^{\prime\prime}} \big)  & G \big( \underline{u^{\prime\prime}} \big) \\ B \big(\underline{u^{\prime\prime}}\big) & E \big( \underline{u^{\prime\prime}} \big) & H \big( \underline{u^{\prime\prime}} \big)  \\ C \big( \underline{u^{\prime\prime}} \big)  &  F \big(\underline{u^{\prime\prime}} \big) & I \big( \underline{u^{\prime\prime}} \big) 
\end{bmatrix}   \text{, } 
\]

\noindent for spatial parameters $\underline{u^{\prime}}, \underline{u^{\prime\prime}}$. The last relation is of the most significance for determining the form of the operator product $G \big( \underline{u} \big) E \big( \underline{u^{\prime}} \big) C \big( \underline{u^{\prime\prime}} \big)$, in which from two applications of the two-dimensional identity,

\begin{align*}
       A \big( \lambda \big) B \big( \lambda^{\prime} \big) = f \big( \lambda , \lambda^{\prime} \big)        B \big( \lambda^{\prime} \big) A \big( \lambda \big) + g \big( \lambda^{\prime} , \lambda \big) B \big( \lambda \big) A \big( \lambda^{\prime} \big)  \text{, }
\end{align*}

\noindent takes the form,

\begin{align*}
       \underline{G \big( \underline{u} \big) E \big( \underline{u^{\prime}} \big) C \big( \underline{u^{\prime\prime}} \big)}  =   f \big( \lambda_{\alpha} , \lambda_r , \lambda_{r^{\prime}} \big)    \bigg[   \bigg[ E \big( \underline{u^{\prime}} \big) C \big( \underline{u^{\prime\prime}} \big)  \bigg]  G \big( \underline{u } \big)     \bigg] + g \big( \lambda_{\alpha} , \lambda_r , \lambda_{r^{\prime}} \big)  \bigg[  \bigg[ E \big( \underline{u} \big) C \big( \underline{u^{\prime}} \big)  \bigg] \\ \times  G \big( \underline{u } \big)     \bigg]    \\ =    f \big( \lambda_{\alpha} , \lambda_r , \lambda_{r^{\prime}} \big)  \bigg[   f \big( \lambda , \lambda^{\prime} \big)     C \big( \underline{u^{\prime\prime}} \big) E \big( \underline{u^{\prime}} \big)  + g \big( \lambda^{\prime} , \lambda \big)     C \big( \underline{u^{\prime}}    \big) E \big( \underline{u^{\prime\prime}} \big)                          \bigg]    G \big( \underline{u} \big)        + g \big( \lambda_{\alpha} , \lambda_r , \lambda_{r^{\prime}} \big)      \\ \times      \bigg[      f \big(   \lambda , \lambda^{\prime} \big)  C \big( \underline{u^{\prime}} \big) E \big( \underline{u} \big)                  +      g \big( \lambda^{\prime} , \lambda \big)   C \big(  \underline{u}     \big) E \big( \underline{u^{\prime}}       \big)    \bigg]                    G \big( \underline{u^{\prime\prime} } \big)                 \text{, }          
\end{align*}

\noindent in three dimensions. Similarly, for the remaining relations within the three-dimensional Yang-Baxter algebra,

\begin{align*}
\underline{ I \big( \underline{u} \big) H \big( \underline{u^{\prime}} \big) G \big( \underline{u^{\prime\prime}} \big)}   =          f \big( \lambda_{\alpha} , \lambda_r , \lambda_{r^{\prime}} \big)  \bigg[   \bigg[ H \big( \underline{u^{\prime}} \big) G \big( \underline{u^{\prime\prime}} \big) \bigg]   I \big( \underline{u} \big)    \bigg]   +  g \big( \lambda_{\alpha} , \lambda_r , \lambda_{r^{\prime}} \big)   \bigg[    \bigg[ H \big( \underline{u^{\prime\prime}} \big) G \big( \underline{u} \big) \bigg] \\ \times  I \big( \underline{u^{\prime} } \big)    \bigg]  \\ =   f \big( \lambda_{\alpha} , \lambda_r , \lambda_{r^{\prime}} \big)   \bigg[       f \big( \lambda , \lambda^{\prime} \big)                G \big( \underline{u^{\prime\prime}} \big) H \big( \underline{u^{\prime}} \big)        + g \big( \lambda^{\prime} , \lambda \big)    G \big( \underline{u^{\prime}} \big) H \big( \underline{u^{\prime\prime}} \big)                                 \bigg]    I \big( \underline{u} \big)       + g \big( \lambda_{\alpha} , \lambda_r  , \lambda_{r^{\prime}} \big) \\ \times  \bigg[    f \big( \lambda , \lambda^{\prime} \big) G \big( \underline{u^{\prime\prime}} \big) H \big(   \underline{u^{\prime}} \big)  +           g \big( \lambda^{\prime} , \lambda \big)                       G \big( \underline{u^{\prime}} \big)  H \big( \underline{u^{\prime\prime}} \big)   \bigg] I \big( \underline{u^{\prime}}  \big)                                     \text{, } \\ \\ 
\underline{  A \big( \underline{u }  \big) D \big( \underline{u^{\prime}} \big) G \big( \underline{u^{\prime\prime}} \big) }   = f \big( \lambda_{\alpha} , \lambda_r , \lambda_{r^{\prime } } \big)  \bigg[     \bigg[                    D \big( \underline{u^{\prime}} \big) G \big( \underline{u^{\prime\prime}} \big)                                  \bigg] A \big( \underline{u }  \big)  \bigg]   +              g \big( \lambda_{\alpha} , \lambda_r , \lambda_{r^{\prime}} \big)    \bigg[     \bigg[                    D \big( \underline{u^{\prime\prime}} \big) G \big( \underline{u} \big)                                  \bigg]  \\ \times  A \big( \underline{u^{\prime} }  \big)  \bigg]    \\  =     f \big( \lambda_{\alpha}  , \lambda_r , \lambda_{r^{\prime}} \big)   \bigg[             f \big( \lambda , \lambda^{\prime} \big)                       G \big(    \underline{u^{\prime\prime}}   \big)    D \big( \underline{u^{\prime}} \big)    + g \big( \lambda^{\prime} , \lambda \big)    G \big( \underline{u^{\prime}} \big) D \big( \underline{u^{\prime\prime}} \big)                             \bigg]   A \big( \underline{u} \big)    +   g \big( \lambda_{\alpha} , \lambda_r , \lambda_{r^{\prime}} \big)  \\ \times   \bigg[          f \big( \lambda , \lambda^{\prime} \big)   G \big( \underline{u} \big)  D \big( \underline{u^{\prime\prime}} \big)    + g \big( \lambda^{\prime} , \lambda \big)  G \big( \underline{u^{\prime\prime}} D \big( \underline{u} \big)        \bigg]    A \big( \underline{u} \big)     \text{, } \\ \\  \underline{A \big( \underline{u } \big) E \big( \underline{u^{\prime}} \big) I \big( \underline{u^{\prime\prime}} \big)}  =   f \big( \lambda_{\alpha}  , \lambda_r , \lambda_{r^{\prime}} \big)          \bigg[   \bigg[ E \big( \underline{u^{\prime}} \big) I \big( \underline{u^{\prime\prime}} \big)                  \bigg]  A \big( \underline{u } \big)  \bigg]     +   g \big( \lambda_{\alpha} , \lambda_r , \lambda_{r^{\prime}} \big)    \bigg[   \bigg[ E \big( \underline{u^{\prime\prime}} \big) I \big( \underline{u} \big)                  \bigg]   \\ \times    A \big( \underline{u^{\prime} } \big) \bigg] \\ =     f \big( \lambda_{\alpha}  , \lambda_r , \lambda_{r^{\prime}} \big)  \bigg[       f \big( \lambda , \lambda^{\prime} \big)     I \big( \underline{u^{\prime\prime}} \big) E \big( \underline{u^{\prime}} \big)     + g \big( \lambda^{\prime} , \lambda \big)   I \big( \underline{u^{\prime}} \big) E \big( \underline{u^{\prime\prime}} \big)    \bigg] A \big( \underline{u} \big)  +   g \big( \lambda_{\alpha} , \lambda_r , \lambda_{r^{\prime}} \big) \\ \times    \bigg[     f \big( \lambda , \lambda^{\prime} \big)     I \big( \underline{u}  \big) E \big( \underline{u^{\prime\prime}} \big)            + g \big( \lambda^{\prime} , \lambda \big)                     I \big( \underline{u^{\prime\prime}} \big)           E \big( \underline{u}  \big)       \bigg]    A \big( \underline{u^{\prime}} \big)    \text{, }
\end{align*}

\noindent after applying the two-dimensional intertwinning relation between $A \big( \lambda \big)$ and $B \big( \lambda^{\prime} \big)$, for the terms,

\begin{align*}
  I \big( \underline{u} \big) \bigg[ H \big( \underline{u^{\prime}} \big) G \big( \underline{u^{\prime\prime}} \big)  \bigg]  \text{, }  \\             A \big( \underline{u} \big)        \bigg[ D \big( \underline{u^{\prime} } \big) G \big( \underline{u^{\prime\prime}} \big)  \bigg]        \text{, } \\    A \big( \underline{u} \big)           \bigg[    E \big( \underline{u^{\prime}} \big) I \big( \underline{u^{\prime\prime}} \big)    \bigg]    \text{, }
\end{align*}

\noindent as was the case for the first relation within the three-dimensional Yang-Baxter algebra,

\begin{align*}
   G \big( \underline{u } \big) \bigg[ E \big( \underline{u^{\prime}} \big) C \big( \underline{u^{\prime\prime}} \big) \bigg]     \text{. }
\end{align*}

\noindent The remaining identites for the three-dimensional Yang-Baxter algebra consist of more straightforward applications of corresponding identities for the two-dimensional Yang-Baxter algebra, some of which include,

\begin{align*}
      A \big( \underline{u} \big) A \big( \underline{u^{\prime}} \big) A \big( \underline{u^{\prime\prime}} \big)  \text{, } \\  B \big( \underline{u} \big) B \big( \underline{u^{\prime}} \big) B \big( \underline{u^{\prime\prime}} \big)  \text{, } \\  C \big( \underline{u} \big) C \big( \underline{u^{\prime}} \big) C \big( \underline{u^{\prime\prime}} \big)  \text{, } \\  D \big( \underline{u} \big) D \big( \underline{u^{\prime}} \big) D  \big( \underline{u^{\prime\prime}} \big)   \text{, }
\end{align*}

\noindent from the three-dimensional Yang-Baxter algebra, resulting from the Poisson structure obtained from evaluating the relations from the following nested Poisson bracket of two tensor products,

\[\
\bigg\{    \begin{bmatrix}
 A \big( \underline{u} 
 \big) & D \big( \underline{u} \big)  & G \big( \underline{u}  \big) \\ B \big( \underline{u}  \big) & E \big( \underline{u}  \big) & H \big( \underline{u}  \big)  \\ C \big( \underline{u}  \big)  &  F \big( \underline{u}  \big) & I \big( \underline{u}  \big) 
\end{bmatrix}\overset{\bigotimes}{,} \bigg\{ \begin{bmatrix}
 A \big( \underline{u^{\prime}} \big) & D \big( \underline{u^{\prime}} \big)  & G \big( \underline{u^{\prime}} \big) \\ B \big( \underline{u^{\prime}}\big) & E \big( \underline{u^{\prime}} \big) & H \big( \underline{u^{\prime}} \big)  \\ C \big( \underline{u^{\prime}} \big)  &  F \big( \underline{u^{\prime}} \big) & I \big( \underline{u^{\prime}} \big) 
\end{bmatrix}  \overset{\bigotimes}{,} \begin{bmatrix}
 A \big( \underline{u^{\prime\prime}} \big) & D \big( \underline{u^{\prime\prime}} \big)  & G \big( \underline{u^{\prime\prime}} \big) \\ B \big( \underline{u^{\prime\prime}}\big) & E \big( \underline{u^{\prime\prime }} \big) & H \big( \underline{u^{\prime\prime}} \big)  \\ C \big( \underline{u^{\prime\prime}} \big)  &  F \big( \underline{u^{\prime\prime}} \big) & I \big( \underline{u^{\prime\prime}} \big) 
\end{bmatrix}  \bigg\}  \text{ } \bigg\} \text{. } 
\]

\bigskip

\noindent The set of relations generated by the embedded system of Poisson brackets above, as a function of the spectral parameters $\underline{u}$, $\underline{u}$, and $\underline{u^{\prime\prime}}$, is a straightforward generalization of the system of Poisson brackets of two-dimensional transfer matrices of the 6-vertex model. As such, the generalized system of Poisson brackets for the 20-vertex model exhibits different characteristics than the two-dimensional counterpart, ranging from: (1) analytical expressions for the maximum, and minimum, eigenvalues of the spectrum of the transfer matrix; (2) relations of the underlying Yang-Baxter algebra, particularly involving whether products of two, or three, operators are taken; (3) projecting to one-dimensional subspaces of the ambient three-dimensional state space, $\textbf{T}$, in comparison to projections to one-dimensional subspaces of the ambient two-dimensional state space, $\textbf{Z}^2$; (4) taking the weak finite volume limit over $\textbf{T}$, in comparison to over $\textbf{Z}^2$. With all of these differences taken into account, below we restate the main result which encapsulates the form of correlation functions for the 20-vertex model. This will be shown to hold from contour integral representations obtained in the next section.

\bigskip

\noindent \textbf{Theorem} (\textit{nonlocal correlations}). The representation for nonlocal correlation functions of the 20-vertex model under domain-wall boundary conditions takes the form (\textit{*}). A closely related representation for the side partition function can be obtained with the same method.

\subsection{Executing the three components of the argument for the 20-vertex model}

\noindent We obtain various representations for partition functions of the 20-vertex model. To provide such an expression in terms of the basis elements $\ket{\Uparrow_{1 , \cdots , N}}, \ket{\Downarrow_{1 , \cdots , N}}, \ket{\Rightarrow_{1 , \cdots , N}}$, observe that the restriction of the transfer matrix to the first, second, and third, degrees of freedom is proportional to,

\begin{align*}
    \textbf{T}^{3D}|_{\ket{\Uparrow}} \propto   \mathrm{span} \big\{ \ket{\Uparrow} \big\}     \text{, } \\  \textbf{T}^{3D}|_{\ket{\Downarrow}} \propto   \mathrm{span} \big\{ \ket{\Downarrow} \big\}     \text{, } \\  \textbf{T}^{3D}|_{\ket{\Rightarrow}} \propto   \mathrm{span} \big\{ \ket{\Rightarrow} \big\}     \text{, }
\end{align*}

\noindent given the basis of $\textbf{T}$ for configurations in the $\ket{\Uparrow}$ state, equals,

\begin{align*}
 Z^{20V}_N \big( \underline{\lambda} \big)  =     \bra{\Uparrow}  \bigg[  \bigg[    \underset{1 \leq k \leq N}{\prod} \big[       A_k \big( \underline{u}  \big)      + \cdots    +     I_k \big( \underline{u} \big)      \big] \bigg]         \ket{\Rightarrow}  +  \bigg[   \underset{1 \leq k \leq N}{\prod} \big[       A_k \big( \underline{u}    \big)    + \cdots    +     I_k \big( \underline{u} \big)      \big]  \bigg]  \ket{\Downarrow}  \bigg]                   \text{, }
\end{align*}

\noindent where,

\begin{align*}
  Z^{20V}_N \big( \underline{\lambda} \big) = Z^{20V}_N \big( \underline{\lambda_1} , \underline{\lambda_2} \big)   \text{, }
\end{align*}

\noindent for the collection of spectral parameters,

\begin{align*}
 \underline{\lambda}  \equiv \big( \lambda_1 , \lambda_2 , \lambda_3 \big) \Longleftrightarrow \left\{\!\begin{array}{ll@{}>{{}}l}  \lambda_1 \text{ is the spectral parameter associated with }    \textbf{T}^{3D}|_{\ket{\Uparrow}}   \text{, } \\ \lambda_2 \text{ is the spectral parameter associated with } \textbf{T}^{3D}|_{\ket{\Rightarrow}}  \text{, } \\ \lambda_3 \text{ is the spectral parameter associated with } \textbf{T}^{3D}|_{\ket{\Downarrow}}   \text{, }   \end{array}\right. 
\end{align*}

\noindent corresponding to vertex configurations of the 20-vertex model for which there is one upwards pointing arrow, and a sideways pointing arrow, and also an upwards pointing arrow, and a downwards pointing arrow. The restriction of the three-dimensional transfer matrix along each degree of freedom of $\textbf{T}$ is given by,

\begin{align*}
 \textbf{T}^{3D}|_{\ket{\Uparrow}} \equiv \underset{\mathcal{R}}{\bigcup} \big\{ 3 \times 3 \text{ } \text{representation } \mathcal{R}  \text{ of } \textbf{T}^{3D} \text{ for which the inner product is identically 1} \\ \text{along the } \ket{\Uparrow}  \text{direction} \big\}   \text{, }  
\end{align*}

\noindent with similar objects for the remaining two degrees of freedom. As previously considered for the 6-vertex model, in {\color{blue}[7]}, the domain-wall partition function under the presence of parameters $r_1 , \cdots , r_s$ can be expressed in terms of a state in terms of up, down, or side, basis elements. As was the case for the computation of correlation functions for the 6-vertex model under domain-wall boundary conditions supported over $\textbf{Z}^2$, {\color{blue}[7]}, the homogenization limit, namely the procedure under which the prefactor of the determinantal representation of the partition function equals $1$, can be adapted for correlation functions of the 20-vertex model which are supported over $\textbf{T}$. In forthcoming arguments, to manipulate the restriction of partition functions,

\begin{align*}
  Z_{20 \text{V}} \equiv \underset{\omega \in \Omega_{20 \text{V}}}{\sum} w \big( \omega \big)   \text{, }
\end{align*}

\noindent for the 20-vertex model, with $Z_{\mathrm{Top}}$, $Z_{\mathrm{Bottom}}$, and $Z_{\mathrm{Side}}$, the following product of operators,

\begin{align*}
    \underset{1 \leq k \leq N}{\prod}  \big[ A_k \big( \underline{u} \big) 
 + D_k \big( \underline{u} \big) \big]     \text{,} \\  \underset{1 \leq k \leq N}{\prod}     \big[   B_k \big( \underline{u } \big) + E_k \big( \underline{u} \big)  \big]     \text{, } \\  \underset{1 \leq k \leq N}{\prod}  \big[    C_k \big( \underline{u} \big)   + F_k \big( \underline{u} \big)       \big]         \text{, }
\end{align*}

\noindent from the three-dimensional transfer matrix are further manipulated.

Specifically, when restricting the partition function to $\ket{\Uparrow_{1 , \cdots , N}}$,$ \ket{\Downarrow_{1 , \cdots , N}}$, or to $ \ket{\Rightarrow_{1 , \cdots , N}}$, the expression above for the partition function, as an object defined over $\textbf{T}$, from the associated basis simplifies to,

\begin{align*}
        Z^{20V}_N \big( \underline{\lambda} \big) |_{\ket{\Uparrow}} =   \bra{\Uparrow}  \bigg[ \bigg[     \underset{1 \leq k \leq N}{\prod}  \big[ A_k \big( \underline{u} \big) 
 + D_k \big( \underline{u} \big) \big]     \bigg] \ket{\Rightarrow}|_{\ket{\Uparrow}} +   \bigg[     \underset{1 \leq k \leq N}{\prod}  I_k \big( \underline{u} \big)          \bigg] \ket{\Downarrow}|_{\ket{\Uparrow}} \text{ } \bigg]     \text{, } \\   \\    Z^{20V}_N \big( \underline{\lambda} \big) |_{\ket{\Downarrow}} =           \bra{\Uparrow}|_{\ket{\Downarrow}} \bigg[                 \bigg[   \underset{1 \leq k \leq N}{\prod}     \big[   B_k \big( \underline{u } \big) + E_k \big( \underline{u} \big)     \big]      \bigg] \ket{\Rightarrow}_{\ket{\Downarrow}} + \bigg[  \underset{1 \leq k \leq N}{\prod}    \big[   I_k \big( \underline{u} \big)    \big]               \bigg]  \ket{\Downarrow}      \bigg]               \text{,  } \\  \\       Z^{20V}_N \big( \underline{\lambda} \big) |_{\ket{\Rightarrow}} = \bra{\Uparrow}|_{\ket{\Rightarrow}}  \bigg[  \bigg[   \underset{1 \leq k \leq N}{\prod}  \big[    C_k \big( \underline{u} \big)   + F_k \big( \underline{u} \big)       \big]          \bigg] \ket{\Rightarrow}         + \bigg[    \underset{1 \leq k \leq N}{\prod}    \big[   I_k \big( \underline{u} \big)    \big]       \bigg] \ket{\Downarrow }|_{\ket{\Rightarrow}}  \bigg]  \text{, }
\end{align*}

\noindent where the restriction of each of the basis states is given by,

\begin{align*}
\ket{\Rightarrow}|_{\ket{\Uparrow}} \equiv \bigg\{ \ket{\Rightarrow} \in \textbf{T} : \ket{\Rightarrow}|_{\ket{\Uparrow}} \equiv  \big[  \frac{\sqrt{3}}{2}, 0, 0   \big]^{\textbf{T}} 
\bigg\}  \text{, } \\   \ket{\Downarrow}|_{\ket{\Uparrow}} \equiv    \bigg\{ \ket{\Downarrow} \in \textbf{T} : \ket{\Downarrow}|_{\ket{\Uparrow}} \equiv  \big[  0, 0, 1   \big]^{\textbf{T}} 
\bigg\}   \text{, }
 \\  \ket{\Uparrow}  \equiv    \bigg\{ \ket{\Uparrow} \in \textbf{T} : \ket{\Uparrow}|_{\ket{\Uparrow}} \equiv  \ket{\Uparrow}
\bigg\}   \text{, } \end{align*}

\noindent with analogous expressions holding for each state when restricting to different degrees of freedom of $\textbf{T}$. In the presence of additional spectral parameters $r^{\prime}_1 , \cdots , r^{\prime}_{s^{\prime}}$, the corresponding domain-wall partition function for the 20-vertex model would read,

\begin{align*}
 \underline{Z^{20V,\mathrm{Bottom}}_{r_1 , \cdots , r_s , r^{\prime}_1 , \cdots , r^{\prime}_s}} \equiv  Z^{20V,\mathrm{Bottom}}_{r_1 , \cdots , r_s , r^{\prime}_1 , \cdots , r^{\prime}_{s^{\prime}}}   =         \bra{\Uparrow}  \bigg\{  \bigg[   \underset{1 \leq \alpha \leq r^{\prime}_s}{\prod}     G_{\alpha} \big( \underline{u} \big)     \bigg]  D_{r^{\prime}_s} \big( \underline{u} \big) \bigg[ \underset{1 \leq \alpha \leq r_s}{\prod}        A_{\alpha} \big( \underline{u} \big)      \bigg]   \cdots   \\  \times \bigg[ \underset{1 \leq \alpha \leq r^{\prime}_1}{\prod}     G_{\alpha} \big( \underline{u} \big)                \bigg]    D_{r^{\prime}_1} \big( \underline{u} \big)    \bigg[    \underset{1 \leq \alpha_s \leq r_1}{\prod}   A_{\alpha} \big( \underline{u} \big)    \bigg]  \bigg\}   \ket{\Rightarrow}   +        \bigg\{  \bigg[ \underset{1 \leq \alpha \leq r^{\prime}_s}{\prod}  G_{\alpha} \big( \underline{u} \big)  \bigg]  H_{r^{\prime}_s} \big( \underline{u} \big) \\ \times  \bigg[  \underset{1 \leq \alpha \leq r_s}{\prod}            A_{\alpha} \big( \underline{u} \big)        \bigg]    \cdots  \times  \bigg[  \underset{1 \leq \alpha \leq r^{\prime}_1}{\prod} G_{\alpha} \big( \underline{u} \big)  \bigg]       H_{r^{\prime}_1} \big( \underline{u} \big) \bigg[ \underset{1 \leq \alpha r_1}{\prod} A_{\alpha} \big( \underline{u} \big)  \bigg]                   \ket{\Downarrow}  \bigg\}            \text{, }
\end{align*}

\noindent corresponding to the product of block representations from $\textbf{T}$. Depending upon the locations of $r_1 , \cdots , r_s, r^{\prime}_1, \cdots$ $, r^{\prime}_{s^{\prime}}$ in $\textbf{T}$, one can readily form other partition functions, such as $Z^{\mathrm{Top}}$, by varying the sequence in which operators from the block representation of $\textbf{T}^{3D}$ are applied. Moreover, in the following computations for the representation of each partition function, before taking the homogeneous limit of all spectral parameters over $\textbf{T}$ we manipulate the determinant,

\[
     \mathcal{D}_{\text{Inhomogeneous}} \equiv   \bigg| \begin{smallmatrix} 
\frac{1}{a ( z_1 , w_2 ) b ( z_1 , w_2 )} - \frac{1}{a ( 1 , z_1 w_2 ) b ( 1 , z_1 w_2 )} & \cdots & \frac{1}{a ( z_N , w_{N+1} ) b ( z_N , w_{N+1} )} - \frac{1}{a ( 1 , z_N w_{N+1} ) b ( 1 , z_N w_{N+1} )}  \\ \frac{1}{a ( z_1 , w_3 ) b ( z_1 , w_3 )} - \frac{1}{a ( 1 , z_1 w_3 ) b ( 1 , z_1 w_3 )}  & \cdots & \frac{1}{a ( z_N , w_{N+2} ) b ( z_N , w_{N+2} )} - \frac{1}{a ( 1 , z_N w_{N+2} ) b ( 1 , z_N w_{N+1} )}  \\ \vdots & \cdots & \vdots \\ \frac{1}{a ( z_1 , w_{N+1} ) b ( z_1 , w_{N+1} )} - \frac{1}{a ( 1 , z_1 w_{N+1} ) b ( 1 , z_1 w_{N+1}  )}  & \cdots & \frac{1}{a ( z_N , w_{2N} ) b ( z_N , w_{2N} )} - \frac{1}{a ( 1 , z_N w_{2N} ) b ( 1 , z_N w_{2N} )} 
\end{smallmatrix} \bigg|  \text{, }
\]

\noindent corresponding to the partition function of the 6-vertex model under U-turn boundary conditions, while in the homogeneous limit, we manipulate the determinant,

\[
    \mathcal{D}_{\text{Homogeneous}} \equiv      \underset{0 \leq u, v \leq n-1}{\mathrm{det}}   \bigg[ \frac{\big( 1 + u^2 \big) \big( 1 + 2u - u^2 \big)}{\big(  1 - u^2 v \big) \big[ \big( 1 -u \big)^2 - v \big( 1 + u \big)^2 \big]} \bigg] 
\]

\noindent corresponding to the homogenization of the partition function for the 20-vertex model. The two determinants, as for their counterparts in the 6-vertex model, can be related to sending all of the inhomogeneities of the vertex model to $0$, or to any other constant, as,

\begin{align*}
           \mathcal{D}_{\text{Inhomogeneous}}    \underset{\lambda_3 \rightarrow 0}{\underset{\lambda_2 \rightarrow 0}{\underset{\lambda_1 \rightarrow 0}{\longrightarrow}}}      \mathcal{D}_{\text{Homogeneous}}        \text{. }
\end{align*}

\noindent We make use of this expression for the top partition function of the 20-vertex model to provide the desired contour integral representation for nonlocal correlations given in $(\textit{*})$, from which we provide similar computations for the remaining partition functions over $\textbf{T}$.

\subsubsection{General description of the method}

In the following three subsections, we implement computations associated with the each that has been described at the beginning of this section. That is, beginning with an inhomogeneous determinantal representation for the partition function, we pass to a homogeneous limit from suitable representation of partition functions. By making use of a partitioning of the entire state space for the 20-vertex model over $\textbf{T}$, we specify the collection of surfaces, along with the residues of each surface, which determine the support of each contour integral. Moreover, higher-dimensional representations for correlation functions supported over $\textbf{T}$ can be obtained by incorporating the determinantal representation of Di Francesco for the function $h^{\prime}$. The determinantal representation for the 20-vertex partition function, in comparison to that for the 6-vertex partition function, not only depends upon the pentagonal lattice, which was introduced in \textit{2.1} when introducing the QISM framework, but also upon the homogenization limit of spectral parameters. Albeit the fact that homogenization limits of the spectral parameters appearing in contour integral representations can be taken over lower-dimensional subspaces, we combine lower-dimensional representations together, after which the homogenization limit is taken over $\textbf{T}$.

\subsubsection{\underline{(1)}}

\noindent \textit{Proof of Lemma 2}. For the first step of computing correlation functions for the 20-vertex model from contour integral representations, we first obtain the representation for $Z^{\mathrm{Top}}$ - the restriction of the 20-verterx partition function to the top sublattice of $\textbf{T}$.

To compute the bottom partition function from the 20-vertex model, we apply the generalization of,

\begin{align*}
   A \big( \lambda_r \big) \bigg[ \underset{1 \leq \beta \leq r-1}{\prod}  B \big( \lambda_{\beta} \big) \bigg]   \text{, }
\end{align*}

\noindent with,

\begin{align*}
      A \big( \lambda_r , \lambda_{r^{\prime}} \big)      \bigg[  \underset{1 \leq \beta^{\prime} \leq r^{\prime}-1}{\underset{1 \leq \beta \leq r-1}{\prod}}    B \big(   \lambda_{\beta} , \lambda_{\beta^{\prime}} \big)   \bigg]  \text{, }
\end{align*}

\noindent corresponding to the product between $A$ and $B$ operators, from block representations of the three-dimensional transfer matrix. In the presence of different operators rather than $A$ and $B$,

\begin{align*}
      D \big( \lambda_r , \lambda_{r^{\prime}} \big)      \bigg[  \underset{1 \leq \beta^{\prime} \leq r^{\prime}-1}{\underset{1 \leq \beta \leq r-1}{\prod}}    G \big(   \lambda_{\beta} , \lambda_{\beta^{\prime}} \big)   \bigg]  \text{, }
\end{align*}

\noindent for $G \big( \underline{u} \big) \equiv G \big( \lambda_{\beta} , \lambda_{\beta^{\prime}} \big)$, and $D \big( \underline{u} \big) \equiv D \big( \lambda_r , \lambda_{r^{\prime}} \big)$, would take the form,

\begin{align*}
    \bigg[  \underset{1 \leq \alpha^{\prime} \leq r^{\prime}}{\underset{1 \leq \alpha \leq r}{\sum}}   \frac{g \big( \lambda_{\alpha} , \lambda_{r} , \lambda_{r^{\prime}} \big) }{f \big( \lambda_{\alpha} , \lambda_{r} , \lambda_{r^{\prime}} \big)}      \bigg]    \bigg[          \underset{1 \leq \beta^{\prime} \leq r^{\prime}-1}{\underset{1 \leq \beta \leq r}{\underset{\beta^{\prime} \neq \alpha^{\prime}}{\underset{\beta \neq \alpha}{\prod} }}}                 f \big( \lambda_{\beta} , \lambda_r , \lambda_{r^{\prime}} \big)            \bigg]              \bigg[    \underset{1 \leq \beta^{\prime} \leq r^{\prime}-1}{\underset{1 \leq \beta \leq r-1}{\underset{\beta^{\prime} \neq \alpha^{\prime}}{\underset{\beta \neq \alpha}{\prod}}}}     G \big( \lambda_{\beta} , \lambda_{\beta^{\prime}} \big)                 \bigg]  D \big( \lambda_{r} , \lambda_{r^{\prime}} \big)            \text{, }
\end{align*}

\noindent for the same choice of functions introduced for the original modification to the identity with the operators $A$ and $B$ at the end of \textit{3.1}. Generally speaking, taking the products of operators from block representations of the transfer matrix, in two, and three, dimensions alike allows for contour integral representations of correlation functions obtained for lower-dimensional subspaces of $\textbf{Z}^2$, and of $\textbf{T}$. For correlation functions supported over $\textbf{T}$, one obtains contour integral representations for correlation functions supported over $\big\{u, v\big\}$, and over $\big\{ v, w \big\}$. After having obtained the correlation functions for the partition functions $Z_{\mathrm{Top}}$, $Z_{\mathrm{Bottom}}$, and $Z_{\mathrm{Side}}$, the correlation function supported over the entirety of $\textbf{T}$ can be formulated. Finally, by combining inhomgeneous contour integral representations together, a homogenization procedure can be performed for passing to the homogeneous limit of the three-dimensional correlation functions. 

Before performing the forthcoming computation for the bottom partition function, and in the process taking the homogeneous limit of the determinantal representation under domain-walls, observe that the inhomogeneous determinantal representation,

\[
\bigg| \begin{smallmatrix} 
\frac{1}{a ( z_1 , w_2 ) b ( z_1 , w_2 )} - \frac{1}{a ( 1 , z_1 w_2 ) b ( 1 , z_1 w_2 )} & \cdots & \frac{1}{a ( z_N , w_{N+1} ) b ( z_N , w_{N+1} )} - \frac{1}{a ( 1 , z_N w_{N+1} ) b ( 1 , z_N w_{N+1} )}  \\ \frac{1}{a ( z_1 , w_3 ) b ( z_1 , w_3 )} - \frac{1}{a ( 1 , z_1 w_3 ) b ( 1 , z_1 w_3 )}  & \cdots & \frac{1}{a ( z_N , w_{N+2} ) b ( z_N , w_{N+2} )} - \frac{1}{a ( 1 , z_N w_{N+2} ) b ( 1 , z_N w_{N+1} )}  \\ \vdots & \cdots & \vdots \\ \frac{1}{a ( z_1 , w_{N+1} ) b ( z_1 , w_{N+1} )} - \frac{1}{a ( 1 , z_1 w_{N+1} ) b ( 1 , z_1 w_{N+1}  )}  & \cdots & \frac{1}{a ( z_N , w_{2N} ) b ( z_N , w_{2N} )} - \frac{1}{a ( 1 , z_N w_{2N} ) b ( 1 , z_N w_{2N} )} 
\end{smallmatrix} \bigg| 
\]

\noindent implies that the general expression for the bottom partition function takes the form,

\begin{align*}
 \bigg| \underset{1 \leq i < j \leq N}{\underset{i,j}{\mathrm{span}}} \bigg[  
            \frac{1}{a ( z_i , w_j ) b ( z_i , w_j )  \underset{1 \leq i < j \leq N}{\prod} ( z_i - z_j  ) ( w_j - w_i  )   \underset{1 \leq i \leq j \leq N}{\prod} ( 1 - z_i z_j ) ( 1 - w_i w_j ) } \\ - \frac{1}{a ( 1 , z_1 w_1 ) b ( 1 , z_1 w_1 ) \underset{1 \leq i < j \leq N}{\prod} ( z_i - z_j  ) ( w_j - w_i  )    \underset{1 \leq i \leq j \leq N}{\prod} ( 1 - z_i z_j ) ( 1 - w_i w_j ) }   \bigg]  \bigg|  \end{align*}

\noindent after normalizing each entry of the inhonogeneous determinantal representation, which equals, {\color{blue}[15]},

\begin{align*}
\frac{\mathcal{D}_{\text{Inhomogeneous}}}{\underset{1 \leq i < j \leq N}{\prod} \big( z_i - z_j \big) \big( w_j - w_i \big) \underset{1 \leq i \leq j \leq N}{\prod} \big( 1 - z_i z_j \big) \big( 1 - w_i w_j \big)  }   \text{, }                \end{align*}

\noindent from which the top partition function of the 20-vertex model, under domain-wall boundary conditions, takes a very similar form to that provided for the 6-vertex model from {\color{blue}[7]}, in which,

\begin{align*}
 \underline{Z^{\mathrm{Top}}_{r_1 , \cdots , r_s , r^{\prime}_1 , \cdots , r^{\prime}_{s^{\prime}}} } =  \bigg[         c^{s+s^{\prime}} a^{( s + s^{\prime} )  (2N-2)}          \bigg[ \underset{1 \leq j \leq s}{\prod}   t^{r_j}              \bigg] \bigg[ \underset{1 \leq k \leq s^{\prime}}{\prod}      \big( t^{\prime}\big)^{r^{\prime}_k}      \bigg]    \bigg] \bigg\{  \underset{\mathscr{S}_1}{\oint}  \times \cdots \times     \underset{\mathscr{S}_{s+s^{\prime}}}{\oint}       \bigg\{   \bigg[ \underset{1\leq j \leq s}{\prod}     \frac{w^{r_j-1}}{\big( w_j - 1 \big)^s}                \bigg] \\ \times    \bigg[ \underset{1\leq k \leq s^{\prime}}{\prod}               \frac{\big( w^{\prime}\big)^{r^{\prime}_k-1}}{\big( w^{\prime}_k - 1 \big)^{s^{\prime}}}     \bigg]               \bigg[ \underset{1 \leq j < k \leq s}{\prod}     \big[ \big( w_j - w_k \big) \big( t^2 w_j w_k - 2 \Delta t w_j +1 \big)  \big]               \bigg] \bigg[ \underset{1 \leq j^{\prime} < k^{\prime} \leq s^{\prime}}{\prod}         \big[ \big( w^{\prime}_j - w^{\prime}_k \big) \\ \times  \big( t^2 w^{\prime}_j w^{\prime}_k - 2 \Delta t w^{\prime}_j   +1 \big)  \big]              \bigg] \bigg\}   \frac{d^s w}{\big( 2 \pi i \big)^s} \frac{d^{s^{\prime}} w^{\prime} }{\big( 2 \pi i \big)^{s^{\prime}}}  \bigg\}   \text{, }
\end{align*}

\noindent In comparison to the contour integral representation for the bottom partition function of the 6-vertex model, that of the 20-vertex model can be realized as the product of partition functions,

\begin{align*}
       Z^{\mathrm{Top},6V}_{r_1 , \cdots , r_s} Z^{\mathrm{Top}, 6V}_{r^{\prime}_1 , \cdots , r^{\prime}_{s^{\prime}}}          \text{, } 
\end{align*}

\noindent given the collection of surfaces corresponding to the representation of the top partition function,

\begin{align*}
       \mathscr{S}^{\mathrm{Top}} \equiv \underset{1 \leq i \leq s + s^{\prime}}{\bigcup} \mathscr{S}_i  =    \underset{\mathcal{S}}{\bigcup}  \big\{ \text{surface } \mathcal{S}  \text{ containing a residue at } \big( 1 , 1 \big) \big\}    \text{. } 
\end{align*}

\noindent The expression for $Z^{\mathrm{Top}}_{r_1 , \cdots , r_s , r^{\prime}_1 , \cdots , r^{\prime}_{s^{\prime}}}$ arises from a straightforward extension of the computation reproduced in {\color{blue}[7]}, in which along a single degree of freedom over $\textbf{Z}^2$, instead of over $\textbf{T}$, in the presence of inhomogeneities at $r_1 , \cdots , r_s$ the representation takes the form,

\begin{align*}
 \underline{Z^{\mathrm{Top}}_{r_1, \cdots, r_s, r^{\prime}_1, \cdots, r^{\prime}_{s^{\prime}}}}  =  \bigg[ c^{s} a^{ s (N-1) } \bigg[ \underset{1 \leq j \leq s}{\prod}   t^{r_j}              \bigg]  \bigg] \underset{\mathscr{C}_1}{\oint}  \times \cdots \times     \underset{\mathscr{C}_s}{\oint}         \bigg\{   \bigg[ \underset{1\leq j \leq s}{\prod}     \frac{w^{r_j-1}}{\big( w_j - 1 \big)^s}                \bigg]           \\ \times         \bigg[ \underset{1 \leq j < k \leq s}{\prod}     \big[ \big( w_j - w_k \big)  \big( t^2 w_j w_k - 2 \Delta t w_j +1 \big)  \big]               \bigg]  \bigg\} \frac{d^s w}{\big( 2 \pi i \big)^s}     \text{, }
\end{align*}

\noindent given the collection of contours,

\begin{align*}
 \mathscr{C}^{\mathrm{Top}} \equiv   \underset{1 \leq i \leq s}{\bigcup} \mathscr{C}_i   =  \underset{1 \leq i \leq s}{\bigcup}  \big\{ \text{contours } \mathscr{C} \text{ containing a residue at } 1 \big\} \text{, }
\end{align*}

\noindent from which we conclude the argument. \boxed{}

\bigskip

\noindent With the contour integral representation provided in the statement of \textbf{Lemma} \textit{2}, below we obtain the contour integral representation provided in the statement of \textbf{Lemma} \textit{3}. In comparison to the contour integral representation provided in \textbf{Lemma} \textit{2}, that provided in \textbf{Lemma} \textit{3} incorporates contributions from the inhomogeneous determinantal representation. In comparison to the inhomogeneous determinantal representation of the partition function for the 6-vertex model, the representation for the partition function of the 20-vertex model depends upon more factors, which determines the choice of parameters $\epsilon_j$, and $\epsilon^{\prime}_j$ which are introduced in the perturbative expansion of contour integral representations.

\bigskip

\noindent \textit{Proof of Lemma 3}. To obtain the bottom partition function of the 20-vertex model from the top partition function, before applying the intertwining operation between operators from the block representation of the transfer matrix, observe that the desired homogeneous determinantal representation takes the form,

\begin{align*}
 \underline{Z^{\text{Inhomogeneous}, 20V}_{r_1 , \cdots , r_s , r^{\prime}_1 , \cdots , r^{\prime}_{s^{\prime}}, z_i, z_j, w_i, w_j}}   \propto     \bigg[ \frac{\big( a b \big)^{N ( N-s)}}{\underset{1 \leq i \leq N-s-1}{\prod} i! \underset{1 \leq j \leq N-1}{\prod} j!} \bigg]  \bigg[ \frac{(ab)^{N(N-s^{\prime}})}{\underset{1 \leq k \leq N- s^{\prime}-1}{\prod} k! \underset{1 \leq l \leq N-1}{\prod} l!} \bigg]    \\ \times \frac{\mathcal{D}_{\text{Inhomogeneous}}}{ \bigg[ \underset{1 \leq i < j \leq N}{\prod} \big( z_i - z_j \big) \big( w_j - w_i \big) \bigg] \bigg[  \underset{1 \leq i \leq j \leq N}{\prod} \big( 1 - z_i z_j \big) \big( 1 - w_i w_j \big) \bigg]  } \\ \times \bigg[ \underset{1 \leq j^{\prime} \leq s^{\prime}}{\underset{1 \leq j \leq s}{\prod}}            \bigg[    \frac{\mathrm{sin}^{N-r_j} \big( \epsilon_j \big) \mathrm{sin}^{r_j - 1} \big( \epsilon_j - 2 \eta \big) }{ \mathrm{sin}^{N-s} \big(     \epsilon_j + \lambda - \eta     \big) }        \bigg] \bigg[   \frac{\mathrm{sin}^{N-r^{\prime}_j} \big( \epsilon_{j^{\prime}} \big) \mathrm{sin}^{r_{j^{\prime}}-1} \big( \epsilon_{j^{\prime}} - 2 \eta \big)  }{ \mathrm{sin}^{N-s^{\prime} } \big(    \epsilon_{j^{\prime}} + \lambda^{\prime} - \eta      \big)  }   \bigg]              \bigg]  \\ 
 \times \bigg[  \underset{1 \leq j^{\prime} < k^{\prime} \leq s^{\prime}}{\underset{1 \leq j < k \leq s}{\prod}} \bigg[              \mathrm{sin}^{-1} \big( \epsilon_j - \epsilon_k + 2 \eta \big)                   \bigg]  \bigg[     \mathrm{sin}^{-1}  \big( \epsilon_{j^{\prime}} - \epsilon_{k^{\prime}} + 2 \eta \big)       \bigg] \bigg]        \text{, }
\end{align*}

\noindent for sufficiently small parameters $\epsilon_j$, and $\epsilon_{j^{\prime}}$. Perturbatively, for $\epsilon_j$ and $\epsilon_{j^{\prime}}$ taken sufficiently small, the inhomgeneous representation for the above partition function of the 20-vertex model can be expressed by replacing the ratio of the inhomogeneous determinant, $\mathcal{D}_{\mathrm{Inhomogeneous}}$, with a determinantal representation that has a prefactor being identically $1$. Besides the new homogeneous determinantal representation appearing in the contour integral for the partition function, the remaining terms,

\begin{align*}
    \bigg[ \frac{\big( a b \big)^{N ( N-s)}}{\underset{1 \leq i \leq N-s-1}{\prod} i! \underset{1 \leq j \leq N-1}{\prod} j!} \bigg]  \bigg[ \frac{(ab)^{N(N-s^{\prime}})}{\underset{1 \leq k \leq N- s^{\prime}-1}{\prod} k! \underset{1 \leq l \leq N-1}{\prod} l!} \bigg] \text{, } 
\end{align*}

\noindent corresponding to the product of $a$, and $b$, weights introduced in the 20-vertex weight function,

\begin{align*}
    \bigg[    \frac{\mathrm{sin}^{N-r_j} \big( \epsilon_j \big) \mathrm{sin}^{r_j - 1} \big( \epsilon_j - 2 \eta \big) }{ \mathrm{sin}^{N-s} \big(     \epsilon_j + \lambda - \eta     \big) }        \bigg]     \text{, }
\end{align*}

\noindent corresponding to the ratio of sine functions raised to $N-r_j$, $r_j-1$, or $N-s$, and,

\begin{align*}
     \bigg[   \frac{\mathrm{sin}^{N-r^{\prime}_j} \big( \epsilon_{j^{\prime}} \big) \mathrm{sin}^{r_{j^{\prime}}-1} \big( \epsilon_{j^{\prime}} - 2 \eta \big)  }{ \mathrm{sin}^{N-s^{\prime} } \big(    \epsilon_{j^{\prime}} + \lambda^{\prime} - \eta      \big)  }   \bigg]       \text{, } \\ \\   \underset{1 \leq j^{\prime} < k^{\prime} \leq s^{\prime}}{\underset{1 \leq j < k \leq s}{\prod}} \bigg[              \mathrm{sin}^{-1} \big( \epsilon_j - \epsilon_k + 2 \eta \big)                   \bigg]   \bigg[     \mathrm{sin}^{-1}  \big( \epsilon_{j^{\prime}} - \epsilon_{k^{\prime}}  + 2 \eta \big)       \bigg]     \text{, }
\end{align*}

\noindent corresponding to sine functions that are dependent upon $\epsilon_j$, and $\epsilon_{j^{\prime}}$.

In the homogeneous limit, the desired determinantal representation takes the form,

\begin{align*}
 \underline{Z^{\text{Homogeneous}, 20V}_{r_1 , \cdots , r_s , r^{\prime}_1 , \cdots , r^{\prime}_{s^{\prime}}, z_i, z_j, w_i, w_j}}   \equiv    Z^{20V}_{r_1 , \cdots , r_s , r^{\prime}_1 , \cdots , r^{\prime}_{s^{\prime}}}  =   \bigg[ \frac{\big( a b \big)^{N ( N-s)}}{\underset{1 \leq i \leq N-s-1}{\prod} i! \underset{1 \leq j \leq N-1}{\prod} j!} \bigg]  \bigg[ \frac{(ab)^{N(N-s^{\prime}})}{\underset{1 \leq k \leq N- s^{\prime}-1}{\prod} k! \underset{1 \leq l \leq N-1}{\prod} l!} \bigg]    \end{align*}

 \begin{align*} \times     \underset{0 \leq u, v \leq n-1}{\mathrm{det}}   \bigg[ \frac{\big( 1 + u^2 \big) \big( 1 + 2u - u^2 \big)}{\big(  1 - u^2 v \big) \big[ \big( 1 -u \big)^2 - v \big( 1 + u \big)^2 \big]} \bigg]           \bigg[ \underset{1 \leq j^{\prime} \leq s^{\prime}}{\underset{1 \leq j \leq s}{\prod}}            \bigg[    \frac{\mathrm{sin}^{N-r_j} \big( \epsilon_j \big) \mathrm{sin}^{r_j - 1} \big( \epsilon_j - 2 \eta \big) }{ \mathrm{sin}^{N-s} \big(     \epsilon_j + \lambda - \eta     \big) }        \bigg]  \\ \times \bigg[   \frac{\mathrm{sin}^{N-r^{\prime}_j} \big( \epsilon_{j^{\prime}} \big) \mathrm{sin}^{r_{j^{\prime}}-1} \big( \epsilon_{j^{\prime}} - 2 \eta \big)  }{ \mathrm{sin}^{N-s^{\prime} } \big(    \epsilon_{j^{\prime}} + \lambda^{\prime} - \eta      \big)  }   \bigg]             \bigg]   \bigg[  \underset{1 \leq j^{\prime} < k^{\prime} \leq s^{\prime}}{\underset{1 \leq j < k \leq s}{\prod}} \bigg[              \mathrm{sin}^{-1} \big( \epsilon_j - \epsilon_k + 2 \eta \big)                   \bigg]   \bigg[     \mathrm{sin}^{-1}  \big( \epsilon_{j^{\prime}} - \epsilon_{k^{\prime}}  + 2 \eta \big)       \bigg] \bigg]        \text{. }
\end{align*}

\noindent Before taking the homogeneous limit of the determinantal representation to obtain an expression of the form indicated above, introduce $ v_{r,r^{\prime}} \big( i,j \big) \equiv       v_{r_s,r^{\prime}_{s^{\prime}}}\big(i , j\big)$, where,

\begin{align*}
    v_{r,r^{\prime}} \big( i,j \big) =    \bigg[    \underset{1 \leq i \leq N}{\prod}        \big( p - p^{-1} w_i \big) a \big( qz_i , q^{-1}  z^{-1}_i \big) c \big( z_i , w_i \big)  \bigg]    \bigg[          \underset{1 \leq i \leq j \leq N}{\prod} a \big( z_i , w_j \big) b \big( z_i , w_j \big)  \\ \times  a \big( 1 , z_i w_j \big)  b \big( 1 , z_i w_j \big)    \bigg]          \text{, }
\end{align*}

\noindent which corresponds to preactors from the inhomogeneous determinantal representation, the homogeneous limit of which is given by the representation, {\color{blue}[7]},

\begin{align*}
   \underset{0 \leq u, v \leq n-1}{\mathrm{det}}   \bigg[ \frac{\big( 1 + u^2 \big) \big( 1 + 2u - u^2 \big)}{\big(  1 - u^2 v \big) \big[ \big( 1 -u \big)^2 - v \big( 1 + u \big)^2 \big]} \bigg]       \text{. }
\end{align*}

\noindent To take a homogenized limit of inhomogeneous determinantal representations for the partition function, and hence for the contour integral representations for correlations, we make use of the functions $h$, and $H$, that were previously introduced. Such functions, under domain-wall boundary conditions, also appear in contour integral representations for partition functions and correlation functions of the 6-vertex model. However, in comparison to the 6-vertex determinantal representation, the 20-vertex determinantal representation for correlations, and the partition function under domain-walls incorporates contributions from a product over,

\begin{align*}
   z_i - z_j \text{, }
\end{align*}

\noindent a subtraction over two coordinates $i$ and $j$, of $\textbf{T}$, and over,

\begin{align*}
 w_i - w_j    \text{, }
\end{align*}

\noindent a subtraction of the two corresponding weights over $i$ and $j$. When performing a homogenization procedure on the inhomogeneous determinantal representation, these terms have the effect of normalizing the inhomgeneous determinant, $\mathcal{D}_{\mathrm{Inhomogeneous}}$, so that the determinant is not too large when $z_i$ and $z_j$ are far apart, and, otherwise, ensuring that the normalization to $\mathcal{D}_{\mathrm{Inhomogeneous}}$ is large when $z_i$ and $z_j$ are very close to each other, in absolute value. The homogeneous representation above, as seen from the expressions for $Z^{\mathrm{Inhomogeneous}}$, and for $Z^{\mathrm{Homogeneous}}$, is related to the study of,

\begin{align*}
     Z^{\mathrm{Homogeneous}, 20V}_{r_1, \cdots , r_s, r^{\prime}_1, \cdots , r^{\prime}_{s^{\prime}}}  \propto Z^{\mathrm{Inhomogeneous},\mathrm{6V}}_{r_1 , \cdots , r_s}   \propto   \frac{\mathcal{D}_{\text{Inhomogeneous}}}{ \bigg[ \underset{1 \leq i < j \leq N}{\prod} \big( z_i - z_j \big) \big( w_j - w_i \big) \bigg] \bigg[ \underset{1 \leq i \leq j \leq N}{\prod} \big( 1 - z_i z_j \big) \big( 1 - w_i w_j \big) \bigg]   } \\    \equiv H_{N,s}                              \text{, }
\end{align*}

\noindent which is proportional to,

\begin{align*}
       \underline{Z^{\mathrm{Bottom}, 20V}_{r_1, \cdots , r_s, r^{\prime}_1, \cdots, r^{\prime}_{s^{\prime}}}} =        \mathcal{D}^{\prime}_{\text{Inhomogeneous}}  \bigg[    \underset{1 \leq j^{\prime}\leq s^{\prime}}{\underset{1 \leq j \leq s}{\prod}}      v_{r,r^{\prime}} \big( i,j \big)  \bigg] \bigg[         \underset{1 \leq j^{\prime} < k^{\prime} \leq s^{\prime}}{\underset{1 \leq j < k \leq s}{\prod}}   \bigg[              \mathrm{sin}^{-1} \big( \epsilon_j - \epsilon_k + 2 \eta \big)                   \bigg] \\ \times  \bigg[     \mathrm{sin}^{-1}  \big( \epsilon_{j^{\prime}} - \epsilon_{k^{\prime}} + 2 \eta \big)       \bigg]       \bigg]         \text{. }
\end{align*}

\noindent In comparison to the expression for $ \mathcal{D}_{\text{Inhomogeneous}}$, $ \mathcal{D}^{\prime}_{\text{Inhomogeneous}}$ takes the form, {\color{blue}[7]},

\begin{align*}
     \mathcal{D}^{\prime}_{\text{Inhomogeneous}} = \frac{\mathcal{D}_{\text{Inhomogeneous}}}{       \bigg[        \underset{1 \leq i \leq N}{\prod}  z^{n-1}_i \bigg] \bigg[ \underset{1 \leq i < j \leq N}{\prod} ( z_i - z_j ) ( w_j - w_i )  \bigg] \bigg[   \underset{1 \leq i < j \leq N}{\prod} ( 1 - z_i z_j ) ( 1 - w_i w_j ) \bigg] }      \text{. }
\end{align*}

\noindent The determinantal representation for $Z^{\mathrm{Bottom},20V}$ above takes on the following asymptotically proportional form,

\begin{align*}
   Z^{\mathrm{Bottom}, 20V}_{r_1, \cdots , r_s, r^{\prime}_1, \cdots, r^{\prime}_{s^{\prime}}}  \underset{\epsilon^{\prime}_1 , \cdots , \epsilon^{\prime}_s \rightarrow 0}{\underset{\epsilon_1 , \cdots , \epsilon_s \rightarrow 0}{\propto}}   \mathcal{D}_{\mathrm{Homogeneous}}      \text{. }
\end{align*}

\noindent Beginning in the next section, to implement the second step of the three items for the computation of the bottom partition function, observe that the expression for the bottom partition function, after applying the intertwinning relation to operators $A\big( \underline{u} \big), B \big( \underline{u} \big), \cdots, I \big( \underline{u}\big)$, of the transfer matrix, has a prefactor to the partition function,

\begin{align*}
  Z^{20V}_{\textbf{T}}  \text{, }
\end{align*}

\noindent which, when restricted to a finite volume of $\textbf{T}$, 

\begin{align*}
  Z^{20V}_{n}  \bigg|_{T \subsetneq \textbf{T}} \equiv   Z^{20V}_{\textbf{T}_n}  \bigg|_{T \subsetneq \textbf{T}}  \equiv \underset{\omega \in ( \Omega_{20 \text{V}} \cap T ) }{\sum} w \big( \omega \big)   \text{, }
\end{align*}

\noindent takes the form,

\begin{align*}
    Z^{20V}_{\textbf{T}}  \bigg|_{T \subsetneq \textbf{T}}         \equiv    Z^{20V}_{\textbf{T}} \bigg|_{\textbf{T} \backslash [  [N-s,   s + 1]  \times   [N-s^{\prime},  s^{\prime}  + 1]  \times [s,s^{\prime}]]  }     \equiv Z^{20V}_{\textbf{T} \backslash [  [N-s,   s + 1]  \times   [N-s^{\prime},  s^{\prime}  + 1]  \times [ s, s^{\prime} ]]}               \text{, }
\end{align*}

\noindent The prefactor to the restricted partition function takes the form,

\begin{align*}
   \bigg\{  {\underset{2 \leq j \leq s}{\prod}} \bigg[  \underset{\alpha_{\underline{i}+\underline{1}} \neq \alpha_{\underline{i}} }{\underset{1 \leq  \alpha_{\underline{i}+\underline{1}} \leq r_{\underline{j}} }{\sum}}  \bigg[    \underset{\underline{1} \leq \underline{i} \leq \underline{s}}{\prod}    \bigg[         \underset{s+1 \leq z^{\prime} \leq N}{\prod }   a  \big( \underline{i}, z^{\prime} \big)            \bigg] \underset{\underline{1} \leq \underline{i} \leq \underline{s}}{\prod} \frac{g \big( \lambda_{\alpha_{\underline{i}}}, \lambda_{r_j}        \big)}{f \big( \lambda_{\alpha_{\underline{i}}}, \lambda_{r_j} \big) }            \bigg]    \\ \times       \underset{1 \leq j^{\prime} \leq s}{\prod} \bigg[        \underset{1 \leq \beta_{j^{\prime}} \leq r_{j^{\prime}}}{\prod}   \frac{g \big( \lambda_{\alpha_{\underline{1}}}, \lambda_{\beta_{j^{\prime}}} \big)}{ f \big( \lambda_{\alpha_{\underline{1}}}, \lambda_{\beta_{j^{\prime}}} \big)}                               \bigg] 
 \bigg]                 \bigg\}    {\underset{2 \leq k \leq s^{\prime} }{\prod}}  \bigg[   \underset{\alpha_{\underline{j}+1} \neq \alpha_{\underline{j}} }{\underset{1 \leq  \alpha_{\underline{i}+1} \leq r^{\prime}_{\underline{k}} }{\sum}}      \bigg[                    \underset{\underline{1} \leq \underline{j}  \leq \underline{s^{\prime}} }{\prod}    \bigg[         \underset{s^{\prime} + 1 \leq z^{\prime\prime} \leq N}{\prod}  a   \big( j , z^{\prime\prime} \big)      \bigg]   \\ \times \underset{\underline{1} \leq \underline{k} \leq \underline{s^{\prime}}}{\prod} \frac{g \big( \lambda_{\alpha_{\underline{i}}}, \lambda_{r_j}        \big)}{f \big( \lambda_{\alpha_{\underline{i}}}, \lambda_{r_j} \big) }            \bigg]   \underset{1 \leq k^{\prime} \leq s^{\prime}}{\prod} \bigg[        \underset{1 \leq \beta_{k^{\prime}} \leq r_{k^{\prime}}}{\prod}   \frac{g \big( \lambda_{\alpha_{\underline{1}}}, \lambda_{\beta_{k^{\prime}}} \big)}{ f \big( \lambda_{\alpha_{\underline{1}}}, \lambda_{\beta_{k^{\prime}}} \big)}                    \bigg] 
\bigg]                 \bigg]         \text{, }
\end{align*}

\noindent from the fact that, for the first spectral parameter when $i \equiv 0$, and when $\underline{j} \equiv 0$, yields the superposition,

\begin{align*}
   \bigg\{  \underset{2 \leq j \leq s^{\prime}}{\prod} \bigg[  \underset{\alpha_{\underline{i}+\underline{1}} \neq \alpha_{\underline{i}} }{\underset{1 \leq  \alpha_{\underline{i}+\underline{1}} \leq r_{\underline{j}} }{\sum}}  \bigg[    \underset{\underline{1} \leq \underline{i} \leq \underline{s}}{\prod}    \bigg[         \underset{s+1 \leq z^{\prime} \leq N}{\prod }   a  \big( \underline{i}, z^{\prime} \big)            \bigg] \underset{\underline{1} \leq \underline{i} \leq \underline{s}}{\prod} \frac{g \big( \lambda_{\alpha_{\underline{i}}}, \lambda_{r_j}        \big)}{f \big( \lambda_{\alpha_{\underline{i}}}, \lambda_{r_j} \big) }            \bigg]   \\ \times           \underset{1 \leq j^{\prime} \leq s}{\prod} \bigg[        \underset{1 \leq \beta_{j^{\prime}} \leq r_{j^{\prime}}}{\prod}   \frac{g \big( \lambda_{\alpha_{\underline{1}}}, \lambda_{\beta_{j^{\prime}}} \big)}{ f \big( \lambda_{\alpha_{\underline{1}}}, \lambda_{\beta_{j^{\prime}}} \big)}                             \bigg] 
\bigg]       +   \underset{1 \leq j \leq 2}{\prod} \bigg[  \underset{\alpha_{\underline{i}+\underline{1}} \neq \alpha_{\underline{i}} }{\underset{1 \leq  \alpha_{\underline{i}+\underline{1}} \leq r_{\underline{j}} }{\sum}}  \bigg[    \underset{\underline{1} \leq \underline{i} \leq \underline{s}}{\prod}    \bigg[         \underset{s+1 \leq z^{\prime} \leq N}{\prod }   a  \big( \underline{i}, z^{\prime} \big)            \bigg]   \\
 \times    \underset{\underline{1} \leq \underline{i} \leq \underline{s}}{\prod} \frac{g \big( \lambda_{\alpha_{\underline{i}}}, \lambda_{r_j}        \big)}{f \big( \lambda_{\alpha_{\underline{i}}}, \lambda_{r_j} \big) }            \bigg]        \underset{1 \leq j^{\prime} \leq s}{\prod} \bigg[        \underset{1 \leq \beta_{j^{\prime}} \leq r_{j^{\prime}}}{\prod}   \frac{g \big( \lambda_{\alpha_{\underline{1}}}, \lambda_{\beta_{j^{\prime}}} \big)}{ f \big( \lambda_{\alpha_{\underline{1}}}, \lambda_{\beta_{j^{\prime}}} \big)}                             \bigg] 
 \bigg]  \bigg\}   \bigg\{ {\underset{2 \leq k \leq s^{\prime} }{\prod}}  \bigg[   \underset{\alpha_{\underline{j}+1} \neq \alpha_{\underline{j}} }{\underset{1 \leq  \alpha_{\underline{i}+1} \leq r^{\prime}_{\underline{k}} }{\sum}}      \bigg[                    \underset{\underline{1} \leq \underline{j}  \leq \underline{s^{\prime}} }{\prod}  \\ \times    \bigg[         \underset{s^{\prime} + 1 \leq z^{\prime\prime} \leq N}{\prod}  a   \big( \underline{j} , z^{\prime\prime} \big)      \bigg]   \underset{\underline{1} \leq \underline{k} \leq \underline{s^{\prime}}}{\prod} \frac{g \big( \lambda_{\alpha_{\underline{i}}}, \lambda_{r_j}        \big)}{f \big( \lambda_{\alpha_{\underline{i}}}, \lambda_{r_j} \big) }            \bigg]    \underset{1 \leq k^{\prime} \leq s^{\prime}}{\prod} \bigg[        \underset{1 \leq \beta_{k^{\prime}} \leq r_{k^{\prime}}}{\prod}   \frac{g \big( \lambda_{\alpha_{\underline{1}}}, \lambda_{\beta_{k^{\prime}}} \big)}{ f \big( \lambda_{\alpha_{\underline{1}}}, \lambda_{\beta_{k^{\prime}}} \big)}                          \bigg] 
\bigg]  \\  +  {\underset{1 \leq k \leq 2}{\prod}}  \bigg[   \underset{\alpha_{\underline{j}+1} \neq \alpha_{\underline{j}} }{\underset{1 \leq  \alpha_{\underline{i}+1} \leq r^{\prime}_{\underline{k}} }{\sum}}      \bigg[                    \underset{\underline{1} \leq \underline{j}  \leq \underline{s^{\prime}} }{\prod}     \bigg[         \underset{s^{\prime} + 1 \leq z^{\prime\prime} \leq N}{\prod}  a   \big( \underline{j} , z^{\prime\prime} \big)      \bigg]    \underset{\underline{1} \leq \underline{k} \leq \underline{s^{\prime}}}{\prod} \frac{g \big( \lambda_{\alpha_{\underline{i}}}, \lambda_{r_j}        \big)}{f \big( \lambda_{\alpha_{\underline{i}}}, \lambda_{r_j} \big) }            \bigg]  \\ \times   \underset{1 \leq k^{\prime} \leq s^{\prime}}{\prod} \bigg[        \underset{1 \leq \beta_{k^{\prime}} \leq r_{k^{\prime}}}{\prod}   \frac{g \big( \lambda_{\alpha_{\underline{1}}}, \lambda_{\beta_{k^{\prime}}} \big)}{ f \big( \lambda_{\alpha_{\underline{1}}}, \lambda_{\beta_{k^{\prime}}} \big)}                          \bigg] 
 \bigg]                 \bigg\}         \text{, }
\end{align*}

\noindent corresponding to the application of the intertwinning operation provided for operators in the product representaton of the 20-vertex model transfer matrix. Altogether, taking the product over $0 \leq j \leq s$, and also over $1 \leq k \leq s^{\prime}$, yields,

\begin{align*}
    {\underset{1 \leq j \leq s}{\prod}} \bigg[  \underset{\alpha_{\underline{i}+\underline{1}} \neq \alpha_{\underline{i}} }{\underset{1 \leq  \alpha_{\underline{i}+\underline{1}} \leq r_{\underline{j}} }{\sum}}  \bigg[    \underset{\underline{1} \leq \underline{i} \leq \underline{s}}{\prod}    \bigg[         \underset{s+1 \leq z^{\prime} \leq N}{\prod }   a  \big( \underline{i}, z^{\prime} \big)            \bigg] \underset{\underline{1} \leq \underline{i} \leq \underline{s}}{\prod} \frac{g \big( \lambda_{\alpha_{\underline{i}}}, \lambda_{r_j}        \big)}{f \big( \lambda_{\alpha_{\underline{i}}}, \lambda_{r_j} \big) }            \bigg]   \\ \times        \underset{1 \leq j^{\prime} \leq s}{\prod} \bigg[        \underset{1 \leq \beta_{j^{\prime}} \leq r_{j^{\prime}}}{\prod}   \frac{g \big( \lambda_{\alpha_{\underline{1}}}, \lambda_{\beta_{j^{\prime}}} \big)}{ f \big( \lambda_{\alpha_{\underline{1}}}, \lambda_{\beta_{j^{\prime}}} \big)}                              \bigg] 
 \bigg]                 \bigg]     {\underset{1 \leq k \leq s^{\prime} }{\prod}}  \bigg[   \underset{\alpha_{\underline{j}+1} \neq \alpha_{\underline{j}} }{\underset{1 \leq  \alpha_{\underline{i}+1} \leq r^{\prime}_{\underline{k}} }{\sum}}      \bigg[                    \underset{\underline{1} \leq \underline{j}  \leq \underline{s^{\prime}} }{\prod}    \bigg[         \underset{s^{\prime} + 1 \leq z^{\prime\prime} \leq N}{\prod}  a   \big( j , z^{\prime\prime} \big)      \bigg]   \\ \times  \underset{\underline{1} \leq \underline{k} \leq \underline{s^{\prime}}}{\prod} \frac{g \big( \lambda_{\alpha_{\underline{i}}}, \lambda_{r_j}        \big)}{f \big( \lambda_{\alpha_{\underline{i}}}, \lambda_{r_j} \big) }            \bigg]  \underset{1 \leq k^{\prime} \leq s^{\prime}}{\prod} \bigg[        \underset{1 \leq \beta_{k^{\prime}} \leq r_{k^{\prime}}}{\prod}   \frac{g \big( \lambda_{\alpha_{\underline{1}}}, \lambda_{\beta_{k^{\prime}}} \big)}{ f \big( \lambda_{\alpha_{\underline{1}}}, \lambda_{\beta_{k^{\prime}}} \big)}                   \bigg] 
  \bigg]                  \bigg]      \text{. }
\end{align*}

\noindent The expression obtained above for the determinantal representation after applying the associated intertwinning operation is related from the fact that the representation takes the form, {\color{blue}[15]},

\begin{align*}
   \underset{0 \leq u, v \leq n-1}{\mathrm{det}}   \bigg[ \frac{\big( 1 + u^2 \big) \big( 1 + 2u - u^2 \big)}{\big(  1 - u^2 v \big) \big[ \big( 1 -u \big)^2 - v \big( 1 + u \big)^2 \big]} \bigg]   \text{, }   
\end{align*}

\noindent after taking the homogeneous limit, in which the prefactor,

\begin{align*}
  \mathcal{M}^{\prime}_{i,j} \equiv      \mathcal{M}^{\prime} \big( i , j \big)  =       \frac{1}{a \big(  z_i ,  w_j  \big)  b \big( z_i , w_j   \big) } - \frac{1}{a \big( 1 , z_i w_j 
 \big)  b \big( 1 , z_i w_j   \big) }    \text{, } 
\end{align*}

\noindent corresponds to the prefactor,

\begin{align*}
    \frac{\underset{1 \leq \alpha \leq k}{\prod} \text{ } \underset{1 \leq k \leq N}{\prod} a \big( \lambda_{\alpha} , \nu_k \big) b \big( \lambda_{\alpha} , \nu_k \big) }{\underset{1 \leq a < \beta \leq N}{\prod} d \big( \lambda_{\beta} , \lambda_{\alpha} \big) \text{ } \underset{1 \leq i < k \leq N}{\prod} d \big( \nu_j , \nu_k \big)}  \text{, }
\end{align*}

\noindent for the 6-vertex model, from which we conclude the argument. \boxed{}

\bigskip

\noindent In the next section, besides performing the computation for obtaining the bottom partition function from the top partition function, to obtain the remaining partition functions using a similar approach, observe, from the representation of,

\begin{align*}
       Z^{\mathrm{Top}}_{r_1 , \cdots , r_s , r^{\prime}_1 , \cdots , r^{\prime}_{s^{\prime}}}       \text{, } 
\end{align*}

\noindent that one may write a similar contour integral representation for nonlocal correlations for the side domain-wall partition function,

\begin{align*}
    Z^{\mathrm{Inhomogeneous}, (\mathrm{Side})_1}_{\lambda_1 , \cdots , \lambda_s, r^{\prime}_1 , \cdots , r^{\prime}_{s^{\prime}}} \equiv   Z^{(\mathrm{Side})_1}_{\lambda_1 , \cdots , \lambda_s, r^{\prime}_1 , \cdots , r^{\prime}_{s^{\prime}}}  \text{, }
\end{align*}

\noindent will be shown to equal,

\begin{align*}
       \underline{Z^{(\mathrm{Side})_1}_{\lambda_1 , \cdots , \lambda_s , r^{\prime}_1 , \cdots , r^{\prime}_{s^{\prime}}}}  =  \bigg[ c^{s+s^{\prime}} a^{( s + s^{\prime} )  (2N-2)} \bigg[ \underset{1 \leq j \leq s}{\prod}   t^{\lambda_j}              \bigg] \bigg[ \underset{1 \leq k \leq s^{\prime}}{\prod}      \big( t^{\prime}\big)^{r^{\prime}_k}      \bigg]   \text{ } \bigg] \bigg\{ \underset{\mathscr{S}_1}{\oint}  \times \cdots \times     \underset{\mathscr{S}_{s+s^{\prime}}}{\oint} \\ \times         \bigg\{   \bigg[ \underset{1\leq j \leq s}{\prod}     \frac{w^{\lambda_j-1}}{\big( w_j - 1 \big)^s}                \bigg]    \bigg[ \underset{1\leq k \leq s^{\prime}}{\prod}               \frac{\big( w^{\prime}\big)^{r^{\prime}_k-1}}{\big( w^{\prime}_k - 1 \big)^{s^{\prime}}}     \bigg]               \bigg[ \underset{1 \leq j < k \leq s}{\prod}     \big[ \big( w_j - w_k \big) \big( t^2 w_j w_k - 2 \Delta t w_j +1 \big)  \big]               \bigg] \\ \times  \bigg[ \underset{1 \leq j^{\prime} < k^{\prime} \leq s^{\prime}}{\prod}         \big[ \big( w^{\prime}_j - w^{\prime}_k \big)  \big( t^2 w^{\prime}_j w^{\prime}_k - 2 \Delta t w^{\prime}_j   +1 \big)  \big]              \bigg]  \bigg\}   \frac{d^s w}{\big( 2 \pi i \big)^s} \frac{d^{s^{\prime}} w^{\prime} }{\big( 2 \pi i \big)^{s^{\prime}}}    \bigg\}    \text{, } 
\end{align*}

\noindent for the collection of surfaces,

\begin{align*}
       \mathscr{S}^{(\mathrm{Side})_1} \equiv \underset{1 \leq i \leq s + s^{\prime}}{\bigcup} \mathscr{S}_i  =    \underset{\mathcal{S}}{\bigcup}  \big\{ \text{surface } \mathcal{S}  \text{ containing a residue at } \big( i , i \big) \big\}    \text{. } 
\end{align*}

\noindent To demonstrate that the remaining contour integral representation from $\textbf{Lemma}$ \textit{1} holds, we adapt computations from previous arguments provided in $\textbf{Lemma}$ \textit{2}. That is, the contour integral representation in $\textbf{Lemma}$ \textit{2} of the partition function can be used to obtain a homogenized determinantal representation. However, in comparison to the perturbative expansion in parameters $\epsilon$ and $\epsilon^{\prime}$
 that are taken to be sufficiently small, the following application of the perturbative expansion takes into account the collection of residues given by,

 \begin{align*}
       \mathscr{S}^{(\mathrm{Side})_2} \equiv \underset{1 \leq i \leq s^{\prime} + s^{\prime\prime}}{\bigcup} \mathscr{S}_i  =    \underset{\mathcal{S}}{\bigcup}  \big\{ \text{surface } \mathcal{S}  \text{ containing a residue at } \big( i, i \big)  \big\}    \text{. } 
\end{align*}

\noindent \textit{Proof of Lemma 1}. In comparison to the contour integral representation for nonlocal correlations of the top partition function, those of the side partition function depend upon a common set of spectral parameters $r^{\prime}_1, \cdots, r^{\prime}_{s^{\prime}}$, but instead on the collection of spectral parameters $\lambda_1, \cdots, \lambda_s$ instead of on $r_1, \cdots, r_s$, which are reflected through the terms,

\begin{align*}
    \underset{1 \leq j \leq s}{\prod} t_{\lambda_j}  \text{, } \\ 
    \underset{1 \leq j \leq s}{\prod}  \frac{w^{\lambda_j - 1}_j }{\big( w_j -1 \big)^s }   \text{. }
\end{align*}

\noindent With a suitable collection of orthogonal polynomials, straightforward polynomials analagous to those introduced in {\color{blue}[7]}, and following the same sequence of steps presented earlier in this subsection, one can take the homogeneous limit of the inhomogeneous determinantal representation to obtain a determinantal representation that is transformed to a contour integral one, which takes the form,

\begin{align*}
    \bigg[ \frac{\big( a b \big)^{N ( N-s)}}{\underset{1 \leq i \leq N-s-1}{\prod} i! \underset{1 \leq j \leq N-1}{\prod} j!} \bigg]  \bigg[ \frac{(ab)^{N(N-s^{\prime}})}{\underset{1 \leq k \leq N- s^{\prime}-1}{\prod} k! \underset{1 \leq l \leq N-1}{\prod} l!} \bigg]   \\ \times   \frac{\mathcal{D}_{\text{Inhomogeneous}}}{ \bigg[ \underset{1 \leq i < j \leq N}{\prod} \big( z_i - z_j \big) \big( w_j - w_i \big) \bigg] \bigg[  \underset{1 \leq i \leq j \leq N}{\prod} \big( 1 - z_i z_j \big) \big( 1 - w_i w_j \big) \bigg]  }  \\ \times  \bigg[ \underset{1 \leq j^{\prime} \leq s^{\prime}}{\underset{1 \leq j \leq s}{\prod}}            \bigg[    \frac{\mathrm{sin}^{N-\lambda_j} \big( \epsilon_j \big) \mathrm{sin}^{\lambda_j - 1} \big( \epsilon_j - 2 \eta \big) }{ \mathrm{sin}^{N-s} \big(     \epsilon_j + \lambda - \eta     \big) }        \bigg]   \bigg[   \frac{\mathrm{sin}^{N-r^{\prime}_j} \big( \epsilon_{j^{\prime}} \big) \mathrm{sin}^{r_{j^{\prime}}-1} \big( \epsilon_{j^{\prime}} - 2 \eta \big)       }{ \mathrm{sin}^{N-s^{\prime} } \big(    \epsilon_{j^{\prime}} + \lambda^{\prime} - \eta      \big)  } \bigg]             \bigg]  \\ \times  \bigg[  \underset{1 \leq j^{\prime} < k^{\prime} \leq s^{\prime}}{\underset{1 \leq j < k \leq s}{\prod}} \bigg[              \mathrm{sin}^{-1} \big( \epsilon_j  - \epsilon_k + 2 \eta \big)                   \bigg]  \bigg[     \mathrm{sin}^{-1}  \big( \epsilon_{j^{\prime}} - \epsilon_{k^{\prime}}  + 2 \eta \big)       \bigg]  \bigg]       \text{, }
\end{align*}

\noindent corresponding to terms from the determinantal representation of the top partition function,

\begin{align*}
    Z^{\text{Inhomogeneous}, 20V}_{r_1 , \cdots , r_s , r^{\prime}_1 , \cdots , r^{\prime}_{s^{\prime}}, z_i, z_j, w_i, w_j} \text{. }
\end{align*}

\noindent Taking the homogeneous limit amongst the $\lambda$ and $r$ spectral parameters for the side determinantal representation yields,

\begin{align*}
     \bigg[ \frac{\big( a b \big)^{N ( N-s)}}{\underset{1 \leq i \leq N-s-1}{\prod} i! \underset{1 \leq j \leq N-1}{\prod} j!} \bigg]  \bigg[ \frac{(ab)^{N(N-s^{\prime}})}{\underset{1 \leq k \leq N- s^{\prime}-1}{\prod} k! \underset{1 \leq l \leq N-1}{\prod} l!} \bigg]   \underset{0 \leq i, j \leq n-1}{\mathrm{det}}   \bigg[ \frac{\big( 1 + u^2 \big) \big( 1 + 2u - u^2 \big)}{\big(  1 - u^2 v \big) \big[ \big( 1 -u \big)^2 - v \big( 1 + u \big)^2 \big]} \bigg]  \\ \times \bigg[ \underset{1 \leq j^{\prime} \leq s^{\prime}}{\underset{1 \leq j \leq s}{\prod}}            \bigg[    \frac{\mathrm{sin}^{N-\lambda_j} \big( \epsilon_j \big) \mathrm{sin}^{\lambda_j - 1} \big( \epsilon_j - 2 \eta \big) }{ \mathrm{sin}^{N-s} \big(     \epsilon_j + \lambda - \eta     \big) }        \bigg] \bigg[   \frac{\mathrm{sin}^{N-r^{\prime}_j} \big( \epsilon_{j^{\prime}} \big) \mathrm{sin}^{r_{j^{\prime}}-1} \big( \epsilon_{j^{\prime}} - 2 \eta \big)  }{ \mathrm{sin}^{N-s^{\prime} } \big(    \epsilon_{j^{\prime}} + \lambda^{\prime} - \eta      \big)  }   \bigg]              \bigg]  \bigg[  \underset{1 \leq j^{\prime} < k^{\prime} \leq s^{\prime}}{\underset{1 \leq j < k \leq s}{\prod}} \bigg[              \mathrm{sin}^{-1} \big( \epsilon_j \\ - \epsilon_k + 2 \eta \big)                   \bigg]  \bigg[     \mathrm{sin}^{-1}  \big( \epsilon_{j^{\prime}} - \epsilon_{k^{\prime}} + 2 \eta \big)       \bigg]  \bigg]       \text{, }
\end{align*}

\noindent for a factor similar to $v_{r,r^{\prime}}$, which takes the form,

\begin{align*}
       v_{\lambda,r^{\prime}} \big( i,j \big) =    \bigg[    \underset{1 \leq i \leq N}{\prod}        \big( p - p^{-1} w_i \big) a \big( qz_i , q^{-1}  z^{-1}_i \big) c \big( z_i , w_i \big)  \bigg]    \bigg[          \underset{1 \leq i \leq j \leq N}{\prod} a \big( z_i , w_j \big) b \big( z_i , w_j \big) a \big( 1 , z_i w_j \big) \\ \times b \big( 1 , z_i w_j \big)    \bigg]           \text{. }
\end{align*}

\noindent Altogether, one can conclude that the form of the partition function for the remaining side, from the contour integral representation provided for $(\mathrm{Side})_1$, takes the form,

\begin{align*}
       \underline{Z^{(\mathrm{Side})_2, 20V}_{r_1, \cdots , r_s, r^{\prime}_1, \cdots, r^{\prime}_{s^{\prime}}}} =        \mathcal{D}^{\prime}_{\text{Inhomogeneous}}  \bigg[    \underset{1 \leq j^{\prime}\leq s^{\prime}}{\underset{1 \leq j \leq s}{\prod}}      v_{\lambda,r^{\prime}} \big( i,j \big)  \bigg] \bigg[         \underset{1 \leq j^{\prime} < k^{\prime} \leq s^{\prime}}{\underset{1 \leq j < k \leq s}{\prod}}   \bigg[              \mathrm{sin}^{-1} \big( \epsilon_j - \epsilon_k + 2 \eta \big)                   \bigg] \\ \times  \bigg[     \mathrm{sin}^{-1}  \big( \epsilon_{j^{\prime}} - \epsilon_{k^{\prime}} + 2 \eta \big)       \bigg]      \bigg]         \text{, }
\end{align*}

\noindent before taking the homogeneous limit. Finally, as is the case for $Z^{\mathrm{Bottom}}$ presented earlier in the section, the determinantal representation for $Z^{(\mathrm{Side})_2,20V}$ above takes on the following asymptotically proportional form,

\begin{align*}
   Z^{(\mathrm{Side})_2, 20V}_{\lambda_1, \cdots , \lambda_s, r^{\prime}_1, \cdots, r^{\prime}_{s^{\prime}}}  \underset{\epsilon^{\prime}_1 , \cdots , \epsilon^{\prime}_s \rightarrow 0}{\underset{\epsilon_1 , \cdots , \epsilon_s \rightarrow 0}{\propto}}   \mathcal{D}_{\mathrm{Homogeneous}}      \text{, }
\end{align*}

\noindent for the homogeneous determinantal representation,

\begin{align*}
     \mathcal{D}^{\prime}_{\text{Inhomogeneous}} = \frac{\mathcal{D}_{\text{Inhomogeneous}}}{      \bigg[         \underset{1 \leq i \leq N}{\prod}  z^{n-1}_i \bigg] \bigg[ \underset{1 \leq i < j \leq N}{\prod} ( z_i - z_j ) ( w_j - w_i )   \bigg] \bigg[  \underset{1 \leq i < j \leq N}{\prod} ( 1 - z_i z_j ) ( 1 - w_i w_j ) \bigg] }      \text{. }
\end{align*}

\noindent For the side partition function, as alluded to in the Introduction and other earlier sections, one must perform the same steps provided for the bottom partition function, which allows one to study nonlocal correlations of a different boundary of some finite volume over $\textbf{T}$. To this end, write,

\begin{align*}
    \underline{Z^{\mathrm{Side}}_{r^{\prime}_1 , \cdots , r^{\prime}_{s^{\prime}} , r^{\prime\prime}_1 , \cdots , r^{\prime\prime}_{s^{\prime\prime}} }} =  \mathscr{P}    \bigg\{ \underset{\mathscr{S}^{\prime}_1  }{\oint}  \times \cdots \times     \underset{\mathscr{S}^{\prime}_1 }{\oint}   \bigg[ \mathscr{P}_1 \mathscr{P}_2   h^{\prime}_{N,s^{\prime},s^{\prime\prime}}         \frac{\mathrm{d}^{s^{\prime}} z^{\prime}}{ \big( 2 \pi i \big)^{s^{\prime}}}   \mu_0                  \bigg]       \bigg\}        \text{, } \tag{\textit{*}}
\end{align*}

\noindent where,

\begin{align*}
        \mathscr{S}^{\prime}_1 \equiv  \big\{ \text{surface } \mathcal{S} \text{ containing the residue at } \big( 1 , 1 \big)  \big\}   \text{, } \\  \\     \mathscr{P}       \equiv       \mathscr{P} \big( a,  b , c , N \big) =   \bigg[   \frac{Z^{20V}_N}{a^{\frac{s^{\prime} ( 2N -s^{\prime} +1) }{2}+ \frac{s^{\prime\prime} ( 2N -s^{\prime\prime} +1) }{2}} b^{\frac{s^{\prime}(s^{\prime}-3)}{2}+ \frac{s^{\prime\prime}(s^{\prime\prime}-3)}{2}}      c^{s^{\prime}+s^{\prime\prime}} }                      \bigg]         \bigg[ \frac{a}{b} \bigg]^{r^{\prime}_1 + \cdots + r^{\prime}_{s^{\prime}} + r^{\prime\prime}_1 + \cdots + r^{\prime\prime}_{s^{\prime\prime}}}    \text{, } \\   \mathscr{P}_1     \equiv   \underset{1 \leq j \leq s^{\prime\prime}}{\underset{1 \leq i \leq s^{\prime}}{\prod} }  \frac{1}{ z_{r^{\prime}_i} z_{r^{\prime\prime}_j} } \text{, } 
 \end{align*}

        \begin{align*}
      \mathscr{P}_2   \equiv    \bigg[ \underset{1 \leq j < k \leq s^{\prime}}{\prod} \frac{z^{\prime}_j - z^{\prime}_k}{ t^2 z^{\prime}_j z^{\prime}_k - 2 \Delta t z^{\prime}_j + 1 }  \bigg] \bigg[ \underset{1 \leq j^{\prime} < k^{\prime} \leq s^{\prime\prime}}{\prod}  \frac{z^{\prime\prime}_j - z^{\prime\prime}_k}{ t^2 z^{\prime\prime}_j z^{\prime\prime}_k - 2 \Delta t z^{\prime\prime}_j + 1 } \bigg]                                   \text{, }   \\ \\ h^{\prime}_{N,s,s^{\prime}}  \equiv     h^{\prime}_{N,s,s^{\prime}} \big( z^{\prime}_1 , \cdots , z^{\prime}_{s^{\prime}} , z^{\prime\prime}_1 , \cdots , z^{\prime\prime}_{s^{\prime\prime}} \big) =  Z^{\mathrm{20V}}_n  = \underset{0 \leq u, v \leq n-1}{\mathrm{det}}   \bigg[ \frac{\big( 1 + u^2 \big) \big( 1 + 2u - u^2 \big)}{\big(  1 - u^2 v \big) \big[ \big( 1 -u \big)^2 - v \big( 1 + u \big)^2 \big]} \bigg] \text{, }  \\ \\  \mu_0 \equiv      \frac{\mathrm{d}^{s^{\prime\prime}} z  }{ \big( 2 \pi i \big)^{s^{\prime\prime}}}    \text{, }
\end{align*}

\noindent corresponding to the determinantal representation under domain-walls. As was the case for the top and bottom partition functions, with a suitable expression for taking the summation over all possible configurations, the representation that one would like to transform into a contour integral. In comparison to the contour integral representation for the side partition function given above, recall that the first contour integral that was manipulated for the top partition function was defined over a different set of spectral parameters, in addition to surfaces, which took the form,

\begin{align*}
       \underline{Z^{\mathrm{Top}}_{r_1 , \cdots , r_s , r^{\prime}_1 , \cdots , r^{\prime}_s }} =  \mathscr{P} \bigg\{    \underset{\mathscr{S}_1  }{\oint}  \times \cdots \times     \underset{\mathscr{S}_1 }{\oint}   \bigg[ \mathscr{P}_1 \mathscr{P}_2   h^{\prime}_{N,s,s^{\prime}}         \frac{\mathrm{d}^{s^{\prime}} z^{\prime}}{ \big( 2 \pi i \big)^{s^{\prime}}}   \mu_0                  \bigg]   \bigg\}              \text{. }
\end{align*}

\noindent From the objects introduced above for the side partition function of the 20-vertex model supported over $\textbf{T}$, one can then proceed to define, and manipulate in the remaining steps $\underline{(2)}$, and $\underline{(3)}$, 

\begin{align*}
    \bigg[ \frac{\big( a b \big)^{N ( N-s)}}{\underset{1 \leq i \leq N-s-1}{\prod} i! \underset{1 \leq j \leq N-1}{\prod} j!} \bigg]  \bigg[ \frac{(ab)^{N(N-s^{\prime}})}{\underset{1 \leq k \leq N- s^{\prime}-1}{\prod} k! \underset{1 \leq l \leq N-1}{\prod} l!} \bigg]  \\ \times   \frac{\mathcal{D}_{\text{Inhomogeneous}}}{ \bigg[ \underset{1 \leq i < j \leq N}{\prod} \big( z_i - z_j \big) \big( w_j - w_i \big) \bigg] \bigg[  \underset{1 \leq i \leq j \leq N}{\prod} \big( 1 - z_i z_j \big) \big( 1 - w_i w_j \big)  \bigg] }  \\ \times  \bigg[ \underset{1 \leq j^{\prime} \leq s^{\prime}}{\underset{1 \leq j \leq s}{\prod}}            \bigg[    \frac{\mathrm{sin}^{N-\lambda_j} \big( \epsilon_j \big) \mathrm{sin}^{\lambda_j - 1} \big( \epsilon_j - 2 \eta \big) }{ \mathrm{sin}^{N-s} \big(     \epsilon_j + \lambda - \eta     \big) }        \bigg] \\ \times   \bigg[   \frac{\mathrm{sin}^{N-r^{\prime}_j} \big( \epsilon_{j^{\prime}} \big) \mathrm{sin}^{r_{j^{\prime}}-1} \big( \epsilon_{j^{\prime}} - 2 \eta \big)  }{ \mathrm{sin}^{N-s^{\prime} } \big(    \epsilon_{j^{\prime}} + \lambda^{\prime} - \eta      \big)  }   \bigg]               \bigg] \\ \times  \bigg[  \underset{1 \leq j^{\prime} < k^{\prime} \leq s^{\prime}}{\underset{1 \leq j < k \leq s}{\prod}} \bigg[              \mathrm{sin}^{-1} \big( \epsilon_j - \epsilon_k + 2 \eta \big)                   \bigg] \\ \times   \bigg[     \mathrm{sin}^{-1}  \big( \epsilon_{j^{\prime}} - \epsilon_{k^{\prime}} + 2 \eta \big)       \bigg]    \bigg]       \text{, }
\end{align*}

\noindent for some $s^{\prime\prime}$, and $s^{\prime\prime\prime}$, corresponding to the inhomogeneous determinantal representation before taking the homogeneous limit. After taking the homogeneous limit and sending all of the spectral parameters to the same value,  and applying the intertwinning operation to operators $A, B, \cdots, I$ of the product representation,

\[
      \begin{bmatrix}
 A \big( \underline{u} \big) & D \big( \underline{u} \big)  & G \big( \underline{u} \big) \\ B \big(\underline{u} \big) & E \big( \underline{u} \big) & H \big( \underline{u} \big)  \\ C \big( \underline{u} \big)  &  F \big( \underline{u} \big) & I \big( \underline{u} \big) 
\end{bmatrix}   \text{, } 
\]

\noindent one obtains, as was the case for the bottom partition function of the 20-vertex model, a prefactor to the partition function,

\begin{align*}
 Z^{20V}_{n}  \equiv  Z^{20V}_{\textbf{T}_n} \Longleftrightarrow Z^{20V}_{\textbf{T}_n}  \longrightarrow Z^{20V}_{\textbf{T}} \Longleftrightarrow n \longrightarrow + \infty  \text{, }
\end{align*}

\noindent which, when restricted to a finite volume of $\textbf{T}$, takes the form,

\begin{align*}
    Z^{20V}_{\textbf{T}}  \bigg|_{T \subsetneq \textbf{T}}         \equiv    Z^{20V}_{\textbf{T}} \bigg|_{\textbf{T} \backslash [  [N-s^{\prime},   s^{\prime} + 1]  \times   [N-s^{\prime},  s^{\prime\prime}  + 1]  \times [s^{\prime},s^{\prime\prime}]]  }     \equiv Z^{20V}_{\textbf{T} \backslash [  [N-s^{\prime},   s^{\prime} + 1]  \times   [N-s^{\prime\prime},  s^{\prime\prime}  + 1]  \times [ s^{\prime}, s^{\prime\prime} ]]}               \text{, }
\end{align*}

\noindent The prefactor to the restricted partition function takes the form,

\begin{align*}
 {\underset{2 \leq j \leq s}{\prod}} \bigg[  \underset{\alpha_{\underline{i}+\underline{1}} \neq \alpha_{\underline{i}} }{\underset{1 \leq  \alpha_{\underline{i}+\underline{1}} \leq r_{\underline{j}} }{\sum}}  \bigg[    \underset{\underline{1} \leq \underline{i} \leq \underline{s}}{\prod}    \bigg[         \underset{s+1 \leq z^{\prime} \leq N}{\prod }   a  \big( \underline{i}, z^{\prime} \big)            \bigg] \underset{\underline{1} \leq \underline{i} \leq \underline{s}}{\prod} \frac{g \big( \lambda_{\alpha_{\underline{i}}}, \lambda_{r_j}        \big)}{f \big( \lambda_{\alpha_{\underline{i}}}, \lambda_{r_j} \big) }            \bigg]     \\ \times      \underset{1 \leq j^{\prime} \leq s}{\prod} \bigg[        \underset{1 \leq \beta_{j^{\prime}} \leq r_{j^{\prime}}}{\prod}   \frac{g \big( \lambda_{\alpha_{\underline{1}}}, \lambda_{\beta_{j^{\prime}}} \big)}{ f \big( \lambda_{\alpha_{\underline{1}}}, \lambda_{\beta_{j^{\prime}}} \big)}                             \bigg] 
 \bigg]                \bigg]    {\underset{2 \leq k \leq s^{\prime} }{\prod}}  \bigg[   \underset{\alpha_{\underline{j}+1} \neq \alpha_{\underline{j}} }{\underset{1 \leq  \alpha_{\underline{i}+1} \leq r^{\prime}_{\underline{k}} }{\sum}}      \bigg[                    \underset{\underline{1} \leq \underline{j}  \leq \underline{s^{\prime}} }{\prod}  \\ \times      \bigg[         \underset{s^{\prime} + 1 \leq z^{\prime\prime} \leq N}{\prod}  a   \big( \underline{j} , z^{\prime\prime} \big)      \bigg]  \underset{\underline{1} \leq \underline{k} \leq \underline{s^{\prime}}}{\prod} \frac{g \big( \lambda_{\alpha_{\underline{i}}}, \lambda_{r_j}        \big)}{f \big( \lambda_{\alpha_{\underline{i}}}, \lambda_{r_j} \big) }            \bigg]   \underset{1 \leq k^{\prime} \leq s^{\prime}}{\prod} \bigg[        \underset{1 \leq \beta_{k^{\prime}} \leq r_{k^{\prime}}}{\prod}   \frac{g \big( \lambda_{\alpha_{\underline{1}}}, \lambda_{\beta_{k^{\prime}}} \big)}{ f \big( \lambda_{\alpha_{\underline{1}}}, \lambda_{\beta_{k^{\prime}}} \big)}                          \bigg] 
 \bigg]                 \bigg]                \text{, }
\end{align*}

\noindent from the fact that, for the first spectral parameter when $i \equiv 0$, and when $j \equiv 0$, yields the superposition,

\begin{align*}
   \bigg\{  \underset{2 \leq j \leq s^{\prime}}{\prod} \bigg[  \underset{\alpha_{\underline{i}+\underline{1}} \neq \alpha_{\underline{i}} }{\underset{1 \leq  \alpha_{\underline{i}+\underline{1}} \leq r_{\underline{j}} }{\sum}}  \bigg[    \underset{\underline{1} \leq \underline{i} \leq \underline{s}}{\prod}    \bigg[         \underset{s+1 \leq z^{\prime} \leq N}{\prod }   a  \big( \underline{i}, z^{\prime} \big)            \bigg] \underset{\underline{1} \leq \underline{i} \leq \underline{s}}{\prod} \frac{g \big( \lambda_{\alpha_{\underline{i}}}, \lambda_{r_j}        \big)}{f \big( \lambda_{\alpha_{\underline{i}}}, \lambda_{r_j} \big) }            \bigg]      \\ \times      \underset{1 \leq j^{\prime} \leq s}{\prod} \bigg[        \underset{1 \leq \beta_{j^{\prime}} \leq r_{j^{\prime}}}{\prod}   \frac{g \big( \lambda_{\alpha_{\underline{1}}}, \lambda_{\beta_{j^{\prime}}} \big)}{ f \big( \lambda_{\alpha_{\underline{1}}}, \lambda_{\beta_{j^{\prime}}} \big)}                             \bigg] 
\bigg]       +   \underset{1 \leq j \leq 2}{\prod} \bigg[  \underset{\alpha_{\underline{i}+\underline{1}} \neq \alpha_{\underline{i}} }{\underset{1 \leq  \alpha_{\underline{i}+\underline{1}} \leq r_{\underline{j}} }{\sum}}  \bigg[    \underset{\underline{1} \leq \underline{i} \leq \underline{s}}{\prod}    \bigg[         \underset{s+1 \leq z^{\prime} \leq N}{\prod }   a  \big( \underline{i}, z^{\prime} \big)            \bigg]     \\ \times \underset{\underline{1} \leq \underline{i} \leq \underline{s}}{\prod} \frac{g \big( \lambda_{\alpha_{\underline{i}}}, \lambda_{r_j}        \big)}{f \big( \lambda_{\alpha_{\underline{i}}}, \lambda_{r_j} \big) }            \bigg]        \underset{1 \leq j^{\prime} \leq s}{\prod} \bigg[        \underset{1 \leq \beta_{j^{\prime}} \leq r_{j^{\prime}}}{\prod}   \frac{g \big( \lambda_{\alpha_{\underline{1}}}, \lambda_{\beta_{j^{\prime}}} \big)}{ f \big( \lambda_{\alpha_{\underline{1}}}, \lambda_{\beta_{j^{\prime}}} \big)}                             \bigg] 
 \bigg] \bigg\}     \bigg\{ {\underset{2 \leq k \leq s^{\prime} }{\prod}}  \bigg[   \underset{\alpha_{\underline{j}+1} \neq \alpha_{\underline{j}} }{\underset{1 \leq  \alpha_{\underline{i}+1} \leq r^{\prime}_{\underline{k}} }{\sum}}      \bigg[                    \underset{\underline{1} \leq \underline{j}  \leq \underline{s^{\prime}} }{\prod} \\ \times       \bigg[         \underset{s^{\prime} + 1 \leq z^{\prime\prime} \leq N}{\prod}  a   \big( \underline{j} , z^{\prime\prime} \big)      \bigg]  \underset{\underline{1} \leq \underline{k} \leq \underline{s^{\prime}}}{\prod} \frac{g \big( \lambda_{\alpha_{\underline{i}}}, \lambda_{r_j}        \big)}{f \big( \lambda_{\alpha_{\underline{i}}}, \lambda_{r_j} \big) }            \bigg]   \underset{1 \leq k^{\prime} \leq s^{\prime}}{\prod} \bigg[        \underset{1 \leq \beta_{k^{\prime}} \leq r_{k^{\prime}}}{\prod}   \frac{g \big( \lambda_{\alpha_{\underline{1}}}, \lambda_{\beta_{k^{\prime}}} \big)}{ f \big( \lambda_{\alpha_{\underline{1}}}, \lambda_{\beta_{k^{\prime}}} \big)}                          \bigg] 
 \bigg] \\  +   {\underset{1 \leq k \leq 2}{\prod}}  \bigg[   \underset{\alpha_{\underline{j}+1} \neq \alpha_{\underline{j}} }{\underset{1 \leq  \alpha_{\underline{i}+1} \leq r^{\prime}_{\underline{k}} }{\sum}}      \bigg[                    \underset{\underline{1} \leq \underline{j}  \leq \underline{s^{\prime}} }{\prod}   \bigg[         \underset{s^{\prime} + 1 \leq z^{\prime\prime} \leq N}{\prod}  a   \big( \underline{j} , z^{\prime\prime} \big)      \bigg]   \underset{\underline{1} \leq \underline{k} \leq \underline{s^{\prime}}}{\prod} \frac{g \big( \lambda_{\alpha_{\underline{i}}}, \lambda_{r_j}        \big)}{f \big( \lambda_{\alpha_{\underline{i}}}, \lambda_{r_j} \big) }            \bigg] \\ \times   \underset{1 \leq k^{\prime} \leq s^{\prime}}{\prod} \bigg[        \underset{1 \leq \beta_{k^{\prime}} \leq r_{k^{\prime}}}{\prod}   \frac{g \big( \lambda_{\alpha_{\underline{1}}}, \lambda_{\beta_{k^{\prime}}} \big)}{ f \big( \lambda_{\alpha_{\underline{1}}}, \lambda_{\beta_{k^{\prime}}} \big)}                          \bigg] 
 \bigg]             \bigg\}         \text{, }
\end{align*}

\noindent corresponding to the application of the intertwining operation provided for operators in the product representation of the 20-vertex model transfer matrix. Altogether, taking the product over $0 \leq j \leq s$, and also over $1 \leq k \leq s^{\prime}$, yields,

\begin{align*}
    {\underset{1 \leq j \leq s}{\prod}} \bigg[  \underset{\alpha_{\underline{i}+\underline{1}} \neq \alpha_{\underline{i}} }{\underset{1 \leq  \alpha_{\underline{i}+\underline{1}} \leq r_{\underline{j}} }{\sum}}  \bigg[    \underset{\underline{1} \leq \underline{i} \leq \underline{s}}{\prod}    \bigg[         \underset{s+1 \leq z^{\prime} \leq N}{\prod }   a  \big( \underline{i}, z^{\prime} \big)            \bigg] \underset{\underline{1} \leq \underline{i} \leq \underline{s}}{\prod} \frac{g \big( \lambda_{\alpha_{\underline{i}}}, \lambda_{r_j}        \big)}{f \big( \lambda_{\alpha_{\underline{i}}}, \lambda_{r_j} \big) }            \bigg]          \underset{1 \leq j^{\prime} \leq s}{\prod}  \\ \times \bigg[        \underset{1 \leq \beta_{j^{\prime}} \leq r_{j^{\prime}}}{\prod}   \frac{g \big( \lambda_{\alpha_{\underline{1}}}, \lambda_{\beta_{j^{\prime}}} \big)}{ f \big( \lambda_{\alpha_{\underline{1}}}, \lambda_{\beta_{j^{\prime}}} \big)}                           \bigg] 
 \bigg]                 \bigg]    {\underset{1 \leq k \leq s^{\prime} }{\prod}}  \bigg[   \underset{\alpha_{\underline{j}+1} \neq \alpha_{\underline{j}} }{\underset{1 \leq  \alpha_{\underline{i}+1} \leq r^{\prime}_{\underline{k}} }{\sum}}      \bigg[                    \underset{\underline{1} \leq \underline{j}  \leq \underline{s^{\prime}} }{\prod}    \bigg[         \underset{s^{\prime} + 1 \leq z^{\prime\prime} \leq N}{\prod}  a   \big( \underline{j} , z^{\prime\prime} \big)      \bigg] \\ \times    \underset{\underline{1} \leq \underline{k} \leq \underline{s^{\prime}}}{\prod} \frac{g \big( \lambda_{\alpha_{\underline{i}}}, \lambda_{r_j}        \big)}{f \big( \lambda_{\alpha_{\underline{i}}}, \lambda_{r_j} \big) }            \bigg]  \underset{1 \leq k^{\prime} \leq s^{\prime}}{\prod}  \bigg[        \underset{1 \leq \beta_{k^{\prime}} \leq r_{k^{\prime}}}{\prod}   \frac{g \big( \lambda_{\alpha_{\underline{1}}}, \lambda_{\beta_{k^{\prime}}} \big)}{ f \big( \lambda_{\alpha_{\underline{1}}}, \lambda_{\beta_{k^{\prime}}} \big)}                            \bigg] 
 \bigg]                 \bigg]                         \text{. }
\end{align*}

\noindent For the side partition function of the 20-vertex model, the expression above will be shown to equal the restriction of the side partition function, which relies upon taking the summation over $j$, $\alpha$, $i$ and $z^{\prime}$, and also over $k$, $\alpha$, $j$ and $z^{\prime\prime}$. The representation that is obtained after applying the intertwinning operation to,

\begin{align*}
    \underline{Z^{\mathrm{Side}}_{r^{\prime}_1 , \cdots , r^{\prime}_{s^{\prime}} , r^{\prime\prime}_1 , \cdots , r^{\prime\prime}_{s^{\prime\prime}}}}  =  \bigg[         c^{s^{\prime}+s^{\prime\prime}} a^{( s^{\prime} + s^{\prime\prime} )  (2N-2)}          \bigg[ \underset{1 \leq j \leq s^{\prime}}{\prod}   t^{r^{\prime}_j}              \bigg] \bigg[ \underset{1 \leq k \leq s^{\prime\prime}}{\prod}      \big( t^{\prime}\big)^{r^{\prime\prime}_k}      \bigg]   \text{ } \bigg] \bigg\{ \underset{\mathscr{S}^{\prime}_1}{\oint}  \times   \cdots \times     \underset{\mathscr{S}^{\prime}_{s^{\prime}+s^{\prime\prime}}}{\oint}       \\ \times      \bigg\{   \bigg[ \underset{1\leq j \leq s^{\prime}}{\prod}     \frac{w^{r^{\prime}_j-1}}{\big( w_j - 1 \big)^s}                \bigg] 
   \bigg[ \underset{1\leq k \leq s^{\prime\prime}}{\prod}               \frac{\big( w^{\prime}\big)^{r^{\prime\prime}_k-1}}{\big( w^{\prime\prime}_k - 1 \big)^{s^{\prime\prime}}}     \bigg]             \bigg[ \underset{1 \leq j < k \leq s^{\prime}}{\prod}     \big[ \big( w_j - w_k \big) \big( t^2 w_j w_k \\ - 2 \Delta t w_j +1 \big)  \big]               \bigg]   \bigg[ \underset{1 \leq j^{\prime} < k^{\prime} \leq s^{\prime\prime}}{\prod}         \big[ \big( w^{\prime}_j - w^{\prime}_k \big)  \big( t^2 w^{\prime}_j w^{\prime}_k - 2 \Delta t w^{\prime}_j   +1 \big)  \big]              \bigg] \bigg\}  \frac{d^{s^{\prime}} w}{\big( 2 \pi i \big)^s} \\ \times   \frac{d^{s^{\prime\prime}} w^{\prime} }{\big( 2 \pi i \big)^{s^{\prime}}}   \bigg\} \text{, }
\end{align*}

\noindent which, as in the case for the top partition function of the 20-vertex model, factorizes into the product,

\begin{align*}
       Z^{\mathrm{Side},6V}_{r^{\prime}_1 , \cdots , r^{\prime}_{s^{\prime}}} Z^{\mathrm{Side}, 6V}_{r^{\prime\prime}_1 , \cdots , r^{\prime\prime}_{s^{\prime\prime}}}          \text{, } 
\end{align*}

\noindent of partition functions for the 6-vertex model, from which we conclude the argument. \boxed{}

\subsubsection{\underline{(2)}}

\noindent For the second step of the argument, to manipulate the top and bottom partition functions for obtaining the desired representation for nonlocal correlations, from the contour integral expression provided in (1) corresponding to the bottom partition function, the quantity, 

\begin{align*}
   Z^{20V,\mathrm{Bottom}}_{r_1 , \cdots , r_s , r^{\prime}_1 , \cdots , r^{\prime}_s}   \text{, }
\end{align*}

\noindent which can be decomposed into the following contributions, the first of which proportional to applying the intertwining operation to,

\begin{align*}
  \bra{\Uparrow} \bigg[  \bigg[   \underset{1 \leq \alpha \leq r^{\prime}_s}{\prod}     G_{\alpha} \big( \underline{u} \big)     \bigg]  D_{r^{\prime}_s} \big( \underline{u} \big) \bigg[ \underset{1 \leq \alpha \leq r_s}{\prod}        A_{\alpha} \big( \underline{u} \big)      \bigg]   \cdots    \times \bigg[ \underset{1 \leq \alpha \leq r^{\prime}_1}{\prod}     G_{\alpha} \big( \underline{u} \big)                \bigg]      D_{r^{\prime}_1} \big( \underline{u} \big)  \\ \times    \bigg[    \underset{1 \leq \alpha_s \leq r_1}{\prod}   A_{\alpha} \big( \underline{u} \big)    \bigg]  \bigg]   \ket{\Rightarrow}  \text{, }
\end{align*}

\noindent and the second of which is proportional to applying the intertwinning operation to,

\begin{align*}
   \bra{\Uparrow}      \bigg[  \bigg[ \underset{1 \leq \alpha \leq r^{\prime}_s}{\prod}  G_{\alpha} \big( \underline{u} \big)  \bigg]  H_{r^{\prime}_s} \big( \underline{u} \big) \bigg[  \underset{1 \leq \alpha \leq r_s}{\prod}            A_{\alpha} \big( \underline{u} \big)        \bigg] \cdots      \times  \bigg[  \underset{1 \leq \alpha \leq r^{\prime}_1}{\prod} G_{\alpha} \big( \underline{u} \big)  \bigg]       H_{r^{\prime}_1} \big( \underline{u} \big) \\ \times \bigg[ \underset{1 \leq \alpha r_1}{\prod} A_{\alpha} \big( \underline{u} \big)  \bigg]           \bigg]        \ket{\Downarrow}   \text{, }
\end{align*}

\noindent introduced in the previous section, under the action of the operators $A \big( \lambda_r , \lambda_{r^{\prime}} \big)$, and $B \big( \lambda_{\beta} , \lambda_{\beta^{\prime}} \big)$, would read,

\begin{align*}
  {\underset{2 \leq j \leq s}{\prod}} \bigg[  \underset{\alpha_{\underline{i}+\underline{1}} \neq \alpha_{\underline{i}} }{\underset{1 \leq  \alpha_{\underline{i}+\underline{1}} \leq r_{\underline{j}} }{\sum}}  \bigg[    \underset{\underline{1} \leq \underline{i} \leq \underline{s}}{\prod}    \bigg[         \underset{s+1 \leq z^{\prime} \leq N}{\prod }   a  \big( \underline{i}, z^{\prime} \big)            \bigg] \underset{\underline{1} \leq \underline{i} \leq \underline{s}}{\prod} \frac{g \big( \lambda_{\alpha_{\underline{i}}}, \lambda_{r_j}        \big)}{f \big( \lambda_{\alpha_{\underline{i}}}, \lambda_{r_j} \big) }            \bigg]          \underset{1 \leq j^{\prime} \leq s}{\prod} \bigg[        \underset{1 \leq \beta_{j^{\prime}} \leq r_{j^{\prime}}}{\prod}   \frac{g \big( \lambda_{\alpha_{\underline{1}}}, \lambda_{\beta_{j^{\prime}}} \big)}{ f \big( \lambda_{\alpha_{\underline{1}}}, \lambda_{\beta_{j^{\prime}}} \big)}                             \bigg]                 \bigg]                         \text{, }
\end{align*}

\noindent corresponding to the first factor, and,

\begin{align*}
{\underset{2 \leq k \leq s^{\prime} }{\prod}}  \bigg[   \underset{\alpha_{\underline{j}+1} \neq \alpha_{\underline{j}} }{\underset{1 \leq  \alpha_{\underline{i}+1} \leq r^{\prime}_{\underline{k}} }{\sum}}      \bigg[                    \underset{\underline{1} \leq \underline{j}  \leq \underline{s^{\prime}} }{\prod}    \bigg[         \underset{s^{\prime} + 1 \leq z^{\prime\prime} \leq N}{\prod}  a   \big( \underline{j}, z^{\prime\prime} \big)      \bigg]   \underset{\underline{1} \leq \underline{k} \leq \underline{s^{\prime}}}{\prod} \frac{g \big( \lambda_{\alpha_{\underline{i}}}, \lambda_{r_j}        \big)}{f \big( \lambda_{\alpha_{\underline{i}}}, \lambda_{r_j} \big) }            \bigg]  \underset{1 \leq k^{\prime} \leq s^{\prime}}{\prod} \\ \times \bigg[        \underset{1 \leq \beta_{k^{\prime}} \leq r_{k^{\prime}}}{\prod}   \frac{g \big( \lambda_{\alpha_{\underline{1}}}, \lambda_{\beta_{k^{\prime}}} \big)}{ f \big( \lambda_{\alpha_{\underline{1}}}, \lambda_{\beta_{k^{\prime}}} \big)}                             \bigg]                    \bigg]               \text{, }
\end{align*}

\noindent corresponding to the second factor. The forthcoming computations for the second step, similar to those of prefactors of the partition function $Z_{\mathrm{Bottom}}$ in the previous step, are dependent upon contributions from,

\begin{align*}
         \underset{\underline{1} \leq \underline{j}  \leq \underline{s^{\prime}} }{\prod}    \bigg[         \underset{s^{\prime} + 1 \leq z^{\prime\prime} \leq N}{\prod}  a   \big( \underline{j}, z^{\prime\prime} \big)      \bigg]   \underset{\underline{1} \leq \underline{k} \leq \underline{s^{\prime}}}{\prod} \frac{g \big( \lambda_{\alpha_{\underline{i}}}, \lambda_{r_j}        \big)}{f \big( \lambda_{\alpha_{\underline{i}}}, \lambda_{r_j} \big) }           \text{,} \\   \\         \underset{\alpha_{\underline{j}+1} \neq \alpha_{\underline{j}} }{\underset{1 \leq  \alpha_{\underline{i}+1} \leq r^{\prime}_{\underline{k}} }{\sum}}      \bigg[                    \underset{\underline{1} \leq \underline{j}  \leq \underline{s^{\prime}} }{\prod}    \bigg[         \underset{s^{\prime} + 1 \leq z^{\prime\prime} \leq N}{\prod}  a   \big( \underline{j}, z^{\prime\prime} \big)      \bigg]   \underset{\underline{1} \leq \underline{k} \leq \underline{s^{\prime}}}{\prod} \frac{g \big( \lambda_{\alpha_{\underline{i}}}, \lambda_{r_j}        \big)}{f \big( \lambda_{\alpha_{\underline{i}}}, \lambda_{r_j} \big) }            \bigg]    \text{,} \\ \\      \underset{s^{\prime} + 1 \leq z^{\prime\prime} \leq N}{\prod}  a   \big( \underline{j}, z^{\prime\prime} \big)          \text{,} \\ \\    \underset{\underline{1} \leq \underline{k} \leq \underline{s^{\prime}}}{\prod} \frac{g \big( \lambda_{\alpha_{\underline{i}}}, \lambda_{r_j}        \big)}{f \big( \lambda_{\alpha_{\underline{i}}}, \lambda_{r_j} \big) }       \text{,} \\ \\  \underset{1 \leq k^{\prime} \leq s^{\prime}}{\prod} \bigg[        \underset{1 \leq \beta_{k^{\prime}} \leq r_{k^{\prime}}}{\prod}   \frac{g \big( \lambda_{\alpha_{\underline{1}}}, \lambda_{\beta_{k^{\prime}}} \big)}{ f \big( \lambda_{\alpha_{\underline{1}}}, \lambda_{\beta_{k^{\prime}}} \big)}                             \bigg]      \text{,}  
\end{align*}

To extend each one of the expressions to products over $0 \leq j \leq s$, and also to $0 \leq k \leq s^{\prime}$, instead of to $2 \leq j \leq s$, and to $2 \leq k \leq s^{\prime}$, one must add in additional terms, the first of which, for the product over $j$ takes the form,

\begin{align*}
 \underset{\alpha_0 \neq 0}{\underset{0 \leq j \leq s}{\prod}} \bigg[  \underset{\alpha_{\underline{i}+\underline{1}} \neq \alpha_{\underline{i}} }{\underset{1 \leq  \alpha_{\underline{i}+\underline{1}} \leq r_{\underline{j}} }{\sum}}  \bigg[    \underset{\underline{1} \leq \underline{i} \leq \underline{s}}{\prod}    \bigg[         \underset{s+1 \leq z^{\prime} \leq N}{\prod }   a  \big( \underline{i}, z^{\prime} \big)            \bigg] \underset{\underline{1} \leq \underline{i} \leq \underline{s}}{\prod} \frac{g \big( \lambda_{\alpha_{\underline{i}}}, \lambda_{r_j}        \big)}{f \big( \lambda_{\alpha_{\underline{i}}}, \lambda_{r_j} \big) }            \bigg]          \underset{1 \leq j^{\prime} \leq s}{\prod} \bigg[        \underset{1 \leq \beta_{j^{\prime}} \leq r_{j^{\prime}}}{\prod}   \frac{g \big( \lambda_{\alpha_{\underline{1}}}, \lambda_{\beta_{j^{\prime}}} \big)}{ f \big( \lambda_{\alpha_{\underline{1}}}, \lambda_{\beta_{j^{\prime}}} \big)}                             \bigg]                \bigg]    \text{, }
\end{align*}

\noindent and the second of which, for the product over $k$, takes the form,

\begin{align*}
     \underset{\alpha_0 \neq 0}{\underset{0 \leq k \leq s^{\prime} }{\prod}}  \bigg[   \underset{\alpha_{\underline{j}+1} \neq \alpha_{\underline{j}} }{\underset{1 \leq  \alpha_{\underline{i}+1} \leq r^{\prime}_{\underline{k}} }{\sum}}      \bigg[                    \underset{\underline{1} \leq \underline{j}  \leq \underline{s^{\prime}} }{\prod}    \bigg[         \underset{s^{\prime} + 1 \leq z^{\prime\prime} \leq N}{\prod}  a   \big( \underline{j}, z^{\prime\prime} \big)      \bigg]   \underset{\underline{1} \leq \underline{k} \leq \underline{s^{\prime}}}{\prod} \frac{g \big( \lambda_{\alpha_{\underline{i}}}, \lambda_{r_j}        \big)}{f \big( \lambda_{\alpha_{\underline{i}}}, \lambda_{r_j} \big) }            \bigg]  \underset{1 \leq k^{\prime} \leq s^{\prime}}{\prod} \bigg[        \underset{1 \leq \beta_{k^{\prime}} \leq r_{k^{\prime}}}{\prod}   \frac{g \big( \lambda_{\alpha_{\underline{1}}}, \lambda_{\beta_{k^{\prime}}} \big)}{ f \big( \lambda_{\alpha_{\underline{1}}}, \lambda_{\beta_{k^{\prime}}} \big)}                             \bigg]                     \bigg]                                                   \text{. }
\end{align*}

\noindent Altogether, this yields the desired expression for the prefactor to the partition function by taking the product of the two terms above.

\bigskip

\noindent The same arguments can be applied to the contour integral representation of the side partition for the 20-vertex model. That is, following all steps of the computation in the second item yields another representation for the side partition function, which is precisely given by manipulating one side partition function to obtain the partition function for the other side of $\textbf{T}$. 

\subsubsection{\underline{(3)}}

\noindent For the third step of the argument, after having manipulated one partition function to obtain another partition function, we conclude the argument for obtaining the final representation for the 20-vertex model partition function. At the center of the third, and final, step, we express an $s$, and $s+s^{\prime}$ fold product of contour integrals for correlation functions. Furthermore, to pass to the homogenized limit of correlation functions, given contour integral representations for $Z_{\mathrm{Top}}$, $Z_{\mathrm{Bottom}}$ and $Z_{\mathrm{Side}}$, we make use of a suitably defined function $\omega$, which in the case of the 6-vertex model was employed to perturbatively expand contour integral representations up to some sufficiently small parameter $\epsilon$. For the choice of another sufficiently small parameter, the contour integral representations for partition functions, and correlation functions, of the 20-vertex model can be used to pass to a homogenized limit.

In the first instance, given the representation for the bottom partition function, provided in (1), can be manipulated to obtain a contour integral representation of the form for the domain-wall 6-vertex model, {\color{blue}[7]},

\begin{align*}
 \underline{Z^{\mathrm{Top}}_{r_1, \cdots, r_s, r^{\prime}_1, \cdots, r^{\prime}_{s^{\prime}}}}  =  \bigg[ c^{s} a^{ s (N-1) } \bigg[ \underset{1 \leq j \leq s}{\prod}   t^{r_j}              \bigg]  \bigg] \bigg\{  \underset{\mathscr{C}_1}{\oint}  \times \cdots \times     \underset{\mathscr{C}_s}{\oint}         \bigg\{  \bigg[ \underset{1\leq j \leq s}{\prod}     \frac{w^{r_j-1}}{\big( w_j - 1 \big)^s}                \bigg]             \\ \times      \bigg[ \underset{1 \leq j < k \leq s}{\prod}     \big[ \big( w_j  - w_k \big)  \big( t^2 w_j w_k - 2 \Delta t w_j +1 \big)  \big]               \bigg]  \bigg\}  \frac{d^s w}{\big( 2 \pi i \big)^s}     \bigg\} \text{, }
\end{align*}

\noindent for the top partition function, which can be used to obtain the representation,

\begin{align*}
    \underline{Z^{\mathrm{Top}}_{r_1 , \cdots , r_s , r^{\prime}_1 , \cdots , r^{\prime}_{s^{\prime}}}}  =  \bigg[         c^{s+s^{\prime}} a^{( s + s^{\prime} )  (2N-2)}          \bigg[ \underset{1 \leq j \leq s}{\prod}   t^{r_j}              \bigg] \bigg[ \underset{1 \leq k \leq s^{\prime}}{\prod}      \big( t^{\prime}\big)^{r^{\prime}_k}      \bigg]   \text{ } \bigg] \bigg\{ \underset{\mathscr{S}_1}{\oint}  \times \cdots \times     \underset{\mathscr{S}_{s+s^{\prime}}}{\oint}   \\ \times        \bigg\{   \bigg[ \underset{1\leq j \leq s}{\prod}     \frac{w^{r_j-1}}{\big( w_j - 1 \big)^s}                \bigg]      \bigg[ \underset{1\leq k \leq s^{\prime}}{\prod}               \frac{\big( w^{\prime}\big)^{r^{\prime}_k-1}}{\big( w^{\prime}_k - 1 \big)^{s^{\prime}}}     \bigg]               \bigg[ \underset{1 \leq j < k \leq s}{\prod}     \big[ \big( w_j - w_k \big) \big( t^2 w_j w_k \\ - 2 \Delta t w_j +1 \big)  \big]               \bigg]   \bigg[ \underset{1 \leq j^{\prime} < k^{\prime} \leq s^{\prime}}{\prod}         \big[ \big( w^{\prime}_j - w^{\prime}_k \big)  \big( t^2 w^{\prime}_j w^{\prime}_k - 2 \Delta t w^{\prime}_j   +1 \big)  \big]              \bigg] \bigg\}  \frac{d^s w}{\big( 2 \pi i \big)^s} \\ \times \frac{d^{s^{\prime}} w^{\prime} }{\big( 2 \pi i \big)^{s^{\prime}}}  \bigg\}   \text{, }
\end{align*}

\noindent corresponding to the bottom partition function, in which one would like to demonstrate how the prefactor to the 20-vertex partition function, and product terms. Before demonstrating how the series of associations can be formulated between contour integral and determinantal representations, introduce, along the lines of quantities introduced for obtaining the desired contour integral representation of the 6-vertex model, {\color{blue}[7]},

\begin{align*}
     \omega \big( \epsilon_j \big) \equiv    \frac{a}{b}  \frac{\mathrm{sin} \big( \epsilon \big)}{\mathrm{sin} \big( \epsilon - 2 \eta \big) } \text{, } \\     \widetilde{\omega \big( \epsilon_j \big) } \equiv \frac{b}{a}  \frac{\mathrm{sin} \big( \epsilon \big)}{\mathrm{sin} \big( \epsilon + 2 \eta \big) }   \text{ , }
\end{align*}

\noindent for $\epsilon$ taken to be sufficiently small. The purpose of introducing the functions $\omega$ and $\widetilde{\omega}$ above is for transforming certain products over spectral parameters $s$ and $s^{\prime}$, from which the determinantal representation will be transformed into a contour integral representation.

\bigskip

\noindent \textit{Proof of Corollary}. The set of associations for obtaining the desired contour integral representation from the representation of the top partition function above can be related through the series of associations:

\begin{align*}
 \mathrm{det} \big\{  \cdot \big\}   \longleftrightarrow           \underset{\mathscr{S}_1}{\oint} \times \cdots \times \underset{\mathscr{S}_{s+s^{\prime}}}{\oint} \big\{ \cdot \big\}  \frac{d^s w}{\big( 2 \pi i \big)^s} \frac{d^{s^{\prime}} w^{\prime} }{\big( 2 \pi i \big)^{s^{\prime}}}       \text{, } \\ \\  \bigg[   \underset{1 \leq j \leq s}{\prod}  \frac{1}{z^{r_j}}   \bigg] \bigg[      \underset{1 \leq j^{\prime} \leq s^{\prime}}{\prod}  \frac{1}{z^{r_{j^{\prime}}}}    \bigg]      \longleftrightarrow    \bigg[ \underset{1 \leq j \leq s}{\prod}    \frac{\omega \big( \epsilon_{j} \big)^{N-r_j - s + j }  \widetilde{\omega \big( \epsilon_j \big)}^{s-j} }{\big[ \omega \big( \epsilon_j \big) - 1 \big]^{N-s}}   \bigg] \bigg[ \underset{1 \leq j^{\prime} \leq s^{\prime}}{\prod}    \frac{\omega \big( \epsilon_{j^{\prime}} \big)^{N-r_{j^{\prime}} - s + j }  \widetilde{\omega \big( \epsilon_j \big)}^{s-j^{\prime}} }{\big[ \omega \big( \epsilon_{j^{\prime}} \big) - 1 \big]^{N-s^{\prime}}}   \bigg]                          \text{, } \\ \\  \bigg[  \underset{1 \leq j < k \leq s}{\prod}   \frac{z_j - z_k}{ t^2 z_j z_k - 2 \Delta t z_j + 1 }  \bigg] \bigg[ \underset{1 \leq j^{\prime} < k^{\prime} \leq s^{\prime}}{\prod}  \frac{z^{\prime}_j - z^{\prime}_k}{ t^2 z^{\prime}_j z^{\prime}_k - 2 \Delta t z^{\prime}_j + 1 } \bigg]          \longleftrightarrow     \bigg[   \underset{1 \leq j < k \leq s}{\prod}                   \big( \widetilde{\omega \big( \epsilon_j \big) }             \omega \big( \epsilon_k \big) -1 \big)^{-1}          \bigg]   \\ \times       \bigg[   \underset{1 \leq j^{\prime} < k^{\prime} \leq s^{\prime}}{\prod}                       \big( \widetilde{\omega \big( \epsilon_{j^{\prime}} \big) }             \omega \big( \epsilon_{k^{\prime}} \big) -1 \big)^{-1}                      \bigg]                       \text{. } 
\end{align*}

\noindent where parameters $\epsilon_j$ are taken to be sufficiently small. As a result of the fact, from previous arguments, that the top partition function for the 20-vertex model is related to the bottom partition function, the series of associations above, for obtaining the final contour integral representation from the determinantal one, entail:

\begin{align*}
        \underset{\mathscr{S}^{\prime}_1}{\oint} \times \cdots \times \underset{\mathscr{S}^{\prime}_{s+s^{\prime}}}{\oint} \big\{ \cdot \big\}  \frac{d^s w}{\big( 2 \pi i \big)^s} \frac{d^{s^{\prime}} w^{\prime} }{\big( 2 \pi i \big)^{s^{\prime}}}     \longleftrightarrow      \underset{\mathscr{C}^{\prime}_1}{\oint} \times \cdots \times \underset{\mathscr{C}^{\prime}_{s+s^{\prime}}}{\oint} \big\{ \cdot \big\}  \frac{d^s w}{\big( 2 \pi i \big)^s} \frac{d^{s^{\prime}} w^{\prime} }{\big( 2 \pi i \big)^{s^{\prime}}}    \text{, }   \\     \\ Z_N                 \longleftrightarrow        \frac{  a^{s(N-1)+s^{\prime}(N-1)}}{a^{\frac{-s ( 2N -s +1) }{2}-  \frac{s^{\prime} ( 2N -s^{\prime} +1) }{2}} b^{-\frac{s(s-3)}{2} - \frac{s^{\prime}(s^{\prime}-3)}{2}}     }   \bigg[ \frac{a}{b} \bigg]^{-r^{\prime}_1 - \cdots - r^{\prime}_{s^{\prime}} - r^{\prime\prime}_1 - \cdots - r^{\prime\prime}_{s^{\prime\prime}}}   \bigg[  \underset{1 \leq j \leq s}{\prod} t^{r_j} \bigg] \bigg[   \underset{1 \leq j^{\prime} \leq s^{\prime}}{\prod} t^{r^{\prime}_j} \bigg]           \text{, } \\     \\            \bigg[ \underset{1 \leq j < k \leq s}{\prod}     \big[ \big( w_j - w_k \big) \big( t^2 w_j w_k - 2 \Delta t w_j +1 \big)  \big]               \bigg] \bigg[ \underset{1 \leq j^{\prime} < k^{\prime} \leq s^{\prime}}{\prod}         \big[ \big( w^{\prime}_j - w^{\prime}_k \big)   \big( t^2 w^{\prime}_j w^{\prime}_k \ - 2 \Delta t w^{\prime}_j   +1 \big)  \big]              \bigg]                                \\     \longleftrightarrow         \bigg[  \underset{1 \leq j < k \leq s}{\prod}   \frac{z_j - z_k}{ t^2 z_j z_k - 2 \Delta t z_j + 1 }  \bigg] \bigg[ \underset{1 \leq j^{\prime} < k^{\prime} \leq s^{\prime}}{\prod}  \frac{z^{\prime}_j - z^{\prime}_k}{ t^2 z^{\prime}_j z^{\prime}_k - 2 \Delta t z^{\prime}_j + 1 } \bigg]   \text{. } 
\end{align*}

\noindent Explicitly, to establish that the collection of sufficiently small parameters $\epsilon$ are prefactors to contour integrals over the set of surfaces $\underset{1 \leq i \leq s+s^{\prime}}{\bigcup} \mathscr{S}_i$, one must formulate a relationship from,

\begin{align*}
  \omega \big( \epsilon_j \big)   \text{, }
\end{align*}

\noindent and from,

\begin{align*}
 \widetilde{\omega \big( \epsilon_j \big)}   \text{, }
\end{align*}

\noindent which takes the form, {\color{blue}[7]},

\begin{align*}
          \frac{b}{c}  \frac{\mathrm{sin} \big( \epsilon - 2 \eta \big)}{\mathrm{sin} \big( \epsilon + \lambda - \eta \big)} = \big[  \omega \big( \epsilon \big) - 1 \big]^{-1}              \text{. }
\end{align*}

\noindent Hence, for the sufficiently small parameters $\epsilon$, the final desired contour integral representation satisfies,

\begin{align*}
     \underline{Z^{\mathrm{Bottom}}_{r_1 , \cdots , r_s , r^{\prime}_1 , \cdots , r^{\prime}_{s^{\prime}}}}  \propto    \text{ } \underset{\mathscr{S}_1}{\oint}  \times \cdots \times     \underset{\mathscr{S}_{s+s^{\prime}}}{\oint}         \bigg\{  \bigg[ \underset{1\leq j \leq s}{\prod}     \frac{w^{r_j-1}}{\big( w_j - 1 \big)^s}                \bigg]     \bigg[ \underset{1\leq k \leq s^{\prime}}{\prod}               \frac{\big( w^{\prime}\big)^{r^{\prime}_k-1}}{\big( w^{\prime}_k - 1 \big)^{s^{\prime}}}     \bigg]            \bigg[ \underset{1 \leq j < k \leq s}{\prod}     \big[ \big( w_j - w_k \big) \\ \times  \big( t^2 w_j w_k - 2 \Delta t w_j +1 \big)  \big]               \bigg]     \bigg[ \underset{1 \leq j^{\prime} < k^{\prime} \leq s^{\prime}}{\prod}         \big[ \big( w^{\prime}_j - w^{\prime}_k \big) \big( t^2 w^{\prime}_j w^{\prime}_k - 2 \Delta t w^{\prime}_j   +1 \big)  \big]              \bigg] \bigg\}  \\ \times  \frac{d^s w}{\big( 2 \pi i \big)^s} \frac{d^{s^{\prime}} w^{\prime} }{\big( 2 \pi i \big)^{s^{\prime}}}            \text{, }
\end{align*}

\noindent where the constant of proportionality for which there is an exact equality takes the form,

\begin{align*}
                c^{s+s^{\prime}} a^{( s + s^{\prime} )  (2N-2)}          \bigg[ \underset{1 \leq j \leq s}{\prod}   t^{r_j}              \bigg] \bigg[ \underset{1 \leq k \leq s^{\prime}}{\prod}      \big( t^{\prime}\big)^{r^{\prime}_k}      \bigg]           \text{. }
\end{align*}

\noindent To perform a similar computation for the remaining side partition function, making use of the two sets of associations above, the first of which relates the determinantal representation to the contour integral representation, and the second of which relates two contour integral representations to each other, implies that the final desired contour integral representation satisfies the proportionality,

\begin{align*}
     \underline{Z^{\mathrm{Side}}_{r^{\prime}_1 , \cdots , r^{\prime}_{s^{\prime}} , r^{\prime\prime}_1 , \cdots , r^{\prime\prime}_{s^{\prime\prime}}} } \propto    \text{ } \underset{\mathscr{S}_1}{\oint}  \times \cdots \times     \underset{\mathscr{S}_{s+s^{\prime}}}{\oint}         \bigg\{  \bigg[ \underset{1\leq j \leq s^{\prime}}{\prod}     \frac{w^{r_j-1}}{\big( w_j - 1 \big)^{s^{\prime}}}                 \bigg]     \bigg[ \underset{1\leq k \leq s^{\prime\prime}}{\prod}               \frac{\big( w^{\prime}\big)^{r^{\prime}_k-1}}{\big( w^{\prime}_k - 1 \big)^{s^{\prime\prime}}}     \bigg]          \\ \times    \bigg[ \underset{1 \leq j < k \leq s}{\prod}     \big[ \big( w_j - w_k \big)   \big( t^2 w_j w_k  - 2 \Delta t w_j +1 \big)  \big]               \bigg] \bigg[ \underset{1 \leq j^{\prime} < k^{\prime} \leq s^{\prime\prime}}{\prod}         \big[ \big( w_{j^{\prime}} - w_{k^{\prime}} \big) \big( t^2 w_{j^{\prime}} w_{k^{\prime}} \\ - 2 \Delta t w_{j^{\prime}}   +1 \big)  \big]              \bigg] \bigg\}  \frac{d^{s^{\prime}} w}{\big( 2 \pi i \big)^s}  \frac{d^{s^{\prime\prime}} w^{\prime} }{\big( 2 \pi i \big)^{s^{\prime\prime}}}            \text{, }
\end{align*}

\noindent where the constant for which the proportionality above is an equality takes the form,

\begin{align*}
            c^{s^{\prime}+s^{\prime\prime}} a^{( s^{\prime} + s^{\prime\prime} )  (2N-2)}          \bigg[ \underset{1 \leq j^{\prime} \leq s^{\prime}}{\prod}   t^{r_{j^{\prime}}}              \bigg] \bigg[ \underset{1 \leq k^{\prime} \leq s^{\prime\prime}}{\prod}      \big( t^{\prime}\big)^{r^{\prime}_{k^{\prime}}}       \bigg]         \text{. }
\end{align*}

\noindent Hence we have obtained the three desired contour integral representations, from the determinantal representation of the partition function for the 20-vertex model, from which we conclude the argument. \boxed{}

\bigskip

\noindent \textit{Proof of Theorem}. The desired representation follows from combining previous arguments together. Specifically, contour integral representations for the restriction of the partition function are obtained from \textbf{Lemma} \textit{1}, \textbf{Lemma} \textit{2}, and \textbf{Lemma} \textit{3}, while the \textbf{Corollary} establishes the residues that appear in contour integral representations of the Top, Bottom, and Side, partition functions simultaneously. After having taken the homogeneized limit so that the prefactor in the determinantal representation is identically $1$, we obtain the desired contour integral representation, from which we conclude the argument. \boxed{}

\section{The emptiness formation probability (EFP)}

\subsection{Significance}

The EFP has been studied under a wide variety of circumstances in Statistical Mechanics for several models under a wide variety of boundary conditions. Given the fact that various ways to partition $\textbf{Z}^2$, or $\textbf{T}$, for the 6-vertex and 20-vertex models can be used to obtain suitable contour integral representations for correlation functions, representations for partition functions that were manipulated in the previous section, $Z_{\mathrm{Top}}$, $Z_{\mathrm{Side}}$ and $Z_{\mathrm{Bottom}}$, also appear in countour integral representations for the EFP. Over $\textbf{T}$, in comparison to over $\textbf{Z}^2$, the EFP determines whether the orientation of all arrows of a 20-vertex configuration point to the right, versus to the left. The orientation of the arrows of 20-vertex configurations, between the $r$ th and $(r+1)$ th line,  physically represents that no interaction is occurring at the vertex that the arrow is pointing away from. On the other hand, if an arrow at any of the vertices of $\textbf{T}$, between the $r$ th and $(r+1)$ th line, were pointing inwards, then there would be no emptiness formation occurring in the finite volume representation of $\textbf{T}$ restricted to that region.

\subsection{Overview}

As described in the introduction, it is of interest to make use of representations for nonlocal correlations so that the emptiness formation probability (EFP) for the 20-vertex model can be studied. As recapitulated for the 6-vertex model under domain-walls in {\color{blue}[7]}, such a probability determines whether all of the arrows are pointing in the left direction within some restricted finite volume of the square lattice, namely,

\begin{align*}
\textbf{P}^{6V}_Z \big[ \text{all arrows point to the right between the } (r) \text{th} \text{ and } (r+1) \text{th} \text{ vertical lines of } Z \big] = 1 \text{, }
\end{align*}

\noindent for some $Z \subsetneq \textbf{Z}^2$. For the 20-vertex model, such a probability would instead take the form,

\begin{align*}
  \textbf{P}^{20V}_T \big[   \text{all arrows point to the right between the } (r) \text{th} \text{ and } (r+1) \text{th} \text{ vertical lines, and out of the page } \\ \text{ between the } (s) \text{th} \text{ and } (s+1) \text{th} \text{ horizontal lines of } T     \big] = 1   \text{, }
\end{align*}

\noindent for some $T \subsetneq \textbf{T}$. In the case of the 20-vertex model over $\textbf{T}$, a similar probability can be studied in light of the contour integral representation introduced in the statement of the main result of the \textbf{Theorem}. Recall that the representation takes the form,

\begin{align*}
    \underline{Z^{\mathrm{Top}}_{r_1 , \cdots , r_s , r^{\prime}_1 , \cdots , r^{\prime}_s }} =  \mathscr{P}  \bigg\{   \underset{\mathscr{S}_1  }{\oint}  \times \cdots \times     \underset{\mathscr{S}_1 }{\oint}   \bigg[ \mathscr{P}_1 \mathscr{P}_2   H_{N,s}   H_{N,s^{\prime}}         \frac{\mathrm{d}^{s^{\prime}} z^{\prime}}{ \big( 2 \pi i \big)^{s^{\prime}}}   \mu_0                  \bigg]     \bigg\}          \text{, } 
\end{align*}

\noindent for objects $\mathscr{P}$, $\mathscr{P}_1$, $\mathscr{P}_2$, $h^{\prime}$, and $\mu_0$ defined previously. In nearly the same way that the contour integral representation of nonlocal correlations for the domain-wall, 6-vertex partition function is manipulated through a series of computations capitulated in $\underline{(1)}$, $\underline{(2)}$, and $\underline{(3)}$, over $\textbf{T}$ the EFP for the 20-vertex model captures whether additional arrows point outwards from the vertex. As such, EFP for the 20-vertex model can be related to the EFP for the 6-vertex model. As two objects that are taken under domain-wall boundary conditions, the EFP demonstrates how contour integral, and determinantal, representations for probabilistic quantities can be defined and asymptotically evalued with methods from Complex Analysis. Despite the fact that the focus of the last section will be to not perform such evaluations with either the residue theorem, or related methods, it certainly remains of interest to not only further explore how such representations can be evaluated, but also for obtaining related representations to the 20-vertex one, which should be of interest to explore for many other models.

Altogether, denote the object, over $\textbf{T}$, as,

\begin{align*}
 \frac{\big( - 1 \big)^{s+s^{\prime}} Z_s Z_{s^{\prime}}}{s! a^{s(s-1)+ s^{\prime}(s^{\prime}-1)} c^{s+s^{\prime}} }      \underset{1 \leq j^{\prime} \leq s^{\prime}}{\underset{1 \leq j \leq s}{\prod}}       \bigg[       \underset{\mathscr{S}_0}{\oint}  \times \cdots \times       \underset{\mathscr{S}_0}{\oint}            \frac{\big[ t^2 z_j  - 2 \Delta t z_j  +1\big]^{s-1}}{z^r_j \big( z_j - 1 \big)^s }   \bigg]   \bigg\{  \underset{\mathscr{S}^{\prime}_0}{\oint}  \times \cdots \times       \underset{\mathscr{S}^{\prime}_0}{\oint}       \\ 
 \times    \bigg[       \frac{\big[ t^2 z_{j^{\prime}}  - 2 \Delta t z_{j^{\prime}}  +1\big]^{s^{\prime}-1}}{z^r_{j^{\prime}} \big( z_{j^{\prime}} - 1 \big)^{s^{\prime}} }       \bigg]    \bigg[    \underset{1 \leq j < k \leq s}{\prod}                          \frac{z_k - z_j}{t^2 z_j z_k - 2 \Delta t z_j + 1 }   \bigg] \bigg[     \underset{1 \leq j^{\prime} < k^{\prime} \leq s^{\prime}}{\prod}      \frac{z_{k^{\prime}} - z_{j^{\prime}}}{t^2 z_{j^{\prime}} z_{k^{\prime}} - 2 \Delta t z_{j^{\prime}} + 1 }    \bigg]   \\ \times \bigg\{ H_{N,s} \big( z_1 , \cdots , z_n \big) \big) H_{s,s}  \big( u \big( z_1 \big) , \cdots , u \big( z_n \big) \big)    H_{N,s^{\prime}} \big( z^{\prime}_1 , \cdots , z^{\prime}_n \big)  \big) H_{s^{\prime},s^{\prime}} \big( u^{\prime} \big( z^{\prime}_1 \big) , \cdots , u^{\prime} \big( z^{\prime}_n \big) \big)  \bigg\} \\ \times \frac{\mathrm{d}^s z }{\big( 2 \pi i \big)^s }  \frac{\mathrm{d}^{s^{\prime}} z }{\big( 2 \pi i \big)^{s^{\prime}}  } \bigg\}    \text{, }
\end{align*}

\noindent corresponding to the EFP of the 20-vertex model, for the coordinates transformations,

\begin{align*}
u  \big(z_1, \cdots, z_n \big) \equiv   \frac{z-1}{\big( t^2 - 2 \Delta t \big) z+1 }                   \text{, } \end{align*}

\begin{align*}
u{\prime}\big(z^{\prime}_1, \cdots, z^{\prime}_n \big) \equiv         \frac{z^{\prime}-1}{\big( t^2 - 2 \Delta t \big) z^{\prime} +1 }              \text{, }
\end{align*}

\noindent on $z_1, \cdots, z_n$, and on $z^{\prime}_1, \cdots, z^{\prime}_n$. As in the case for the EFP for the 6-vertex model, the contour integral representation above is obtained from applying antisymmetrization,

\begin{align*}
 \underset{s, s^{\prime}}{\underset{z_1, \cdots, z_s, z^{\prime}_1, \cdot, z^{\prime}_{s^{\prime}}}{\mathrm{Asym}}}  \bigg[        \frac{\big[ t^2 z_j  - 2 \Delta t z_j  +1\big]^{s-1}}{z^r_j \big( z_j - 1 \big)^s } 
       \frac{\big[ t^2 z_{j^{\prime}}  - 2 \Delta t z_{j^{\prime}}  +1\big]^{s^{\prime}-1}}{z^r_{j^{\prime}} \big( z_{j^{\prime}} - 1 \big)^{s^{\prime}} }       \bigg] \text{, }
\end{align*}

\noindent to the integrand of the contour integral. To this end, the EFP for the 20-vertex model, after having performed the transformation above on the product of integrand factors over $j,j^{\prime}$, admits the decomposition,

\begin{align*}
        \frac{Z_s Z_{s^{\prime}}}{s! \big(s+1 \big)! a^{s(s-1) + s^{\prime}(s^{\prime}-1)} c^{s+s^{\prime}}}         \bigg[ \underset{1 \leq j \leq s}{\prod}                      \frac{\big[ t^2 z_j  - 2 \Delta t z_j  +1\big]^{s-1}}{z^r_j \big( z_j - 1 \big)^s } 
             \bigg]    \bigg[ \underset{1 \leq j^{\prime} \leq s^{\prime}}{\prod}       \frac{\big[ t^2 z_{j^{\prime}}  - 2 \Delta t z_{j^{\prime}}  +1\big]^{s^{\prime}-1}}{z^r_{j^{\prime}} \big( z_{j^{\prime}} - 1 \big)^{s^{\prime}} }          \bigg]     \\ \times     \underset{1 \leq j^{\prime} < k^{\prime} \leq s^{\prime}}{\underset{1 \leq j <k \leq s}{\prod}}             \big( z_k - z_j \big) \big( z_{k^{\prime} } - z_{j^{\prime}} \big) \bigg\{  H_{s,s} \big( u \big( z_1 \big), \cdots , u \big( z_N \big) \big)  H_{s^{\prime},s^{\prime}} \big( u^{\prime} \big( z_1 \big), \cdots , u^{\prime} \big( z_N \big) \big) \bigg\}    \text{. }
\end{align*}

\noindent Given such a product representation as the one above that is obtained from antisymetrization, one may further transform the contour integral presented at the beginning of the EFP section. The transformation introduces a new mapping of the product terms that are taken under $j,j^{\prime}$, in which the transformed representation takes the form,

\begin{align*}
 \frac{\big( - 1 \big)^{s+s^{\prime}} Z_s Z_{s^{\prime}}}{s! a^{s(s-1)+ s^{\prime}(s^{\prime}-1)} c^{s+s^{\prime}} }         \bigg[       \underset{\mathscr{S}_0}{\oint}  \times \cdots \times       \underset{\mathscr{S}_0}{\oint}  \bigg[         \underset{1 \leq j \leq s}{\prod}     \big( w_j - 1 \big)^{-s}     \text{ } \bigg] \underset{1 \leq r_1 < r_2 < \cdots < r_s \leq r }{\sum}   \bigg[            \underset{1 \leq j \leq s}{\prod}         \frac{w^{r_j}-1}{z^{r_j}_j }       \bigg]        \bigg] \\ \times  \bigg[     \underset{\mathscr{S}^{\prime}_0}{\oint}  \times \cdots \times       \underset{\mathscr{S}^{\prime}_0}{\oint}         \bigg[         \underset{1 \leq j^{\prime} \leq s^{\prime}}{\prod}     \big( w_{j^{\prime}} - 1 \big)^{-s^{\prime}}     \text{ } \bigg] \underset{1 \leq r^{\prime}_1 < r^{\prime}_2 < \cdots < r^{\prime}_s \leq r^{\prime} }{\sum}   \bigg[            \underset{1 \leq j^{\prime} \leq s^{\prime}}{\prod}         \frac{w^{r^{\prime}_{j^{\prime}}}-1}{z^{r^{\prime}_{j^{\prime}}}_{j^{\prime}} }       \bigg]     \bigg] \\ 
 \times     \bigg[    \underset{1 \leq j < k \leq s}{\prod}                          \frac{\big( z_k - z_j\big) \big( t^2 w_j w_k - 2 \Delta t w_j + 1 \big) \big(         w_j - w_k      \big) }{t^2 z_j z_k - 2 \Delta t z_j + 1 }   \bigg] \\ \times \bigg[     \underset{1 \leq j^{\prime} < k^{\prime} \leq s^{\prime}}{\prod}                              \frac{\big( z_{k^{\prime}} - z_{j^{\prime}}\big) \big( t^2 w_{j^{\prime}} w_{k^{\prime}} - 2 \Delta t w_{j^{\prime}} + 1 \big) \big(         w_{j^{\prime}} - w_{k^{\prime}}      \big) }{t^2 z_{j^{\prime}} z_{k^{\prime}} - 2 \Delta t z_{j^{\prime}}  + 1 }     \bigg]   \\ \times \bigg\{ H_{N,s} \big( z_1 , \cdots , z_n \big)   H_{N,s^{\prime}} \big( z^{\prime}_1 , \cdots , z^{\prime}_n \big)  \big)  \bigg\} \frac{\mathrm{d}^s z }{\big( 2 \pi i \big)^s }  \frac{\mathrm{d}^{s^{\prime}} z }{\big( 2 \pi i \big)^{s^{\prime}}  }    \text{. }
\end{align*}

\noindent In comparison to the first contour integral representation for the EFP, the terms,

\begin{align*}
 H_{s,s} \big( u \big( z_1 \big), \cdots , u \big( z_N \big) \big)  \equiv  \underset{1 \leq i \leq N}{\bigcup} H_{s,s} \big( \big\{ u \big( z_i \big) \big\} \big)    \text{, } \\   H_{s^{\prime},s^{\prime}} \big( u^{\prime} \big( z_1 \big), \cdots , u^{\prime} \big( z_N \big) \big)  \equiv  \underset{1 \leq i \leq N}{\bigcup} H_{s^{\prime},s^{\prime}} \big( \big\{ u^{\prime} \big( z_i \big) \big\} \big)      \text{, }
\end{align*}

\noindent would contribute to terms in the product representation over spectral parameters.

\bigskip

\noindent Altogether, to evaluate the second representation for the EFP that was obtained from the first representation of the EFP, as with the first contour integral representation presented in the section, make use of the identity,

\begin{align*}
 \bigg[  \underset{- \infty < r_1 < r_2 < \cdots < r_n \leq r}{\sum} \bigg[        \underset{1 \leq j \leq s}{\prod}     X^{-r_j}_j     \bigg]  \bigg] \bigg[   \underset{- \infty < r^{\prime}_1 < r^{\prime}_2 < \cdots < r^{\prime}_n \leq r^{\prime}}{\sum} \bigg[        \underset{1 \leq j^{\prime} \leq s^{\prime}}{\prod}     X^{-r^{\prime}_j}_j     \bigg]        \bigg] \\   =            \bigg[   \underset{1 \leq j \leq s}{\prod}            \frac{1}{X^{r-s+j} \big( 1 - \underset{1 \leq l \leq j}{\prod}  X_l \big)  }  \bigg] \bigg[                 \underset{1 \leq j^{\prime} \leq s^{\prime}}{\prod}            \frac{1}{X^{r^{\prime}-s^{\prime}+j^{\prime}} \big( 1 - \underset{1 \leq l^{\prime} \leq j^{\prime}}{\prod}  X_{l^{\prime}} \big)  }    \bigg]        \text{, }
\end{align*}

\noindent which is obtained from the one-dimensional version of the identity, when multiplying and dividing by,

\begin{align*}
  \underset{- \infty < r^{\prime}_1 < r^{\prime}_2 < \cdots < r^{\prime}_n \leq r^{\prime}}{\sum} \bigg[   \underset{1 \leq j^{\prime} \leq s^{\prime}}{\prod}     X^{-r_j}_j     \bigg]   \text{. }
\end{align*}

\noindent We conclude the section by obtaining the final expression of the contour integral that can be obtained from applying the two-dimensional symmetrization identity above.

\bigskip

\noindent \textit{Proof of Lemma 4}. From the two-dimensional symmetrization identity applied to the integrand of the second contour integral representation, compute,

\begin{align*}
        \underset{\mathscr{S}_0}{\oint}  \times \cdots \times       \underset{\mathscr{S}_0}{\oint}  \bigg[ \bigg[         \underset{1 \leq j \leq s}{\prod}     \big( w_j - 1 \big)^{-s}     \text{ } \bigg] \underset{1 \leq r_1 < r_2 < \cdots < r_s \leq r }{\sum}   \bigg[            \underset{1 \leq j \leq s}{\prod}         \frac{w^{r_j}-1}{z^{r_j}_j }       \bigg]  \bigg]  =        \underset{\mathscr{S}_0}{\oint}  \times \cdots \times       \underset{\mathscr{S}_0}{\oint}    \\ \times   \bigg[ \bigg[    \underset{1 \leq j \leq s}{\prod}     \big( w_j - 1 \big)^{-s}     \text{ } \bigg]                \bigg[   \underset{1 \leq j \leq s}{\prod}            \frac{w^{r_j}}{z^{r-s+j}_j \big( 1 - \underset{1 \leq l \leq j}{\prod}  z_l \big)  }  \bigg]          \frac{\mathrm{d}^s z }{\big( 2 \pi i \big)^s}  \frac{\mathrm{d}^s z}{\big( 2 \pi i \big)^s }    \text{. }
\end{align*}

\noindent One can obtain a corresponding expression for taking the summation and product over other spectral parameters by applying the symmetrization identity once more. Hence the final contour integral representation takes the form,

\begin{align*}
          \underset{\mathscr{S}^{\prime}_1}{\oint}  \times \cdots \times   \underset{\mathscr{S}^{\prime}_1}{\oint}    \bigg[  \frac{\mathrm{d}^s w }{\big( 2 \pi i \big)^s}  \underset{\mathscr{S}^{\prime}_0}{\oint} \times \cdots \times    \underset{\mathscr{S}^{\prime}_0}{\oint}   \bigg[ \underset{1 \leq j \leq s}{\prod}                \frac{w^r_{j}}{\big( w_j - 1 \big)^s z^{r-s+j}_j \big(  \underset{1 \leq l \leq j}{\prod}  \big[ w_l - z_l \big]     \big) }       \bigg]  \\ 
          \times    \bigg[    \underset{1 \leq j < k \leq s}{\prod}                 \frac{\big( w_j - w_k \big) \big( t^2 w_j w_k - 2 \Delta t w_j + 1 \big) \big( z_k - z_j \big) }{t^2 z_j z_k - 2 \Delta t z_j + 1}           \bigg]   \bigg]      \\                             \times      \underset{\mathscr{S}^{\prime}_1}{\oint}  \times \cdots \times   \underset{\mathscr{S}^{\prime}_1}{\oint}   \bigg[                       \frac{\mathrm{d}^{s^{\prime}}  w }{\big( 2 \pi i \big)^{s^{\prime}}}   \underset{\mathscr{S}^{\prime}_0}{\oint} \times \cdots \times    \underset{\mathscr{S}^{\prime}_0}{\oint}   \bigg[ \underset{1 \leq j^{\prime} \leq s^{\prime}}{\prod}                \frac{w^r_{j^{\prime}}}{\big( w_{j^{\prime}} - 1 \big)^{s^{\prime}} z^{r^{\prime}-s^{\prime}+j^{\prime}}_{j^{\prime}} \big(  \underset{1 \leq l \leq j^{\prime}}{\prod}  \big[ w_l - z_l \big]     \big) }                  \bigg]     \\ \times             \bigg[    \underset{1 \leq j^{\prime} < k^{\prime} \leq s^{\prime}}{\prod}                 \frac{\big( w_{j^{\prime}} - w_{k^{\prime}} \big) \big( t^2 w_{j^{\prime}} w_{k^{\prime}} - 2 \Delta t w_{j^{\prime}} + 1 \big) \big( z_{k^{\prime}} - z_{j^{\prime}} \big) }{t^2 z_{j^{\prime}} z_{k^{\prime}} - 2 \Delta t z_{j^{\prime}} + 1}           \bigg]    \bigg]                                                   \\             \times       \bigg\{ H_{N,s} \big( z_1, \cdots, z_s \big) H_{N,s^{\prime}} \big( z^{\prime}_1, \cdots, z^{\prime}_{s^{\prime}} \big)   \bigg\}     \frac{\mathrm{d}^s z}{\big( 2 \pi i \big)^s } \frac{\mathrm{d}^{s^{\prime}} z^{\prime}}{\big( 2 \pi i \big)^{s^{\prime}} }           \text{, }
\end{align*}

\noindent from which we conclude the argument, as the contour integral representation above can be trivially rearranged to obtain the desired representation given in the statement, from which we conclude the argument. \boxed{}

\section{Data Availability Statement}

The manuscript has no associated data.

\section{Conflicts of Interest Statement}

On behalf of all authors, the corresponding author states that there is no conflict of interest.

\nocite{*}
\bibliography{sn-bibliography}

\end{document}